%% file: csg2_harmonic_v4.tex
\documentclass[11pt, a4paper]{article}
\usepackage{jheppub}

\usepackage{amsmath, amsfonts, amssymb}
\usepackage{booktabs}
\usepackage{mathrsfs}
\usepackage{hyperref}

\include{sugracom}
\newcommand{\bbA}{{\mathbb A}}
\newcommand{\bbD}{{\mathbb D}}
\newcommand{\bnabla}{{\bar \nabla}}
\newcommand{\gU}[1]{\ensuremath{{\rm U}(#1)}}
\newcommand{\SU}[1]{\ensuremath{{\rm SU}(#1)}}
\newcommand{\gSU}[1]{\ensuremath{{\rm SU}(#1)}}

\newcommand{\trv}{{\rm \scriptscriptstyle v}}
\newcommand{\trA}{{\rm \scriptscriptstyle A}}
\newcommand{\trD}{{\rm \scriptscriptstyle D}}
\newcommand{\trQ}{{\rm \scriptscriptstyle Q}}
\newcommand{\trw}{{\rm \scriptscriptstyle w}}
\newcommand{\tru}{{\rm \scriptscriptstyle u}}

\newcommand{\rS}{{S^2}}
\newcommand{\loco}{\vert}
\def\Sp(#1){\ensuremath{\mathrm{Sp}(#1)}}

\numberwithin{equation}{section}

\title{On conformal supergravity and harmonic superspace}

\author{Daniel Butter}
\affiliation{Nikhef Theory Group, \\
Science Park 105, 1098 XG Amsterdam, The Netherlands}
\emailAdd{dbutter@nikhef.nl}

\preprint{Nikhef-2015-031}

\arxivnumber{1508.07718}

\abstract{This paper describes a fully covariant approach to harmonic
superspace. It is based on the conformal superspace
description of conformal supergravity
and involves extending the supermanifold $\cM^{4|8}$ by the
tangent bundle of $\mathbb CP^1$.
The resulting superspace
$\cM^{4|8} \times T \mathbb CP^1$
can be identified in a certain gauge with the
conventional harmonic superspace $\cM^{4|8} \times S^2$.
This approach not only makes the connection to projective superspace
transparent, but simplifies calculations in harmonic
superspace significantly by eliminating the need to deal directly
with supergravity prepotentials. As an application of the covariant
approach, we derive from harmonic superspace the full component action for 
the sigma model of a hyperk\"ahler cone coupled to conformal
supergravity. Further applications are also sketched.\\
}

\begin{document}
\maketitle

\section{Introduction}
$\cN=2$ supersymmetric theories in four dimensions face a particular hurdle
relative to their $\cN=1$ cousins: the general matter hypermultiplet cannot
be off-shell without introducing an infinite number of auxiliary fields.
This understandably makes the direct construction of supersymmetric actions --
a straightforward procedure for $\cN=1$ actions even with supergravity couplings
and higher derivatives -- 
significantly more difficult as conventional $\cN=2$ superspace proves insufficient.
Instead, one requires a more elaborate superspace where infinite sets of
auxiliary fields are encoded in a controlled way so that the most general
off-shell actions of hypermultiplets and vector multiplets
may be described. There are two well-developed options:
harmonic superspace and projective superspace.
Harmonic superspace, developed by Galperin, Ivanov, Kalitzin, Ogievetsky
and Sokatchev \cite{GIKOS, GIOS} exploits
an additional bosonic manifold $S^2$, with the infinite auxiliary fields
appearing in a convergent harmonic expansion.
In contrast, the projective superspace approach, constructed by
Karlhede, Lindstr\"om, and Ro\v{c}ek \cite{KLR, LR88, LR90}
(see also the recent reviews \cite{LR:Prop, Kuzenko:PSLectures})
involves an auxiliary $\mathbb CP^1$ where the hypermultiplet is a holomorphic
function near one of the poles, with the auxiliary fields described by coefficients
in a Taylor expansion.\footnote{A general framework for discussing higher $\cN$ analogues
of harmonic and projective superspace involves the so-called $(\cN,p,q)$ superspaces,
introduced by Hartwell and Howe \cite{HoweHartwell:Survey, HartwellHowe:Npq}. These
emphasize the geometrization of the $R$-symmetry group and the nature of the
superconformal transformations, both of which play an important role here.
As in \cite{Butter:CSG4d.Proj}, our discussion corresponds to the case $(2,1,1)$.}

While $S^2$ and $\mathbb CP^1$ describe the same manifold,
the differing nature of the superfields
has important consequences. For example, the respective action principles
on the two spaces are quite different: harmonic actions involve integrals
over the $S^2$ and are completely specified by their Lagrangians,
while projective actions are defined on a contour in $\mathbb CP^1$,
with different contours corresponding (in principle) to different actions
for the same Lagrangian.
But there are other differences between these two approaches which are less obviously
connected with their auxiliary structures.
Quite early on, prepotential superfields were identified within harmonic superspace
both for gauge theories and supergravity \cite{GIKOS}; these enabled a large
body of supergraph calculations involving vector and hypermultiplets.
In contrast, while projective prepotentials appeared in \cite{LR90}
(see \cite{CSG5d, KT-M:N4_SYM_AdS3}
for a discussion of gauge prepotentials on curved
supermanifolds)
projective supergraph calculations in non-abelian gauge theories
have appeared only relatively recently \cite{GonzalezRey:FR1, GonzalezRey:FR2, GonzalezRey:FR3,
JainSiegel:Hypergraphs1, JainSiegel:Hypergraphs2}, while
supergravity prepotentials remain {\it terra incognita}.
This does not prevent the construction of supergravity actions in projective
superspace; to the contrary, an extremely powerful manifestly covariant
method has been developed over the last few years to address general
supergravity-matter systems and their component reduction, first in five dimensions
\cite{KT-M:5DSugra1, KT-M:5DSugra2, KT-M:5DSugra3} and then in four
\cite{KLRT-M1, KLRT-M2, KT-M:DiffReps, Butter:CSG4d.Proj}, building on the initial work of
\cite{Kuzenko:SPH}. Within harmonic superspace, covariant methods have been explored
in two papers \cite{DelamotteKaplan:HHS, GKS:Sugra}, which addressed
how to derive supergravity prepotentials from a covariant supergeometry, but
further applications of harmonic superspace to supergravity systems have mainly
used prepotentials.

The distinction between a prepotential-based approach and a fully covariant method
can be illustrated with a simple example. Take pure (gauged) supergravity consisting
of a single vector multiplet compensator coupled to a single hypermultiplet compensator.
Using the arctic multiplet $\U^+$ of projective superspace as the hypermultiplet,
the action reads\footnote{The projective action for an arctic multiplet
minimally coupled to a vector multiplet and conformal supergravity first appeared
in 5D \cite{KT-M:5DSugra3} where a different but equivalent action principle was employed.
We are using here the reformulation of projective superspace given in \cite{Butter:CSG4d.Proj}.}
\begin{align}\label{eq:ProjAct1}
S = -\frac{1}{4} \int \rd^4x\,\rd^4\q\, \cE\, W^2 
	+ \frac{1}{2\pi} \oint_\cC \rd \tau \int \rd^4x\, \rd^4\q^+\, \cE^{--}
	\Big(2 i \breve \U^+ \U^+\Big)~.
\end{align}
The arctic multiplets in the second term carry charge $g$ under the vector multiplet;
if the vector prepotential were made explicit, this term would be written
$\breve \U^+ e^{g V} \U^+$.
The important feature is that both terms above are manifestly covariant
and defined in any gauge. In particular, one can make arbitrary
conformal supergravity gauge transformations (with arbitrary superfield parameters)
for both actions, and invariance is ensured using the properties of the
respective chiral and analytic measures $\cE$ and $\cE^{--}$ \cite{Butter:CSG4d.Proj}.
In harmonic superspace, the corresponding action involving the harmonic
hypermultiplet $\widehat Q^+$ is rendered with explicit prepotentials as
\begin{align}\label{eq:HarmAct1}
S = -\frac{1}{4} \int \rd \hat u \, \rd^4 \hat x\, \rd^8 \hat \q\,
	\hat E\, V^{++} V^{--}
	+ 2 \int \rd \hat u \, \rd^4 \hat x \, \rd^4\hat \q^+ \Big(
	\widetilde {\widehat Q}{}^+ (\hat\cD^{++} - i g V^{++}) \widehat Q^+
	\Big)~.
\end{align}
One works in the analytic basis for the coordinates, which we have denoted with
hats. In this gauge, explicit gravitational prepotentials appear within the harmonic
covariant derivative $\hat\cD^{++}$; these can in turn be used to construct the
full superspace measure in the analytic basis, denoted $\hat E$ above.
There is no analytic measure as $\widehat Q^+$ is chosen
to transform as a scalar density under analytic diffeomorphisms.

The advantage of the second action, and a prepotential approach in particular, is the
relative ease with which one can calculate superspace equations of motion
and perform quantum calculations. These were major successes of the harmonic
approach, and require the dependence on the prepotentials to be laid bare;
that dependence is obscured in a covariant formulation.
On the other hand, there are a number of advantages of a covariant formulation.
The first action \eqref{eq:ProjAct1} is constructed in a generic gauge,
and its component reduction can be performed in a manifestly covariant manner
(see e.g. \cite{Butter:HKP, BN:CR} for the two pieces).
To reduce the second action \eqref{eq:HarmAct1} to
components, one must adopt a Wess-Zumino gauge for the various
analytic prepotentials, perform the $\theta$ integrals, and then
reconstitute various composite objects such as the covariant derivative,
the spin connection, etc. While this is possible in
principle, to our knowledge it has never been explicitly undertaken
for all terms in any supergravity action; even the most extensive
component treatment \cite{IV:QuatMetrics} of general supergravity-matter
actions in harmonic superspace restricted to bosonic terms.

Of course, for actions like those discussed above, the question of how easy it is to perform
a component reduction is essentially moot, as the results are well-known.
A more interesting question is how to construct new higher-derivative actions 
for hypermultiplets  coupled to supergravity, and to analyze the dependence of
such higher derivative terms on the underlying prepotentials so that one
may analyze supersymmetric equations of motion, supercurrents, and so forth.
For addressing such questions, it is useful to have a formulation of general
supergravity-matter systems with both a covariant and a prepotential description.

Our major goal in this paper will be to provide a covariant reformulation of
supergravity-matter actions in harmonic superspace so that any action, even
a higher-derivative one, can be addressed in a manifestly covariant way.
Because the prepotential approach for harmonic superspace already exists,
we will begin by seeking a manifestly covariant formulation from the outset.
This will cover some similar ground as \cite{DelamotteKaplan:HHS, GKS:Sugra},
but where these authors were concerned with Einstein supergravity
(with two hidden implicit compensators within the supergeometry),
we will build conformal supergravity into the structure group
of superspace from the very beginning. This so-called conformal superspace approach,
which corresponds to the superconformal tensor calculus in components,
offers significant simplifications to calculations: recent applications have
included constructing previously unknown higher-derivative invariants in 5D
as well as the construction of all off-shell 3D $\cN \leq 6$ conformal supergravity
actions, including auxiliary fields \cite{CSG3d2, KNT-M:CSG3d_6}.\footnote{The 3D $\cN\leq 5$
component actions were constructed for the first time using superspace techniques.
The $\cN=6$ action was constructed first in components \cite{NishimuraTanii:CSG3D_6}
and then in superspace.}
In applying this to harmonic superspace, it will permit us to give not only covariant reformulations of
all harmonic superspace actions, but will also allow the construction of 
the covariant component reduction rule for a general analytic Lagrangian --
novel results not found in any previous formulation of harmonic superspace.

Interestingly, the incorporation of the superconformal algebra directly into the
structure group, and the presence of the $\gSU{2}$ $R$-symmetry group in particular,
will necessitate a reinterpretation of just what harmonic superspace actually is.
Before elaborating further, we should pause to answer the following
question: just why should one focus on harmonic superspace when another
formulation -- projective superspace -- already offers fully developed covariant methods?
The answer is that projective and harmonic superspace are actually not intrinsically different
approaches, but possess a quite non-trivial relation. By fully addressing these issues
within harmonic superspace -- in part using inspiration from projective superspace --
we can learn important lessons about both. Recently, Jain and Siegel have argued to interpret
projective superspace as an analytic continuation of
harmonic superspace \cite{JainSiegel:HarmProj}.\footnote{This analytic
continuation is related to a similar approach in twistor theory \cite{Newman}.
An earlier proposal by Kuzenko \cite{Kuzenko:DP} to relate harmonic and projective superspace
works rather differently.} This has proven to be
a robust scheme and has enabled a direct link between
the harmonic and projective descriptions of vector prepotentials \cite{JainSiegel:HarmProj}
and between their respective descriptions of hyperk\"ahler
sigma models \cite{Butter:HarmProj}. 
The main idea is to ``complexify'' the $S^2$ of harmonic superspace to
the tangent bundle of $\mathbb CP^1$, identified as $\mathbb CP^1 \times \mathbb CP^1$
with the anti-diagonal removed.
The two $\mathbb CP^1$ factors possess different $\SU{2}$ isometry groups.
The first is to be identified
with the $R$-symmetry subgroup of the superconformal group,
while the second is a spectator.
The connection between projective and harmonic superfields
can be described concisely as follows:
\begin{itemize}
\item The natural superfields on (complex) harmonic superspace are biharmonic
functions $\cF^{(n,m)}$ with charge $(n,m)$ under the $\gU{1}_\trv \times \gU{1}_\trw$ subgroup
of $\SU{2}_\trv \times \SU{2}_\trw$ of the form (for $n+m\geq 0$)
\begin{align}\label{eq:BiHarmExp}
\cF^{(n,m)} = (v^+, w^-)^n
	\sum_{k=0}^\infty \cF^{(i_1 \cdots i_{n+m+k} j_1 \cdots j_{k})}
	\frac{v_{i_1}^+}{(v^+, w^-)} \cdots \frac{v_{i_{n+m+k}}^+}{(v^+, w^-)}
	w_{j_1}^- \cdots w_{j_k}^-~.
\end{align}
(A similar series exists for $n+m < 0$.)
The series \eqref{eq:BiHarmExp}
is presumed to converge on the so-called real $S^2$ manifold, corresponding
to the diagonal submanifold of $\mathbb CP^1 \times \mathbb CP^1$ where $v^{i+} \propto w^{i+}$.
The anti-diagonal subset of $\mathbb CP^1 \times\mathbb CP^1$
where $v^{i+} \propto w^{i-}$ must be excluded so that the series exists
(at least asymptotically) everywhere; this implies that we are dealing with
the tangent bundle of $\mathbb CP^1$.

\item Associated with every biharmonic function
$\cF^{(n,m)}$ is an arctic superfield $\U^{(n)}$ and an antarctic
superfield $\breve {\bar\U}^{(n)}$ given by
\begin{align}
\U^{(n)} &= \cF^{(n,m)}\vert_{w_i^- = (1,0)} = (v^{\1})^n \sum_{j=0}^\infty \U_j \z^j~, \eol
\breve {\bar\U}^{(n)} &= \cF^{(n,m)}\vert_{w_i^- = (0,1)} = (v^\2)^n \sum_{j=0}^\infty \tilde\U_j (-\z)^{-j}~.
\end{align}
The arctic nature of $\U^{(n)}$ and the antarctic nature of $\breve {\bar\U}^{(n)}$ are
guaranteed because of the presumed convergence of \eqref{eq:BiHarmExp} in the vicinity of $\rS$.
\end{itemize}

This interpretation of harmonic superspace is actually not particularly revolutionary.
As discussed in the original harmonic superspace literature \cite{GIOS:Conformal},
the superconformal group acts on complex harmonics $u^{i\pm}$,
but one performs the harmonic integrals as if they were real. This suggests
(see the comment in chapter 9 of \cite{GIOS}) that the harmonic
$S^2$ should be reinterpreted as lying within $\mathbb CP^1 \times \mathbb CP^1$.
Guided by these old observations and the requirement to reproduce the harmonic-projective
mapping in curved space, a covariant scheme immediately presents itself.
We will begin with the covariant projective superspace of \cite{Butter:CSG4d.Proj} defined on
the supermanifold $\cM^{4|8} \times \SU{2}$. (As fields and operators had
fixed charges in $\gU{1} \subset \SU{2}$, this effectively became
$\cM^{4|8} \times \mathbb CP^1$.)
The auxiliary $\SU{2}$ was identified with the $\SU{2}$ $R$-symmetry group
and non-trivial $R$-symmetry curvature was encoded in the fibering of the
$\SU{2}$ over $\cM^{4|8}$.
We then extend the auxiliary manifold with an additional
\emph{completely rigid} $\SU{2}$ factor, giving the supermanifold
$\cM^{4|8} \times \SU{2} \times \SU{2}$
(effectively $\cM^{4|8} \times T\mathbb CP^1$).
Over this complex harmonic manifold, we will introduce superfields defined
exactly as in the rigid case sketched above.
The virtue of this approach is that it efficiently meets two
goals. First, it gives a covariant formulation that agrees (as we will show)
with the conventional harmonic superspace description whose harmonics are
naturally complex but integrated on a real $S^2$. Second, it
permits the mapping between harmonic and projective superspace to be lifted to
a general curved supermanifold.

To describe the biharmonic space, we will need 
two sets of harmonics $v^{i\pm}$ and $w^{i\pm}$, or equivalently,
complex harmonics $u^{i\pm}$ and additional coordinates $z^{\pm\pm}$ and $z^0$.
This complex harmonic description has already been employed within the harmonic superspace
literature to describe the target space of quaternionic sigma models, where some of
the harmonics are interpreted as compensator fields for the sigma model \cite{GIO:QK}.
We will be using these complex harmonics to instead describe the auxiliary
harmonic manifold of superspace itself. Once this point of view is adopted, we will see
that many curious features of the conventional harmonic approach reveal
themselves quite naturally.

It is worth observing that biprojective superfields
have already been discussed in
\cite{GatesHullRocek, BuscherLindstromRocek, LindstromIvanovRocek}
to describe extended supersymmetric systems in two dimensions (see also
\cite{T-M:2DSugra, KLT-M:3DSugra} for curved superspace applications of biprojective
superfields in 2D and 3D). In these cases,
the $R$-symmetry group is $\SU{2} \times \SU{2}$, and so conventional
projective and harmonic approaches \emph{already} lead to $\mathbb CP^1 \times \mathbb CP^1$.
Constructing a complex harmonic superspace for these cases would seem to lead to
a quadriharmonic space involving $(\mathbb CP^1)^4$.

This paper is organized as follows. In section \ref{sec:CompS2}, we review
some details of harmonic analysis on $T\mathbb CP^1$. Our approach will use
some of the tools and ideas introduced in \cite{GIO:QK}, but our conventions and emphasis will
differ rather extensively. In section \ref{sec:CHS}, we present a concise discussion of the
covariant harmonic superspace $\cM^{4|8} \times \SU{2} \times \SU{2}$ built
on the covariant projective superspace $\cM^{4|8} \times \SU{2}$.
Covariant action principles, including the covariant component reduction,
will be discussed in section \ref{sec:Actions}. These two sections establish
the self-consistency of our approach. Section \ref{sec:HSS} is somewhat
disconnected and may be omitted for those interested only in the covariant
superconformal approach: there we show that this formulation agrees with the conventional
harmonic superspace in the analytic basis, and we relate it to the existing
covariant approach of \cite{DelamotteKaplan:HHS, GKS:Sugra}.

The main applications are contained within the last two sections.
As a sample calculation, we demonstrate in section \ref{sec:SigmaModels} the covariant
component reduction of a general superconformal sigma model,
reproducing the general hyperk\"ahler cone sigma model coupled to conformal
supergravity \cite{dWKV} just as in projective superspace \cite{Butter:HKP}.
The result of this calculation is not new, but it provides a useful test
that covariant component reductions within harmonic superspace are tractable.
Our interest is actually in exploring higher-derivative actions. A brief discussion
of these applications follows in the concluding section.
Our notation and conventions follow \cite{Butter:CSG4d.Proj}.
A technical appendix addresses aspects of integration on analytic
submanifolds, to which we will refer as needed.

\section{Harmonic analysis on the complexified $S^2$}\label{sec:CompS2}
We begin with a discussion of elements of harmonic analysis on
the complexified $S^2$, which is equivalent to $T \mathbb CP^1$.
The formulation uses the biharmonic approach of \cite{GIO:QK},
although we will use somewhat different conventions and emphasize different aspects.

\subsection{Elements of analysis on a real $S^2$}
Let us briefly review the harmonic description of an $S^2$ manifold \cite{GIOS}.
It is described by harmonics $u^{i+}$ and $u_i^-$ obeying
$u_i^- = (u^{i+})^*$ with $u^{i+} u_i^- = 1$.
These parametrize a group element $\textbf{g}$ of $\SU{2}$,
\begin{align}
\textbf{g} =
\begin{pmatrix}
u^{\1+} & - u_\2^- \\
u^{\2+} & u_\1^-
\end{pmatrix}~, \qquad
\textbf{g}^{-1} = \textbf{g}^\dag~, \qquad \det \textbf{g} = 1~.
\end{align}
The two-sphere is isomorphic to $\SU{2} / \gU{1}$ with
the equivalence relation $u^{i+} \sim e^{i \alpha} u^{i+}$.
That is, the harmonics are in one-to-one correspondence
with real coordinates $X^I = u_j^- (\sigma^I)^j{}_k u^{k+}$ obeying
$\sum_I X^I X^I = 1$, where $\sigma^I$ are the Pauli matrices.

Rather than introduce two derivatives and one tangent space rotation on
$S^2$, it is customary to introduce three $\SU{2}$ derivatives, $D_\tru^{++}$, $D_\tru^0$
and $D_\tru^{--}$ defined as
\begin{align}\label{eq:RSDerv}
D_\tru^{++} \equiv u_i^+ \frac{\partial}{\partial u_i^-}~,\qquad
D_\tru^{--} \equiv u^{i-} \frac{\partial}{\partial u^{i+}}~, \qquad
D_\tru^0 \equiv u^{i +} \frac{\partial}{\partial u^{i+}} -  u_i^- \frac{\partial}{\partial u_i^-} ~,
\end{align}
possessing the commutation relations
$[D_\tru^{++}, D_\tru^{--}] = D_\tru^0$
and $[D_\tru^0, D_\tru^{\pm\pm}] = \pm 2 D_\tru^{\pm\pm}$.
Associated with these are vielbeins
\begin{align}\label{eq:RSViel}
\cU^{++} = u_i^+ \rd u^{i+}~, \qquad
\cU^{--} = u_i^- \rd u^{i-}~, \qquad
\cU^{0} = u_i^- \rd u^{i+} = u_i^+ \rd u^{i-}~,
\end{align}
so that the exterior derivative can be written
$\rd = -\cU^{++} D_\tru^{--} + \cU^0 D_\tru^0 + \cU^{--} D_\tru^{++}$.
We employ the usual superspace conventions for differential forms so that
the exterior derivative acts from the right.

Given some globally defined function $f^{(0)}(u^+, u^-)$,
its integral on $S^2$ is given by
\begin{align}
\int_\rS \rd u\, f^{(0)}
	= \frac{i}{2\pi} \int_\rS \cU^{++} \wedge \cU^{--} f^{(0)}~,
\end{align}
normalized so that
$\int_\rS \rd u\, = \frac{i}{2\pi} \int_\rS \cU^{++} \wedge \cU^{--} = 1$.
The integrand can be interpreted as a closed two-form $\omega = \cU^{++} \wedge \cU^{--} f^{(0)}$
on either $S^2$ or $\gSU{2}$.

\subsection{Analysis on the complexified $S^2$ and twisted biholomorphy}
We define the complexified $S^2$ as the complex affine quadric $Q^2$,
\begin{align}
Q^2 = \Big\{Z^I \in \mathbb C^3 : \sum_{I=1}^3 (Z^I)^2 = 1\Big\}~.
\end{align}
One can show that $Q^2 \subset \mathbb CP^1 \times \mathbb CP^1$ by identifying
\begin{align}
Z^I = \frac{1}{(v, \bar w)} \, \bar w_j (\sigma^I)^j{}_k v^k~, \qquad (v, \bar w) := v^k \bar w_k~,
\end{align}
which defines $v^i$ and $\bar w_j$ up to the identifications
$v^i \sim \l v^i$ and $\bar w_j \sim \widetilde \l \bar w_j$ for
$\l$ and $\widetilde\l$ unrelated complex numbers.
In other words, $Q^2$ can be identified as $\mathbb CP^1 \times \mathbb CP^1$
with the anti-diagonal region $(v,\bar w) = 0$ excised: this is just
the tangent bundle of $S^2$.\footnote{One can prove $Q^2 \cong T \mathbb CP^1$
directly by decomposing $Z^I$ into its real and imaginary parts.}

As with the real $\rS$, it is convenient to identify each
$\mathbb CP^1$ with $\SU{2} / \gU{1}$ and to introduce harmonic
coordinates on the respective $\SU{2}$ groups. Denote the two groups
by $\SU{2}_\trv$ and $\SU{2}_\trw$ with harmonics $v^{i\pm}$ and $w^{i\pm}$ defined as
\begin{align}
v^{i+} = \frac{v^i}{\sqrt{(v, \bar v)}}~, \qquad
v_i^{-} = \frac{\bar v_i}{\sqrt{(v, \bar v)}}~,
\end{align}
and similarly for $w^{i\pm}$.
The corresponding derivatives
$(D_\trv^{\pm\pm}, D_\trv^0)$ and
$(D_\trw^{\pm\pm}, D_\trw^0)$, as well as the vielbeins
$(\cV^{\pm\pm}, \cV^0)$ and
$(\cW^{\pm\pm}, \cW^0)$,
are defined analogously to \eqref{eq:RSDerv} and \eqref{eq:RSViel}.
We write the exterior derivative as
$\rd = \cV^{\ul a} D_{\trv \ul a} + \cW^{\bar a} D_{\trw \bar a}$
where
\begin{align}
\cV^{\ul a} D_{\trv \ul a}
	&=  - \cV^{++} D_\trv^{--} + \cV^0 D_\trv^0 + \cV^{--} D_\trv^{++}
	= \cV^{++} D_{\trv ++} + \cV^0 D_{\trv 0} + \cV^{--} D_{\trv --}
\end{align}
and similarly for $\cW^{\bar a} D_{\trw \bar a}$.

The space $Q^2 \cong T\mathbb CP^1$ is defined above in a twisted biholomorphic manner --
that is, the coordinates $Z^I$ are holomorphic in $v^i$ and anti-holomorphic
in $\bar w_i$. We will be mainly interested in fields that share this feature,
properly interpreted on the harmonic coordinates. Following the
same abuse of nomenclature as in \cite{Butter:CSG4d.Proj}, we will refer to
fields annihilated by $D_\trv^{++}$ as
\emph{holomorphic} on (an open domain of) $\SU{2}_\trv$
and those annihilated by $D_\trw^{--}$ as
\emph{anti-holomorphic} on (an open domain of) $\SU{2}_\trw$. Fields satisfying both
conditions will be called \emph{(twisted) biholomorphic}.
We specialize to such fields $\cF^{(n,m)}$ with charge $(n,m)$ under
$\gU{1}_\trv \times \gU{1}_\trw$, so that
\begin{align}
D_\trv^0 \cF^{(n,m)} = n \cF^{(n,m)}~, \quad
D_\trw^0 \cF^{(n,m)} = m \cF^{(n,m)}~, \quad
D_\trv^{++} \cF^{(n,m)} = D_\trw^{--} \cF^{(n,m)} = 0~.
\end{align}

The natural integration principle is a twisted biholomorphic integral
\begin{align}\label{eq:TBInt}
S = \frac{i}{2\pi} \int_{\cS} \cV^{++} \wedge \cW^{--} \, \omega^{(-2,+2)}(v^+, w^-)
	= \frac{i}{2\pi} \int_{\cS} \omega
\end{align}
where $\omega$ is a two-form and $\cS$ is some closed
two-dimensional surface in $T \mathbb CP^1$. $\omega$ is closed
as a consequence of being twisted biholomorphic. The action is invariant
under an infinitesimal diffeomorphism $\delta_\xi \omega = \rd (\imath_\xi \omega)$
corresponding to a small deformation of the surface $\cS$, and so it depends only
on the homotopy class of the surface provided $\omega$ is non-singular in the interior.
In fact, as we will discuss shortly, there is but one interesting class for
$\cS$.

We will actually need a slightly more general two-form given by
\begin{align}\label{eq:CHStwoform}
\omega = \cV^{++} \wedge \cW^{--} \, \omega^{(-2,2)}
	- \cV^{--} \wedge \cW^{--} \, \omega^{(2,2)}~.
\end{align}
In this case, the closure condition amounts to two requirements.
The first is
\begin{align}
D_\trw^{--} \omega^{(-2,2)} = 0~, \qquad D_\trw^{--} \omega^{(2,2)} = 0~,
\end{align}
equivalent to the condition that $\omega$ is anti-holomorphic
in $w_i^-$. The second requirement
\begin{align}
D_\trv^{++} \omega^{(-2,2)} = D_\trv^{--} \omega^{(2,2)}~,
\end{align}
constrains the $v_i^-$ dependence of $\omega^{(-2,2)}$.

For later use, it will be convenient to establish two analogues of Stokes' formula.
Taking $\L^{(0,2)}$ and $\L^{(-2,0)}$ to be functions annihilated by $D_\trw^{--}$ but
otherwise arbitrary, one can show
\begin{subequations}\label{eq:HarmTD1}
\begin{align}
\int_\cS \cV^{++} \wedge \cW^{--} D_\trv^{--} \L^{(0,2)}
	= \int_\cS \cV^{--} \wedge \cW^{--} D_\trv^{++} \L^{(0,2)}~, \quad
	D_\trw^{--} \L^{(0,2)} = 0~, \label{eq:HarmTD1a} \\
\int_\cS \cV^{++} \wedge \cW^{--} D_\trw^{++} \L^{(-2,0)}
	= - \int_\cS \cV^{++} \wedge \cV^{--} D_\trv^{++} \L^{(-2,0)}~, \quad
	D_\trw^{--} \L^{(-2,0)} = 0~. \label{eq:HarmTD1b}
\end{align}
\end{subequations}
An important special case is when
$\L^{(0,2)}$ and $\L^{(-2,0)}$ are twisted biholomorphic and then the right-hand
sides of \eqref{eq:HarmTD1} vanish.

A special class of diffeomorphisms are the isometries that leave the vielbeins
invariant. An $\SU{2}_\trv$ isometry acts as
\begin{align}\label{eq:SU2vIso}
\delta_\trv(\l) =
	-\l_\trv^{++} D_\trv^{--} + \l_\trv^0 D_\trv^0 + \l_\trv^{--} D_\trv^{++}~, \qquad
\l_\trv^{\pm\pm} = \l^{ij} v_i^\pm v_j^\pm~, \qquad \l^0_\trv = \l^{ij} v_i^+ v_j^-
\end{align}
in terms of harmonic-independent $\l^{ij}$. Similar formulae hold
for an $\SU{2}_\trw$ isometry $\delta_\trw(\rho)$. 
It is interesting to note for these isometries that
\begin{alignat}{2}
\delta_\trv \omega^{(-2,2)} &= D_\trv^{--} (\l_\trv^{--} \omega^{(2,2)} - \l_\trv^{++} \omega^{(-2,2)})~, &\qquad
\delta_\trw \omega^{(-2,2)} &= D_\trw^{++} (\rho_\trw^{--} \omega^{(-2,2)})~, \eol
\delta_\trv \omega^{(2,2)} &= D_\trv^{++} (\l_\trv^{--} \omega^{(2,2)} - \l_\trv^{++} \omega^{(-2,2)})~, &\qquad
\delta_\trw \omega^{(2,2)} &= D_\trw^{++} (\rho_\trw^{--} \omega^{(2,2)})~.
\end{alignat}

\subsection{The emergence of a complex harmonic structure}\label{sec:HA_emergence}
Now let us recover the complex harmonic structure required for harmonic superspace.
The simplest choice of closed surface $\cS$ is the real $\rS = \SU{2} / \gU{1}$
constructed from the diagonal $\SU{2}$ submanifold
of $\SU{2}_\trv \times \SU{2}_\trw$ with $v^{i \pm} = w^{i \pm}$.
Up to small deformations this is the only homotopically non-trivial choice:
any non-contractible $\cS$ in $T \mathbb CP^1$
is continuously deformable into the real $\rS$.
We require an additional assumption that each of the twisted biholomorphic quantities
are globally defined along this submanifold.
This implies that $\cF^{(n,m)}$ possesses an expansion as in \cite{GIO:QK}
\begin{align}
\cF^{(n,m)} = (v^+, w^-)^n
	\sum_{k=0}^\infty \cF^{(i_1 \cdots i_{n+m+k} j_1 \cdots j_{k})}
	\frac{v_{i_1}^+}{(v^+, w^-)} \cdots \frac{v_{i_{n+m+k}}^+}{(v^+, w^-)}\, 
	w_{j_1}^- \cdots w_{j_k}^-~.
\end{align}
A similar expansion applies for $n+m < 0$.
To shed some light on the meaning of this, we follow
\cite{GIO:QK} and introduce new complex harmonics,
\begin{align}
u_i^{+} \equiv u_i^{(0,1)} =  \frac{v_i^{+}}{(v^+, w^-)}~, \qquad
u_i^- \equiv u^{(0,-1)}_i = w_i^-~, \qquad u^{i+} u_i^- = 1~,
\end{align}
and three additional complex coordinates,
\begin{align}
z^{++} &\equiv z^{(0,2)} = \frac{(v^+,w^+)}{(v^+, w^-)}~, \eol
z^{--} &\equiv z^{(0,-2)} = (v^-, w^-) (v^+, w^-)~, \eol
z^0 &\equiv z^{(1,-1)} = (v^+, w^-)~.
\end{align}
We have followed existing convention in labeling the
coordinates by the sums of their $\gU{1}_\trv$ and $\gU{1}_\trw$ charges.
Note in particular that
the complex harmonics and the coordinates $z^{\pm\pm}$ carry only
$\gU{1}_\trw$ charge.
Relative to the conventions of \cite{GIO:QK}, we have exchanged the roles of
$u^{i\pm}$ and $w^{i\pm}$ so that $u^{i\pm}$ is reserved for the complex harmonic
coordinate.

Now the original harmonics are given in terms of the new coordinates as
\begin{align}
v^{i+} = z^0 u^{i+}~, \qquad v_i^- = \frac{1}{z^0} \big(u_i^- + u_i^+ z^{--}\big)~, \quad
w^{i+} = u^{i+} + z^{++} u^{i-}~, \quad
w_i^- = u_i^-~.
\end{align}
In the new variables, the old derivatives become
\begin{align}
D_\trv^{++} &= (z^0)^2 \frac{\pa}{\pa z^{--}}~, \qquad
D_\trv^0 = z^0 \frac{\pa}{\pa z^0}~, \eol
D_\trv^{--} &= \frac{1}{(z^0)^2} \Big(\pa_\tru^{--}
	- \frac{\pa}{\pa z^{++}}
	+ z^{--} z^0 \frac{\pa}{\pa z^0}
	+ (z^{--})^2 \frac{\pa}{\pa z^{--}}\Big)~,
\end{align}
and
\begin{align}
D_\trw^{++} &= \pa_\tru^{++}
	- z^{++} \pa_\tru^0
	- (z^{++})^2 \frac{\pa}{\pa z^{++}}
	+ z^{++} z^0 \frac{\pa}{\pa z^0}
	+ (2 z^{++} z^{--} - 1) \frac{\pa}{\pa z^{--}}~, \eol
D_\trw^0 &= \pa_\tru^0 + 2 z^{++} \pa_{z^{++}} - 2 z^{--} \pa_{z^{--}}
	- z^0 \frac{\pa}{\pa z^0}~, \qquad
D_\trw^{--} = \frac{\pa}{\pa z^{++}}~.
\end{align}
The corresponding vielbeins are 
\begin{align}
\cV^{++} &= (z^0)^2 \cU^{++} ~, \qquad
\cV^{0} = \cU^0 + z^{--} \cU^{++} + \frac{\rd z^0}{z^0}~, \eol
\cV^{--} &= \frac{1}{(z^0)^2} \Big(
	\cU^{--} + 2 z^{--} \cU^0 + (z^{--})^2 \cU^{++} + \rd z^{--}
	\Big)~, \eol
\cW^{++} &= \cU^{++} + 2 z^{++} \cU^0 
	+ (z^{++})^2 \cU^{--} - \rd z^{++} ~, \eol
\cW^{0} &= \cU^0 + z^{++} \cU^{--} ~, \qquad
\cW^{--} = \cU^{--} ~.
\end{align}
We have introduced
\begin{align}
\pa_\tru^{++} = u_{i}^+ \frac{\pa}{\pa u_{i}^-}~, \qquad
\pa_\tru^{--} = u^{i-} \frac{\pa}{\pa u^{i+}}~, \qquad
\pa_\tru^0 = u^{i+} \frac{\pa}{\pa u^{i+}} - u_{i}^- \frac{\pa}{\pa u_{i}^-}~, \\
\cU^{++} = u_i^+ \rd u^{i+}~, \qquad
\cU^{--} = u_i^- \rd u^{i-}~, \qquad
\cU^0 = u_i^- \rd u^{i+} = u_i^+ \rd u^{i-}~.
\end{align}

The $\SU{2}_\trv \times \SU{2}_\trw$ isometry transformations can be rewritten
\begin{align}
\delta_\trv(\l)
	= \l_\tru^{++} (\pa_{z^{++}} - \pa_\tru^{--})
	+ \l_\tru^0 (z^0 \pa_{z^0} + 2 z^{--} \pa_{z^{--}})
	+ \l_\tru^{--} \pa_{z^{--}}~, \eol
\delta_\trw(\rho)
	= - \rho_\tru^{++} \pa_{z^{++}}
	+ \rho_\tru^0 (\pa_\tru^0 - z^0 \pa_{z^0} - 2 z^{--} \pa_{z^{--}})
	+ \rho_\tru^{--} (\pa_\tru^{++} - \pa_{z^{--}})~,
\end{align}
where $\l_\tru^{\pm\pm} = \l^{ij} u_i^\pm u_j^\pm$ and similarly for $\rho$.
These special diffeomorphisms are induced by infinitesimal
general coordinate transformations\footnote{We denote an infinitesimal
(passive) general coordinate transformation by 
$\delta^* x^m = -\xi^m$. The corresponding (active) diffeomorphism induced on
a scalar field $f(x)$ is always written $\delta f(x) = \xi^m \pa_m f(x)$.
This notation is opposite that employed in \cite{GIOS}.}
\begin{gather}
\delta_\trv^* u^{i+} = \l_\tru^{++} u^{i-}~, \qquad \delta_\trv^* u^{i-} = 0~, \eol
\delta_\trv^* z^{++} = -\l_\tru^{++}~, \qquad
\delta_\trv^* z^0 = - \l_\tru^0 z^0~, \qquad
\delta_\trv^* z^{--} = - \l_\tru^{--} - 2 z^{--} \l_\tru^0~,
\end{gather}
and
\begin{gather}
\delta_\trw^* u^{i+} = -\rho_\tru^0 u^{i+}~, \qquad \delta_\trw^* u^{i-} = -\rho_\tru^{--} u^{i+} - \rho_\tru^0~, \eol
\delta_\trw^* z^{++} = \rho_\tru^{++}~, \qquad
\delta_\trw^* z^0 = \rho_\tru^0 z^0~, \qquad
\delta_\trw^* z^{--} = \rho_\tru^{--} + 2 z^{--} \rho_\tru^0~.
\end{gather}
The $\gSU{2}_\trv$ transformations with parameters $\l^{ij}$, which will be identified with
the $\gSU{2}_R$ gauge transformations of conformal supergravity in the central basis,
are generated by asymmetric transformations of the complex harmonics.
The diagonal isometry group generated by $\rho^{ij} = \l^{ij}$ corresponds to the external
group of automorphisms $\gSU{2}_A$ on $u^{i\pm}$, taking $\delta^* u^{i \pm} = \l^i{}_j u^{j\pm}$
and leaving $z^{\pm\pm}$ and $z^0$ invariant.

We have denoted the derivatives of the complex harmonics by simple partial
derivatives $\pa_\tru^{\pm\pm}$ and $\pa_\tru^0$
to emphasize that they are not covariant with respect to
$\gSU{2}_\trv \times \gSU{2}_\trw$. Following \cite{GIO:QK}, one can introduce
covariant derivatives $\mathscr{D}^{\pm\pm}$ and $\mathscr{D}^0$ defined by
\begin{align}
\mathscr{D}^{\pm\pm} := D_\trv^{\pm\pm} + D_\trw^{\pm\pm}~, \qquad
\mathscr{D}^0 := D_\trv^0 + D_\trw^0~.
\end{align}
These obey the usual algebra
$[\mathscr{D}^{++}, \mathscr{D}^{--}] = \mathscr{D}^0$ and
$[\mathscr{D}^0, \mathscr{D}^{\pm\pm}] = \pm 2\, \mathscr{D}^{\pm\pm}$
and act on $D_{\trv \ul a}$ and $D_{\trw \bar a}$ as external automorphisms, e.g.
$[\mathscr{D}^{\pm\pm}, D_\trv^{\mp\mp}] = \pm D_\trv^0$ and
$[\mathscr{D}^{0}, D_\trv^{\pm\pm}] = \pm 2\, D_\trv^0$.
Note that the $\gSU{2}_\trv$ derivatives were denoted $Z^{\pm\pm}$ and $Z^0$ in \cite{GIO:QK}.

In the complex harmonic coordinates, twisted biholomorphic functions 
are independent of $z^{++}$ and $z^{--}$, while their dependence on
$z^0$ is constrained to a single overall factor,
\begin{align}\label{eq:DefCHF}
\cF^{(n,m)} &= (z^0)^n
	\sum_{k=0}^\infty \cF^{(i_1 \cdots i_{n+m+k} j_1 \cdots j_{k})}
	u_{i_1}^+ \cdots u_{i_{n+m+k}}^+
	u_{j_1}^-\cdots u_{j_k}^-
	= (z^0)^n \cF^{(n+m)}(u^\pm)~,
\end{align}
where $\cF^{(n+m)}(u^\pm)$ is a convergent expansion. Observe that
\begin{align}
\mathscr{D}^{++} \cF^{(n,m)} = D_\trw^{++} \cF^{(n,m)}
	= \pa_\tru^{++} \cF^{(n,m)} - m z^{++} \cF^{(n,m)}
\end{align}
is not twisted biholomorphic unless $m=0$. Similarly,
$\mathscr{D}^{--} \cF^{(n,m)} = D_\trv^{--} \cF^{(n,m)}$
is not twisted biholomorphic unless $n$ vanishes.

\subsection{Complex harmonic integration}
In the remainder of this paper, we will primarily work with the harmonics
$v^{i\pm}$ and $w^{i\pm}$, but it is enlightening to rewrite some of the previous
formulae using the complex harmonics.
For example, the integral \eqref{eq:TBInt}
becomes, using $\omega^{(-2,+2)}(v^+, w^-) = (z^0)^{-2} \omega(u^\pm)$,
\begin{align}
S = \frac{i}{2\pi} \int_\cS \cU^{++} \wedge \cU^{--} \omega(u^\pm)~.
\end{align}
Because $\omega$ is closed, the integral is unchanged if we continuously deform $\cS$ to $\rS$.
This is apparent in the above form as the integrand is manifestly independent of the coordinates
$z^{++}$, $z^{--}$ and $z^0$, so we may certainly choose
$z^{++} = z^{--} = 0$ and $z^0=1$. This recovers the usual notion of harmonic integration.

If instead we have the more general two-form \eqref{eq:CHStwoform},
it is convenient to rewrite
\begin{align}
\omega^{(-2,2)}(v^+, w^-) = \frac{1}{(z^0)^2} \,\omega(u^\pm, z^{--})~, \qquad
\omega^{(2,2)}(v^+, w^-) = (z^0)^2 \,\omega^{+4}(u^\pm, z^{--})~.
\end{align}
The components are each independent of $z^{++}$ and are required to obey 
\begin{align}
\frac{\pa}{\pa z^{--}} \omega(u^\pm, z^{--})
	= \Big(\pa_\tru^{--}
	+ 2 z^{--} 
	+ (z^{--})^2 \frac{\pa}{\pa z^{--}}
	\Big)\omega^{+4}(u^\pm, z^{--})~.
\end{align}
The action principle is now a bit more complicated for a general surface $\cS$,
\begin{align}
S = \frac{i}{2\pi} \int_\cS \cU^{++} \wedge \cU^{--} \omega
	+ \frac{i}{2\pi} \int_\cS \cU^{--} \wedge 
		\Big(2 z^{--} \cU^0 + (z^{--})^2 \cU^{++} + \rd z^{--}\Big) \omega^{+4}~.
\end{align}
For these two-forms, it is more convenient to use the original expressions
with $v_i^\pm$ and $w_i^\pm$.

One may introduce analogues of Stokes' theorem just as before, but as we will mainly
be working with the original harmonics $v^\pm$ and $w^\pm$, reformulating
\eqref{eq:HarmTD1} for complex harmonics will not be necessary. However, it is useful
to note that when the integrands are twisted biholomorphic,
\begin{align}\label{eq:CHarmTD2}
\int_\cS \cU^{++} \wedge \cU^{--} \,\pa_\tru^{--} \L^{++}(u^\pm) = 0~, \qquad
\int_\cS \cU^{++} \wedge \cU^{--} \,\pa_\tru^{++} \L^{--}(u^\pm) = 0~.
\end{align}
Using these identities, we can prove a number of results that establish that complex
harmonic integration works exactly as real harmonic integration. First, one can show that
\begin{align}
\frac{i}{2\pi} \int_\cS \cU^{++} \wedge \cU^{--}
	u_{(i_1}^+ \cdots u_{i_\ell}^+ u_{j_1}^- \cdots u_{j_\ell)}^- = 0 \qquad \text{for} \quad
		\ell\geq 1~,
\end{align}
for \emph{any} closed surface $\cS$.
This follows by choosing $\L^{--} = u_{(i_1}^+ \cdots u_{i_{\ell-1}}^+ u_{i_\ell}^- u_{j_1}^- \cdots u_{j_\ell)}^-$ in \eqref{eq:CHarmTD2}. Similar identities with unequal numbers
of symmetrized positive and negative harmonics can be established.
We would like to also impose the normalization condition
$\frac{i}{2\pi} \int_\cS \cU^{++} \wedge \cU^{--} = 1$.
This obviously holds when $\cS = \rS$ -- we chose the overall normalization of the integral
to ensure this -- and holds more generally because the integrand is closed.
It follows that
\begin{align}\label{eq:HarmIntNorm}
\frac{i}{2\pi} \int_\cS \cU^{++} \wedge \cU^{--}
	\equiv \frac{i}{2\pi} \int_\cS \cV^{++} \wedge \cW^{--} (z^0)^{-2}
	= \frac{i}{2\pi} \int_\cS \cV^{++} \wedge \cV^{--} = 1~.
\end{align}
These results will prove crucial when performing component reductions in superspace.

\section{Complex harmonic superspace on $\cM^{4|8} \times \SU{2}_\trv \times \SU{2}_\trw$}
\label{sec:CHS}
Now we are prepared to introduce the first main result of this paper: the construction
of complex harmonic superspace on the supermanifold
$\cM^{4|8} \times \SU{2}_\trv \times \SU{2}_\trw$.  This approach is
based on the projective superspace $\cM^{4|8} \times \SU{2}_\trv$
elaborated upon in \cite{Butter:CSG4d.Proj}, with
the underlying structure of supergravity on $\cM^{4|8}$
described by $\cN=2$ conformal superspace \cite{CSG4d_2}.

Conformal superspace is a recent approach in the superspace literature that
gauges the entire superconformal group including
dilatations, special conformal transformations, and $S$-supersymmetry;
it is precisely the superspace version of the superconformal tensor calculus \cite{dWvHvP:Structure}.
In contrast to other superspace formulations gauging at most the Lorentz and $R$-symmetry
groups, conformal superspace proves to be quite economical and simple to work
with, as seen in the rather simple algebra of covariant spinor derivatives.

\subsection{Construction in the central basis}
Let us begin by recalling the salient details of \cite{Butter:CSG4d.Proj}.
That superspace, which we can identify as $\cM^{7|8} = \cM^{4|8} \times \SU{2}_\trv$,
involves local coordinates $z^{\ul M} = (z^M, y^{\ul m})$ and a vielbein
$E_{\ul M}{}^{\ul A}$ given in block form as
\begin{align}
E_{\ul M}{}^{\ul A} =
\begin{pmatrix}
E_M{}^A & E_M{}^{\trv \ul a} \\[0.2em]
E_{\ul m}{}^A & E_{\ul m}{}^{\trv \ul a} 
\end{pmatrix}~,
\end{align}
where the tangent space index $A = (a, \ul\alpha \pm) = (a, \alpha \pm, \dot\alpha \pm)$
is associated with $\cM^{4|8}$
and $\ul a = (\pm\pm,  0)$ is associated with $\SU{2}_\trv$.
The covariant derivatives $\nabla_{\ul A} = (\nabla_a, \nabla_{\ul\alpha \pm}, \nabla_{\trv\pm\pm}, \nabla_{\trv 0})$ are defined by the relation
\begin{align}
\pa_{\ul M} = E_{\ul M}{}^{\ul A} \nabla_{\ul A}
	+ \frac{1}{2} \Omega_{\ul M}{}^{ab} M_{ba}
	+ A_{\ul M} \bbA
	+ B_{\ul M} \bbD
	+ F_{\ul M}{}^A K_A~,
\end{align}
involving the Lorentz generator $M_{ab}$, the dilatation generator $\bbD$,
the $\gU{1}_R$ generator $\bbA$, and the special (super)conformal
generators $K_A = (K_a, S_{\ul\alpha \pm})$.
The Lorentz, dilatation, and $\gU{1}_R$ generators are normalized as
\begin{alignat}{3}
[M_{ab}, \nabla_\gamma^\pm] &= {(\sigma_{ab})_\gamma}^{\beta} \nabla_\beta^\pm, &\quad
[M_{ab}, \bar \nabla^\dgamma{}^\pm] &= {(\bsigma_{ab})^\dgamma}_{\dbeta} \bar \nabla^\dbeta{}^\pm~, &\quad
[M_{ab}, \nabla_c] &= \eta_{bc} \nabla_a - \eta_{ac} \nabla_b~, \eol {}
{}
[\bbD, \nabla_\alpha^\pm] &= \frac{1}{2} \nabla_\alpha^\pm~, &\quad
[\bbD, \bar \nabla^\dalpha{}^\pm] &= \frac{1}{2} \bar \nabla^{\dalpha \pm}~, &\quad
[\bbD, \nabla_a] &= \nabla_a~, \eol {}
[\bbA, \nabla_\alpha^\pm] &= -i \nabla_\alpha^\pm~, &\quad
[\bbA, \bar \nabla^\dalpha{}^\pm] &= +i \bar \nabla^\dalpha{}^\pm~.
\end{alignat}
We use the following prescription for raising the $\pm$ tangent space indices,
\begin{align}
\nabla_{\ul\alpha \mp} = \pm \nabla_{\ul\alpha}^\pm~, \qquad
\nabla_{\trv \mp\mp} = \pm \nabla_\trv^{\pm \pm}~, \qquad
S_{\ul\alpha \mp} = \mp S_{\ul\alpha}^\pm~, \qquad
\nabla_{\trv 0} = \nabla_\trv^0~,
\end{align}
so that they corresponded to the $\nabla_\trv^0$ charge of the operators.
The algebra of the special superconformal generators
with the spinor derivatives generates the $\gSU{2}_\trv$ derivatives,
\begin{align}
\{S_\beta^\pm, \nabla_\alpha^\pm\} &= \pm 4 \eps_{\beta \alpha} \nabla_\trv^{\pm\pm}, \qquad\qquad
\{\bar S^{\dbeta \pm}, \bar \nabla^{\dalpha \pm}\} = \mp 4 \eps^{\dbeta \dalpha} \nabla_\trv^{\pm\pm}~, \eol
\{S_\beta^\mp, \nabla_\alpha^\pm\} &= \pm (2 \eps_{\beta \alpha} \bbD - 2 M_{\beta \alpha}
     - i \eps_{\beta \alpha} \bbA) - 2 \eps_{\beta \alpha} \nabla_\trv^0 ~,\eol
\{\bar S^{\dbeta \mp}, \bar \nabla^{\dalpha \pm}\} &= \mp (2 \eps^{\dbeta \dalpha} \bbD - 2 M^{\dbeta \dalpha}
     + i \eps^{\dbeta \dalpha} \bbA) + 2 \eps^{\dbeta \dalpha} \nabla_\trv^0 ~.
\end{align}
This identifies $\gSU{2}_\trv$ with the superconformal group $\gSU{2}_R$.
The remaining relations between these generators and their action on the covariant
derivatives can be found in \cite{Butter:CSG4d.Proj}.

In the central basis (or central gauge), the vielbein decomposes as
$E^A = \rd z^M\, E_M{}^A$ and
$E^{\trv \ul a} \equiv \cV^{\ul a} = \rd y^{\ul m} \cV_{\ul m}{}^{\ul a} + \rd z^M\, \cV_M{}^{\ul a}$,
or in block form
\begin{align}
E_{\ul M}{}^{\ul A} =
\begin{pmatrix}
E_M{}^A & \cV_M{}^{\ul a} \\[0.2em]
0 & \cV_{\ul m}{}^{\ul a} 
\end{pmatrix}~.
\end{align}
The components of $E^A$ correspond to the vielbein on $\cM^{4|8}$, while
the vielbein $E^{\trv \ul a}$, which we rename to $\cV^{\ul a}$ in the central
basis for convenience, decomposes into the vielbein $\cV_{\ul m}{}^{\ul a}$ on
$\gSU{2}_\trv$ and the $\gSU{2}_R$ connection $\cV_M{}^{\ul a}$ on $\cM^{4|8}$.
The covariant derivatives may then be grouped into the $\cM^{4|8}$ covariant derivative
\begin{align}\label{eq:CovDCentral}
\nabla_{A} &= E_A{}^M \Big(
	\pa_M - \cV_M{}^{\ul a} D_{\ul a}
	- \frac{1}{2} \Omega_M{}^{bc} M_{cb}
	- A_M \bbA
	- B_M \bbD
	- F_M{}^B K_B \Big)~,
\end{align}
with $E_A{}^M = (E_M{}^A)^{-1}$
and the $\SU{2}_\trv$ covariant derivative
\begin{align}
\nabla_{\ul a} &\equiv D_{\ul a} = \cV_{\ul a}{}^{\ul m} \pa_{\ul m}~, \qquad
D_{\trv++} \equiv -D_\trv^{--}~, \qquad D_{\trv--} \equiv D_\trv^{++}~, \qquad
D_{\trv 0} \equiv D_\trv^0~,
\end{align}
with $\cV_{\ul a}{}^{\ul m} = (\cV_{\ul m}{}^{\ul a})^{-1}$.
The constraints chosen on the curvatures imply that
\begin{align}\label{eq:CovSpCentral}
\nabla_{\ul\alpha}^\pm = v_i^\pm \nabla_{\ul\alpha}{}^i~,
\end{align}
with the connections in the covariant derivatives $\nabla_{\ul\alpha}{}^i$ and $\nabla_a$
essentially independent of the harmonics $v^{i\pm}$.

Now let us extend this curved superspace to complex harmonic superspace.
Beginning in the central basis, we attach the \emph{completely decoupled}
space $\SU{2}_\trw$ with local coordinates $y^{\bar m}$ and 
vielbein $\cW_{\bar m}{}^{\bar a}$, obeying
$\pa_{\bar m} = \cW_{\bar m}{}^{\bar a} D_{\bar a}$,
for covariant $\SU{2}_\trw$  derivatives
$D_{\trw \bar a} = (D_{\trw ++}, D_{\trw --}, D_{\trw 0}) = (-D_{\trw}^{--}, D_\trw^{++}, D_\trw^0)$.
Denoting the full set of coordinates by $z^\cM$, the full vielbein $E_\cM{}^\cA$ is
in block form
\begin{align}\label{eq:CBViel}
E_{\cM}{}^{\cA} =
\begin{pmatrix}
E_M{}^A & \cV_M{}^{\ul a} & 0 \\[0.2em]
0 & \cV_{\ul m}{}^{\ul a}  & 0 \\[0.2em]
0 & 0 & \cW_{\bar m}{}^{\bar a}
\end{pmatrix}~.
\end{align}
The other connections are even simpler,
\begin{align}
\Omega_{\cM}{}^{ab} &= (\Omega_M{}^{ab}, 0, 0)~, \quad
A_{\cM} = (A_M, 0, 0)~, \quad \text{etc.}
\end{align}
and the covariant derivative $\nabla_{\cA} = (\nabla_A, D_{\trv \ul a}, D_{\trw \bar a})$
is given by
\begin{align}
\pa_{\cM} = E_{\cM}{}^{\cA} \nabla_{\cA}
	+ \frac{1}{2} \Omega_{\cM}{}^{ab} M_{ba}
	+ A_{\cM} \bbA
	+ B_{\cM} \bbD
	+ F_{\cM}{}^A K_A~.
\end{align}
In the central basis, there is
a clear distinction between $\SU{2}_\trv$ and $\SU{2}_\trw$.
The first is identified with the $\SU{2}$ $R$-symmetry group, while the second
remains decoupled.
In particular, one finds for $\nabla_A = (\nabla_a, \nabla_{\ul\alpha \pm})$
the same expression \eqref{eq:CovDCentral} in the central basis. The
covariant derivatives $\nabla_{\trv\ul a}$ and $\nabla_{\trw \bar a}$
retain their flat forms, $D_{\trv \ul a} = (D_{\trv \pm\pm}, D_{\trv 0})$
and $D_{\trw \bar a} = (D_{\trw \pm\pm}, D_{\trw 0})$.

Of course, there is no barrier to going to a basis (or gauge) where the vielbein
and connections take a more general form. We retain the same algebra of
covariant derivatives given in \cite{Butter:CSG4d.Proj}, and append 
the $\SU{2}_\trw$ covariant derivatives
$\nabla_{\trw \bar a} = (\nabla_{\trw{++}}, \nabla_{\trw{--}}, \nabla_{\trw 0})
	= (-\nabla_{\trw}^{--}, \nabla_{\trw}^{++}, \nabla_{\trw}^0)$, which
commute with all the other generators and covariant derivatives.
The resulting algebra of covariant spinor derivatives in any gauge is
\begin{gather}\label{eq:AnalyticCommutes}
\{\nabla_{\ul \alpha}^\pm, \nabla_{\ul \beta}^\pm\} = 0~, \\
\{\nabla_\alpha^\pm, \bnabla_{\dbeta}^\mp\} = \mp 2i \nabla_{\alpha \dbeta}~, \quad
\{\nabla_\alpha^\pm, \nabla_\beta^\mp\} = \pm 2 \eps_{\alpha \beta} \bar \cW~, \quad
\{\bnabla^{\dalpha \pm}, \bnabla^{\dbeta \mp}\} = \pm 2 \, \eps^{\dalpha \dbeta} \cW~.
\label{eq:defCurvSpinor}
\end{gather}
The first equation implies the existence of covariantly analytic
superfields. The operator $\cW$ appearing in the latter equations is constructed
from a single complex superfield $W_{\alpha\beta}$,
\begin{subequations}
\begin{align}
\cW &= \frac{1}{2} W^{\alpha \beta} M_{\beta \alpha}
	+ \frac{1}{4} \nabla^{\beta +} W_{\beta}{}^\alpha S_{\alpha}^-
	- \frac{1}{4} \nabla^{\beta -} W_{\beta}{}^\alpha S_{\alpha}^+
     + \frac{1}{4} \nabla^{\dalpha \beta} W_\beta{}^\alpha K_{\alpha \dalpha}~, \\
\bar \cW &= \frac{1}{2} \bar W_{\dalpha \dbeta} M^{\dbeta \dalpha}
     + \frac{1}{4} \bar \nabla_{\dbeta}^- \bar W^{\dbeta}{}_\dalpha \bar S^{\dalpha +}
	- \frac{1}{4} \bar \nabla_{\dbeta}^+ \bar W^{\dbeta}{}_\dalpha \bar S^{\dalpha -}
     + \frac{1}{4} \nabla_{\alpha \dbeta} \bar W^\dbeta{}_\dalpha K^{\dalpha \alpha}~.
\end{align}
\end{subequations}
$W_{\alpha\beta}$ is covariantly independent of the harmonics and is the single
curvature superfield of conformal superspace \cite{CSG4d_2}. The remaining curvatures may be
compactly written
\begin{align}
[\nabla_{\beta}^\pm, \nabla_{\alpha \dalpha}] &= -2 \eps_{\beta \alpha} \bar\cW_\dalpha^\pm~, \qquad
[\bar\nabla_{\dbeta}^\pm, \nabla_{\alpha \dalpha}] = -2 \eps_{\dbeta \dalpha} \cW_\alpha^\pm~, \eol{}
[\nabla_{\beta \dbeta}, \nabla_{\alpha \dalpha}] &= -\cF_{\beta \dbeta\, \alpha \dalpha}
     = -2 \eps_{\dbeta \dalpha} \cF_{\sym{\beta \alpha}}
     + 2 \eps_{\beta \alpha} \cF_{\sym{\dbeta \dalpha}}~.
\end{align}
The spinor operators $\cW_{\ul\alpha}^\pm$ and anti-selfdual and selfdual components of
$\cF_{ba}$ are given by
\begin{align}
\cW_{\ul\alpha}^{\pm} = -\frac{i}{2} [\nabla_{\ul\alpha}^\pm,\cW]~, \qquad
\cF_{\sym{\beta \alpha}} = \frac{1}{4} \{\nabla_{(\beta}^+,[\nabla_{\alpha)}^-, \cW]\}~, \qquad
\cF_{\sym{\dbeta \dalpha}} = \frac{1}{4} \{\bar\nabla_{(\dbeta}^+,[\bar\nabla_{\dalpha)}^-, \bar \cW]\}.
\end{align}
Explicit expressions for these can be found in \cite{Butter:CSG4d.Proj}.
The simplicity of these relations is one of the main advantages
of conformal superspace. 

In a general gauge, an arbitrary covariant diffeomorphism and gauge transformation
may be written
\begin{align}
\delta &= \xi^\cA \nabla_\cA 
	+ \frac{1}{2} \l{}^{ab} M_{ba}
	+ \L_\trD \bbD
	+ \L_\trA \bbA
	+ \eta^{\ul \alpha +} S_{\ul \alpha}^-
	- \eta^{\ul \alpha -} S_{\ul \alpha}^+
	+ \eps^a K_a
\end{align}
in terms of arbitrary parameters $\xi^\cA$, $\l^{ab}$, $\L_\trA$, $\L_\trD$,
$\eps^a$ and $\eta^{\ul \alpha \pm}$. We remind that
\begin{align}
\xi^\cA \nabla_\cA &=
	\xi^a \nabla_a
	+ \xi^{\ul \alpha -} \nabla_{\ul\alpha}^+
	- \xi^{\ul \alpha +} \nabla_{\ul\alpha}^-
	+ \xi_\trv^{--} \nabla_\trv^{++} + \xi_\trv^0 \nabla_\trv^0 - \xi_\trv^{++} \nabla_\trv^{--}
	\eol & \quad
	+ \xi_\trw^{--} \nabla_\trw^{++} + \xi_\trw^0 \nabla_\trw^0 - \xi_\trw^{++} \nabla_\trw^{--}~.
\end{align}
The charges on each parameter refer to their $\gSU{2}_\trv$ charge, except for the
local $\gSU{2}_\trw$ parameters $\xi_\trw^{\pm\pm}$.
In practice, one should restrict to gauges connected to the
central basis by complex harmonic gauge transformations, that is,
gauge transformations that are at most twisted biholomorphic on
$\SU{2}_\trv \times \SU{2}_\trw$. This means that the vielbein and connections
will generally be constrained so that $\nabla_\trw^{++}$ and $\nabla_\trv^{--}$
acquire additional connections while the other covariant harmonic derivatives remain
relatively simple.

In the central basis, the harmonic dependence on the parameters is restricted
to maintain the block form \eqref{eq:CBViel},
\begin{alignat}{2}
\xi^{\ul \alpha \pm} &= v_i^\pm \xi^{\ul \alpha i}~, &\qquad
\eta^{\ul \alpha \pm} &= v_i^\pm \eta^{\ul \alpha i}~, \eol
\xi_\trv^{\pm\pm} &\equiv \l_\trv^{\pm\pm} = v_i^\pm v_j^\pm \l^{ij}~, &\qquad
\xi_\trv^{0} &\equiv \l_\trv^{0} = v_i^+ v_j^- \l^{ij}~, \eol
\xi_\trw^{\pm\pm} &\equiv \rho_\trw^{\pm\pm} = w_i^\pm w_j^\pm \rho^{ij}~, &\qquad
\xi_\trw^{0} &\equiv \rho_\trw^{0} = w_i^+ w_j^- \rho^{ij}~,
\end{alignat}
with the other parameters harmonic-independent. Moreover, the absence of
$\gSU{2}_\trw$ connections means that $\rho^{ij}$ are always constants in the
central basis. In fact, we may refrain from ever performing $\gSU{2}_\trw$
diffeomorphisms.

This is perhaps a good place to emphasize again that the derivatives
$\nabla_{\ul\alpha}^\pm$ are given in the central basis by \eqref{eq:CovSpCentral}
and \emph{not}, as one might otherwise expect, by $u_i^\pm \nabla_{\ul\alpha}{}^i$.
That is, the algebra they obey with the $\SU{2}_\trv$ and $\SU{2}_\trw$ derivatives
can be written
\begin{alignat}{3}
[D_\trv^{\pm\pm}, \nabla_{\ul\alpha}^\pm] &= 0~, &\qquad
[D_\trv^{\pm\pm}, \nabla_{\ul\alpha}^\mp] &= \nabla_{\ul\alpha}^\pm~, &\qquad
[D_\trv^0, \nabla_{\ul\alpha}^\pm] &= \pm \nabla_{\ul\alpha}^\pm~, \eol {}
[D_\trw^{\pm\pm}, \nabla_{\ul\alpha}^\pm] &= 0~, &\qquad
[D_\trw^{\pm\pm}, \nabla_{\ul\alpha}^\mp] &= 0~, &\qquad
[D_\trw^0, \nabla_{\ul\alpha}^\pm] &= 0~.
\end{alignat}
The closest analogues of the conventional harmonic derivative relations are
\begin{align}
[\mathscr{D}^{\pm\pm}, \nabla_{\ul\alpha}^\pm] = 0~, \qquad
[\mathscr{D}^{\pm\pm}, \nabla_{\ul\alpha}^\mp] = \nabla_{\ul\alpha}^\pm~, \qquad
[\mathscr{D}^0, \nabla_{\ul\alpha}^\pm] = \pm \nabla_{\ul\alpha}^\pm~,
\end{align}
using the derivatives $\mathscr{D}:= D_\trv + D_\trw$ defined in
section \ref{sec:HA_emergence}. These commutators hold in any gauge,
replacing $D_\trv \rightarrow \nabla_\trv$ and $D_\trw \rightarrow \nabla_\trw$.

\subsection{Covariant primary analytic superfields}
Because the covariant spinor derivatives obey the conditions \eqref{eq:AnalyticCommutes},
the superspace admits analytic superfields $\Psi$ obeying $\nabla_{\ul\alpha}^+ \Psi = 0$.
We are interested only in superfields that are also \emph{primary},
$S_{\ul\alpha}^{\pm} \Psi = K_a \Psi = 0$.
Consistency with the operator algebra implies that $\Psi$ is a Lorentz scalar,
invariant under ${\rm U}(1)_R$, and obeys\footnote{These conditions were discussed in
chapter 9 of the monograph \cite{GIOS}. They are also the conditions required for
covariant projective multiplets \cite{Kuzenko:SPH}.}
\begin{align}\label{eq:AnalyticConsequences}
\nabla_\trv^0 \Psi = \bbD\Psi, \qquad \nabla_\trv^{++} \Psi = 0~.
\end{align}
In other words, $\Psi$ must have a $\gU{1}_\trv$ charge equal to its conformal dimension -- 
for definiteness, let us denote both quantities by $n$ -- and $\Psi$
must be holomorphic on an open domain of $\SU{2}_\trv$.
We may further choose
this open domain to be the vicinity of the diagonal $\SU{2}$ of $\SU{2}_\trv \times \SU{2}_\trw$
and restrict $\Psi$ to be a twisted biholomorphic scalar $\cF^{(n,m)}$ with charges
$(n,m)$ under $\gU{1}_\trv \times \gU{1}_\trw$.
A general conformal supergravity transformation of such a superfield is
\begin{align}\label{eq:deltaPrimary}
\delta \cF^{(n,m)} =
	\xi^A \nabla_A \cF^{(n,m)}
	+ n (\L_\trD + \l_\trv^0) \cF^{(n,m)}
	- \l_\trv^{++} D_\trv^{--} \cF^{(n,m)}
\end{align}
when written in the central basis. In terms of the complex harmonic
coordinates, one finds
\begin{align}
\delta \cF^{(n,m)} = \xi^A \nabla_A \cF^{(n,m)}
	+ n (\L_\trD + \l_\tru^0) \cF^{(n,m)}
	- \l_\tru^{++} \pa_\tru^{--} \cF^{(n,m)}~.
\end{align}
Below we will summarize the various types of multiplets commonly encountered
in harmonic superspace (see e.g. \cite{GIOS} for further details and references)
and discuss their twisted biholomorphic description in the central basis.

\subsubsection*{$\cO(n)$ multiplets}
In flat harmonic superspace, one can introduce complex
$\cO(n)$ multiplets \cite{KLT:O(n), KL:O(n)} that obey
$D^{++} \cH^{(n)} = D_{\ul\alpha}^+ \cH^{(n)} = 0$.
The generalization to curved harmonic superspace is straightforward:
we need twisted biholomorphic analytic superfields
$\cH^{(n,m)}$ obeying the additional
restriction $D_\trw^{++} \cH^{(n,m)} = 0$. As a consequence of
the twisted biholomorphy, one finds the integrability condition
$[D_\trw^{++}, D_\trw^{--}] \cH^{(n,m)} = m \cH^{(n,m)} = 0$,
so we are restricted to superfields
$\cH^{(n,0)}$ with
\begin{align}\label{eq:O(n)Constraint}
\cH^{(n,0)} = \cH^{j_1 \cdots j_n} v_{j_1}^+ \cdots v_{j_n}^+~, \qquad
\nabla_{\ul\alpha}^{(i} \cH^{j_1 \cdots j_n)} = 0~.
\end{align}
As a consequence of \eqref{eq:AnalyticConsequences}, $\cH^{(n,0)}$
must have weight $n$ under dilatations.
If $n$ is an even integer, it is possible to impose a
reality condition. The most familiar such multiplet is the $\cO(2)$
multiplet, or tensor multiplet, $\cG^{++} = \cG^{ij} v_i^+ v_j^+$,
which plays a major role as a compensator in one of the off-shell
formulations of $\cN=2$ Poincar\'e supergravity \cite{dWPvP:ImpTensor}.
It possesses the same form in either complex harmonic or projective
superspace, and the same holds for the general complex $\cO(n)$ multiplets.

\subsubsection*{Relaxed hypermultiplets}
We next consider the general class of so-called relaxed hypermultiplets.
In flat harmonic superspace, these are given by analytic superfields $\cR^{+q}$
obeying $(D^{++})^p \cR^{+q} = 0$ for some set of integers $p$ and $q$
(see \cite{GIOS} for a discussion and further references).
Their generalization in curved harmonic superspace involves 
twisted biholomorphic superfields $\cR^{(n,m)}$ with $n+m=q$ and the
constraint $(D_\trw^{++})^p \cR^{(n,m)} = 0$.
As a consequence of the twisted biholomorphic condition, one finds
$m = 1-p$. 

To understand this condition, it helps to specialize to the case
where $n+m=2$ and $p=2$. Here one finds a superfield
$\cR^{(3,-1)}$ with
\begin{align}\label{eq:RelHyp1}
\cR^{(3,-1)}
	= (z^0)^3 \Big(\cR^{(ij)} u_i^+ u_j^+
		+ \cR^{(ijk\ra )} u_i^+ u_j^+ u_k^+ u_{\ra}^-\Big)~.
\end{align}
We have denoted the index
of the negative harmonic with a Roman index to distinguish it from
the others. This is because the above expression can
be rewritten
\begin{align}\label{eq:RelHyp2}
\cR^{(3,-1)}
	= \cR^{ijk \ra} v_i^+ v_j^+ v_k^+ w_{\ra}^-~,
\end{align}
where $\cR^{ijk \ra} = \cR^{(ijk) \ra}$ is symmetric in
its first three indices only. The expression \eqref{eq:RelHyp1}
is recovered by decomposing
$\cR^{ijk \ra} = -\cR^{(ij} \eps^{k) \ra} + \cR^{(ijk \ra)}$.
The form \eqref{eq:RelHyp2} is advantageous for several reasons:
the constraint is manifestly satisfied for
$p=2$, the analyticity condition amounts to
$\nabla_{\ul\alpha}^{(l} \cR^{ijk) \ra} = 0$, and
the transformation \eqref{eq:deltaPrimary} leads to
\begin{align}
\delta \cR^{ijk \ra} 
	= \xi^A \nabla_A \cR^{ijk \ra}
		+ 3 \L_\trD \cR^{ijk \ra}
		+ 3 \l^{(i}{}_l \cR^{jk)l \ra}~.
\end{align}
This is consistent with the simple interpretation that
$\cR^{ijk \ra}$ possesses three $\SU{2}_R$ indices
associated with the isometric action on $\SU{2}_\trv$,
and an additional external $\SU{2}$ index
associated with $\SU{2}_\trw$.
The superfield $\cR^{ijk\ra}$ is just a globally
defined $\cO(3)$ superfield $\cR^{(3) \ra}$ in projective
superspace with an extra inert index. It is naturally embedded into
complex harmonic superfield by writing $\cR^{(3,-1)} \equiv \cR^{(3) \ra} w_{\ra}^-$.

In like fashion, the general relaxed hypermultiplet $\cR^{(n, 1-p)}$ obeying
$(D_\trw^{++})^p \cR^{(n,1-p)} = 0$, is associated with a
harmonic-independent superfield $\cR^{i_1 \cdots i_n {\ra}_1 \cdots {\ra}_{p-1}}$.
This can be interpreted as a projective superspace
$\cO(n)$ multiplet $\cR^{(n) \ra_1 \cdots \ra_{p-1}}$ with $p-1$ symmetric
external $\SU{2}$ indices.

\subsubsection*{The nonlinear multiplet}
Our third example is the nonlinear multiplet. Within real
harmonic superspace, it is given by an analytic superfield $N^{++}$
obeying the constraint $D^{++} N^{++} = -(N^{++})^2$.
Because the complex harmonic version of this analytic superfield
must be weight zero under dilatations, it must have vanishing
$D^0_\trv$ charge, and so it should be given by an
analytic multiplet $N^{(0,2)}$.
As usual, analyticity implies that $D_\trv^{++} N^{(0,2)} = 0$, so
the constraint must be given by $D_\trw^{++}$. In fact, it turns
out two constraints are needed,
\begin{align}\label{eq:NLCon}
D_\trw^{++} N^{(0,2)} = - (N^{(0,2)})^2~, \qquad
D_\trw^{--} N^{(0,2)} = 1~.
\end{align}
The second implies that $N^{(0,2)}$ is not twisted biholomorphic, but instead
possesses some dependence on $w^{i +}$.
This leads the conformal supergravity transformation
$\delta N^{(0,2)} = \xi^A \nabla_A N^{(0,2)} - \l_\trv^{++} D_\trv^{--} N^{(0,2)}$
to take an unusual form when written in terms of complex harmonic coordinates:
\begin{align}\label{eq:NLTrans2}
\delta N^{(0,2)}&= \xi^A \nabla_A N^{(0,2)} - \l_\tru^{++} \pa_\tru^{--} N^{(0,2)}
		- \l_\tru^{++} ~.
\end{align}
The inhomogeneous term appears also in the conventional harmonic superspace description
of this multiplet: there it arises as consistency condition for the constraint.

It is well-known that the nonlinear multiplet possesses a formulation in terms of a harmonic-independent
superfield $L^{\ra i}$. In complex harmonic superspace, it is encoded as
\begin{align}\label{eq:NLDef}
N^{(0,2)} \equiv \frac{L^{\ra +} w_{\ra}^+}{L^{\rb +} w_\rb^-}
	= \frac{L^{\ra i} w_\ra^+ v_i^+}{L^{\rb j} w_\rb^- v_j^+}~.
\end{align}
One can easily confirm the constraints \eqref{eq:NLCon}.
In terms of the complex harmonic coordinates, this expression becomes
\begin{align}
N^{(0,2)} = z^{++} +
	\frac{L^{\ra i} u_\ra^+ u_i^+}{L^{\rb j} u_\rb^- u_j^+}~.
\end{align}
The first term may be understood as generating the inhomogeneous term in
\eqref{eq:NLTrans2}.
Following \cite{GIOS}, we take $L^{\ra i}$ to be normalized as
$L^{\ra i} L_{\ra j} = \delta^i_j$ and $L^{\ra i} L_{\rb i} = \delta^\ra_\rb$
and raise/lower the indices in the same way.
The analyticity condition and transformation rule becomes
\begin{align}
L^{\ra (k} \nabla_{\ul\alpha}^i L_\ra^{j)} = 0 ~, \qquad
\delta L^{\ra i} = \xi^A \nabla_A L^{\ra i} + \l^i{}_j L^{\ra j}~.
\end{align}
As with the relaxed hypermultiplet, these conditions indicate that
the index $i$ of $L^{\ra i}$ is an $\SU{2}_R$ index,
while $\ra$ is an external index, consistent
with \eqref{eq:NLDef}. This form of the nonlinear multiplet frequently appears
as a compensator in $\cN=2$ supergravity
\cite{FV:N2Sugra, dWvH:N2Sugra, dWvHvP:Structure, dWPvP:ImpTensor}.

\subsubsection*{The $Q^+$ hypermultiplet}
Now we turn to the $Q^+$ hypermultiplet, which is the general matter
multiplet of harmonic superspace as well as the general compensating multiplet
of supergravity \cite{GIOS:Sugra}.
In the complex harmonic description,
it possesses charge $(1,0)$ under $\gU{1}_\trv \times \gU{1}_\trw$.
Assuming that $Q^+$ is twisted biholomorphic, 
it is easy to see that the charge assignments are consistent with the free
hypermultiplet equation of motion $D_\trw^{++} Q^+ = 0$: in that
case the free on-shell $Q^+$ is an $\cO(1)$ superfield.
Because of the importance of this multiplet, we will make a few further
comments that are obvious generalizations of its conventional description
in harmonic superspace. Its general off-shell version can be expanded
in the central basis as
\begin{align}\label{eq:Q+exp}
Q^+ &= \sum_{n=0}^\infty Q^{(i_1 \cdots i_{n+1} j_1 \cdots j_n)}
	v_{i_1}^+ \cdots v_{i_{n+1}}^+ 
	w_{j_1}^- \cdots w_{j_{n}}^-\, (z^0)^{-n}  \eol
&= z^0 \sum_{n=0}^\infty Q^{(i_1 \cdots i_{n+1} j_1 \cdots j_n)}
	u_{i_1}^+ \cdots u_{i_{n+1}}^+
	u_{j_1}^- \cdots u_{j_{n}}^-~.
\end{align}
We have not attempted here to maintain any distinction between the
$\SU{2}_\trv$ and $\SU{2}_\trw$ indices because the presence of
the $z^0$ factors renders the distinction
meaningless; the various terms in this expansion will mix under
$\SU{2}_R$. To see this, we note that the
transformation property of $Q^+$ may be written
\begin{align}
\delta Q^+ = \xi^A \nabla_A Q^+ + \L_\trD Q^+
	+ \l_\tru^0 Q^+
	- \l_\tru^{++} \pa_\tru^{--} Q^+~.
\end{align}
This implies for the leading term $Q^i$ in \eqref{eq:Q+exp} the
transformation
\begin{align}
\delta Q^i = \xi^A \nabla_A Q^i + \L_\trD Q^i
	+ \l^i{}_j Q^j - \frac{1}{2} \l^{jk} Q_{jk}{}^i~.
\end{align}
For the free on-shell hypermultiplet, all the higher terms vanish,
leaving an $\cO(1)$ multiplet.

\subsubsection*{The $\omega$ hypermultiplet}
The $\omega$ hypermultiplet is a variant version of the hypermultiplet,
which can take several forms. We discuss here its simplest version, which can
be constructed in conventional harmonic superspace
from a pseudoreal doublet $Q^{\ra +} = (Q^+, \widetilde{Q}^+)$ involving
a $Q^+$ hypermultiplet and its conjugate $\widetilde{Q}^+$ as
$\omega = u_{\ra}^- Q^{\ra +}$. If $Q^+$ is free, then $\omega$
obeys the free equation of motion $(D^{++})^2 \omega = 0$ and corresponds
to a relaxed hypermultiplet of Weyl weight 1.

In complex harmonic superspace, this version of $\omega$ becomes a
twisted biholomorphic analytic superfield $\omega^{(1,-1)}$.
Its $\gU{1}_\trv$ charge is implied by its Weyl weight, while
its $\gU{1}_\trw$ charge is implied if we assume that
the free equation of motion should be $(D_\trw^{++})^2 \omega^{(1,-1)} = 0$.
Each of these properties is consistent with the choice
$\omega^{(1,-1)} = w_\ra^- Q^{\ra +}$. Such a multiplet is manifestly
twisted biholomorphic with each of the requisite weights.
The general $\omega^{(1,-1)}$ hypermultiplet transforms as
\begin{align}
\delta \omega^{(1,-1)} = \xi^A \nabla_A \omega^{(1,-1)}
	- \l_\tru^{++} \pa_\tru^{--} \omega^{(1,-1)}
	+ (\L_\trD + \l_\tru^0) \omega^{(1,-1)}~.
\end{align}

\subsubsection*{Abelian vector multiplet}
Finally, we turn to the vector multiplet. For simplicity, our attention here
will be on the abelian case, but the non-abelian version is a straightforward extension.
Recall that the abelian vector multiplet is described by an
analytic prepotential $V^{++}$ constructed in terms of a bridge
superfield $B$ via $V^{++} = D^{++} B$. The bridge $B$ is globally defined
on $\rS$ but not analytic. Both the bridge and $V^{++}$ must have vanishing Weyl
weight.

In complex harmonic superspace, the vector multiplet is described
by a twisted biholomorphic analytic superfield $V^{(0,2)}$.
Its $\gU{1}_\trv$ charge must vanish, consistent with its Weyl weight.
Obviously, $V^{(0,2)}$ should be interpreted as a connection for
the complex harmonic derivative $\nabla_\trw^{++}$ in the analytic basis.
This implies that it should be related to a bridge superfield $B$ via
$V^{(0,2)} = D_\trw^{++} B$.
The bridge $B$ should be twisted biholomorphic
with vanishing harmonic charges.
To confirm this interpretation, we note that the
transformation of $V^{(0,2)}$ reproduces the
transformation in conventional harmonic superspace,
\begin{align}
\delta V^{(0,2)} &= \xi^A \nabla_A V^{(0,2)} - \l_\trv^{++} D_\trv^{--} V^{(0,2)}
	= \xi^A \nabla_A V^{(0,2)} - \l_\tru^{++} \pa_\tru^{--} V^{(0,2)}~.
\end{align}
In contrast, we expect $V^{--} \equiv V^{(-2,0)}$ to be the connection for the
complex harmonic derivative $\nabla_\trv^{--}$ in the analytic basis and given by
$V^{(-2,0)} = D_\trv^{--} B$.
It is easy to check that $V^{(-2,0)}$ is also twisted biholomorphic, though it
is not analytic. It transforms as
\begin{align}
\delta V^{(-2,0)} &= \xi^A \nabla_A V^{(-2,0)} - \l_\trv^{++} D_\trv^{--} V^{(-2,0)}
	- 2 \l_\trv^{0} V^{(-2,0)} \eol
	&= \xi^A \nabla_A V^{(-2,0)} - \l_\tru^{++} \pa_\tru^{--} V^{(-2,0)}
	- 2 \l_\tru^0 V^{(-2, 0)}~.
\end{align}
The differences in the covariant transformation laws for $V^{(-2,0)}$
and $V^{(0,2)}$ are naturally explained by their $\gU{1}_\trv \times \gU{1}_\trw$
charges in the complex harmonic approach.

\section{Superspace action principles on $\cM^{4|8} \times \SU{2}_\trv \times \SU{2}_\trw$}
\label{sec:Actions}
In this section, we will address both the full harmonic superspace and analytic superspace actions,
discuss how to relate one to the other, and provide the component reduction formula
for the analytic action. A few specific examples will also be discussed.

\subsection{Full superspace}
The natural twisted biholomorphic integral over full superspace is given by
\begin{align}\label{eq:FullSuperActionCentral}
\frac{i}{2\pi} \int \rd^4x\, \rd^4\q\, \rd^4\bar\q\, E\,
	\int_\cS \cV^{++} \wedge \cW^{--} \mathscr{L}^{(-2,2)}~,
\end{align}
where the first integral is evaluated in the central basis and the second
is over the closed surface $\cS$ homotopic to the real $\rS$.
The harmonic charges of the Lagrangian are chosen to counter the measure factor.
The Lagrangian must have vanishing dilatation and $\gU{1}_R$ weights
and be a conformal primary. This expression can be generalized to any gauge,
\begin{align}\label{eq:FullSuperAction}
\frac{i}{2\pi} \int_\cS \rd^2 \z \int \rd^4x \, \rd^4\q^+\, \rd^4\q^- \, E^{(2, -2)} \mathscr{L}^{(-2,2)}
\end{align}
with the measure
\begin{align}
E^{(2,-2)} = \sdet \begin{pmatrix}
E_M{}^A & E_M{}^{\trv ++} & E_M{}^{\trw --} \\[0.2em]
E_\z{}^A & E_\z{}^{\trv ++} & E_\z{}^{\trw --} \\[0.2em]
E_{\tilde\z}{}^A & E_{\tilde\z}{}^{\trv ++} & E_{\tilde\z}{}^{\trw --}
\end{pmatrix}~.
\end{align}
The complex coordinates $\z$ and $\tilde \z$ parametrize $\cS$ with
$(\z)^* \neq \tilde \z$ in general,
and $E_\z{}^{\cA}$ and $E_{\tilde \z}{}^{\cA}$
are the pullback of the vielbein. The charge assignments of
the measure $E^{(2,-2)}$ correspond to its weight under
covariant $\nabla_\trv^0$ and $\nabla_\trw^0$ diffeomorphisms.
We have written the Grassmann coordinates suggestively as $\q^+$ and $\q^-$,
but in a general gauge they possess no meaningful charge or relation
to the harmonics. Using e.g. the results of Appendix B of \cite{Butter:CSG4d.Proj}
(see the summary in Appendix \ref{App:AI&D})
one may confirm that \eqref{eq:FullSuperAction} is a gauge and
diffeomorphism-invariant quantity.

It is useful to know when a quantity is a total derivative in the covariant
approach. In Appendix \ref{App:AI&D}, we show that the covariant expression
$\mathscr{L}^{(-2,2)} = \nabla_\trw^{++} V^{(-2,0)} + \nabla_\trv^{--} V^{(0,2)}$
is a total derivative for any twisted biholomorphic conformal primary superfields
$V^{(-2,0)}$ and $V^{(0,2)}$ with vanishing Weyl and $\gU{1}_R$ weights.
Other expressions which appear to be total derivatives such as $\nabla_A V^A$
generally fail to be primary -- and so are not permitted as
covariant integrands -- or leave residual connections when integrated by parts.

\subsection{Analytic superspace}
We will be particularly interested in the action principle for analytic superspace.
In the analytic basis, its form is well-known:
\begin{align}\label{eq:AnalyticSuperActionAB}
\int_\rS \rd u\, \int \rd^4\hat x \, \rd^4\hat\q^+\, \hat{\mathscr{L}}^{+4}~.
\end{align}
The analytic Lagrangian $\hat{\mathscr{L}}^{+4}$ must transform as a scalar density,
ensuring that the action is invariant under analytic gauge transformations.
The integration is performed over the real $S^2$ manifold.
To describe the same action principle in a general gauge, we propose
\begin{align}\label{eq:AnalyticSuperAction}
	\frac{i}{2\pi} \int_\cS \rd^2 \z \int \rd^4x \, \rd^4\q^+\, \cE^{(-2, -2)} \mathscr{L}^{(2,2)}~.
\end{align}
The Lagrangian $\mathscr{L}^{(2,2)}$ is now a scalar function rather than a scalar
density and the analytic measure is given by
\begin{align}
\cE^{(-2,-2)} = 
\sdet \begin{pmatrix}
E_m{}^a & E_m{}^{\ul \alpha +} & E_m{}^{\trv++} & E_m{}^{\trw --}\\[0.3em]
E_{\ul\mu +}{}^a & E_{\ul\mu +}{}^{\ul \alpha +} & E_{\ul\mu +}{}^{\trv++} & E_{\ul\mu +}{}^{\trw --} \\[0.3em]
E_\z{}^a & E_\z{}^{\ul \alpha +} & E_\z{}^{\trv++} & E_\z{}^{\trw --} \\[0.3em]
E_{\tilde\z}{}^a & E_{\tilde\z}{}^{\ul \alpha +} & E_{\tilde\z}{}^{\trv++} & E_{\tilde\z}{}^{\trw --}
\end{pmatrix}~.
\end{align}
We have labeled the measure again with its weights under covariant $\nabla_\trv^0$
and $\nabla_\trw^0$ diffeomorphisms, and we reiterate that the charge assignment of
the Grassmann coordinates is not meaningful in a general gauge. Later on,
when we return to the analytic basis, we will find a different notion of charge
for these objects that concurs with \eqref{eq:AnalyticSuperActionAB}.
One can show that \eqref{eq:AnalyticSuperAction} is gauge-invariant
provided $\mathscr{L}^{(2,2)}$ is an analytic twisted biholomorphic conformal primary
of Weyl weight two. To establish the equivalence of \eqref{eq:AnalyticSuperActionAB} and
\eqref{eq:AnalyticSuperAction} requires a more elaborate discussion, which will be
postponed until section \ref{sec:HSS}.

It is a straightforward exercise (by e.g. generalizing the argument in
\cite{Butter:CSG4d.Proj}) to show that any full superspace integral
\eqref{eq:FullSuperAction} can be written as an analytic superspace integral
\begin{align}\label{eq:FtoA}
\frac{i}{2\pi} \!\int_\cS \rd^2 \z \!\int \rd^4x \, \rd^8\q\, E^{(2, -2)} \mathscr{L}^{(-2,2)}
	= \frac{i}{2\pi} \!\int_\cS \rd^2 \z \!\int \rd^4x \, \rd^4\q^+\, \cE^{(-2, -2)} (\nabla^+)^4 \mathscr{L}^{(-2,2)}~.
\end{align}
The superfield $(\nabla^+)^4 \mathscr{L}^{(-2,2)}$ obeys all the requirements
of an analytic Lagrangian $\mathscr{L}^{(2,2)}$.
Similarly, one can lift any analytic action to a full superspace action.
Following the same procedure as in \cite{Butter:CSG4d.Proj}, we introduce a
real nowhere-vanishing harmonic-independent superfield $\Omega$ of Weyl weight two, writing
\begin{align}\label{eq:AtoF}
\frac{i}{2\pi} \int_\cS \rd^2 \z \int \rd^4x \, \rd^4\q^+\, \cE^{(-2, -2)} \mathscr{L}^{(2,2)}
&= \frac{i}{2\pi} \int_\cS \rd^2 \z \int \rd^4x \, \rd^8\q\, E^{(2, -2)} \frac{\Omega}{(\nabla^+)^4 \Omega} \mathscr{L}^{(2,2)}~.
\end{align}
Now let us choose $\Omega = W \bar W$ for a vector multiplet $W$ and then adopt the
gauge where $W=1$. This effects the conversion of conformal superspace \cite{CSG4d_2} to
$\gSU{2}$ superspace \cite{Grimm, KLRT-M1} and lets one rewrite \eqref{eq:AtoF} as
\begin{align}
\frac{i}{2\pi} \!\int_\cS \rd^2 \z \!\int \rd^4x \, \rd^8\q\,  \frac{E^{(2, -2)}}{(S^{++})^2} \mathscr{L}^{(2,2)}
&= \frac{i}{2\pi} \!\int \rd^4x \, \rd^8\q\, E\, \!\int_\cS \cV^{++} \wedge \cW^{--} \frac{\mathscr{L}^{(2,2)}}{(S^{++})^2} ~,
\end{align}
where $S^{++}$ is a torsion superfield of $\rm SU(2)$ superspace 
and the right-hand side is written in the central basis.
If we restrict to the real $S^2$, this simplifies still further to
\begin{align}
\int \rd^4x \, \rd^8\q\, E\, \int_\rS \rd u\, \frac{1}{(S^{++})^2} \mathscr{L}^{+4}~.
\end{align}
This is a convenient formulation of curved harmonic superspace in the central basis
using $\SU{2}$ superspace and is inspired by an analogous
formula in projective superspace \cite{KLRT-M1}.

In Appendix \ref{App:AI&D}, we briefly discuss how to show that
$\mathscr{L}^{(2,2)} = \nabla_\trw^{++} V^{(2,0)}$
is a total derivative when $V^{(2,0)}$ is an analytic twisted biholomorphic
conformal primary.
As an exercise, one may show this by introducing a prepotential for $V^{(2,0)}$
as $V^{(2,0)} = (\nabla^+)^4 V^{(-2,0)}$ and then 
observing that $\nabla_\trw^{++} V^{(-2,0)}$ is a total derivative as
a full superspace Lagrangian. Note that a similar quantity,
$\nabla_\trv^{--} V^{(4, 2)}$, is not an allowed analytic Lagrangian as it is
not twisted biholomorphic.

\subsection{Analytic superspace component action}
Now we turn to deriving the component form of the analytic superspace
action \eqref{eq:AnalyticSuperAction}.
Upon integration over the Grassmann
coordinates, the final form of the action should be
\begin{align}
S = \frac{i}{2\pi} \int_{\cM^4 \times \cS} \cJ~,
\end{align}
for some closed six-form $\cJ$ integrated over the product of $4D$ spacetime
$\cM^4$ and the auxiliary manifold $\cS$.
Here it helps to recall the projective superspace result \cite{Butter:CSG4d.Proj}, where
\begin{align}
-\frac{1}{2\pi} \oint_\cC \rd \tau \int \rd^4x \, \rd^4\q^+\, \cE^{--} \mathscr{L}^{++}
	= -\frac{1}{2\pi} \int_{\cM^4 \times \cC} \cJ_P~,
\end{align}
in terms of a five-form $\cJ_P$, which was quite complicated in a general
gauge. Its leading term was
\begin{align}
\cJ_P = e^0 \wedge e^1 \wedge e^2 \wedge e^3 \wedge e^{\trv ++}  (\nabla^-)^4 \mathscr{L}^{++}\loco
	+ \cdots
\end{align}
with the subleading terms each involving a five-form multiplied by a certain
number of covariant derivatives of $\mathscr{L}^{++}$. Here we have written
$e^a = E^a \loco\!\loco$ and $e^{\trv ++} = E^{\trv ++}\loco\!\loco$ as the double-bar projections
(setting $\q = \rd \q = 0$) of the corresponding vielbeins. 
Keeping in mind that complex
harmonic superspace can be understood as projective superspace combined
with an additional $\mathbb CP^1$ manifold, one can make the guess that $\cJ$ should be
given by inserting the harmonic Lagrangian into $\cJ_P$ and taking the wedge product
with $e^{\trw --} = E^{\trw --}\loco\!\loco$, that is,
\begin{align}\label{eq:HarmJProposal}
\cJ = \cJ_P[\mathscr{L}^{(2,2)}] \wedge e^{\trw --}
	=  e^0 \wedge e^1 \wedge e^2 \wedge e^3 \wedge e^{\trv ++} \wedge e^{\trw --} (\nabla^-)^4 \mathscr{L}^{(2,2)}\loco
	+ \cdots
\end{align}
This turns out to be the correct answer.

There are two approaches to checking this result. The first is simply to repeat the normal
coordinate calculation given in Appendix C of \cite{Butter:CSG4d.Proj}. The main difference
is that one encounters the volume six-form $\hat e^{(2,-2)} = \hat e^{++} \wedge e^{\trw --}$
where $\hat e^{++}$ is the volume five-form when restricted to the bosonic
body $\cM^4 \times \mathbb CP^1$ of projective superspace. Viewed as a superform,
its $\q^+$ expansion is responsible for giving the subleading terms in the component action.
But because $e^{\trw --}$ has a trivial $\q^+$ expansion when written in Grassmann normal
coordinates, no new features are encountered. This reproduces \eqref{eq:HarmJProposal}.

A less direct approach is to observe that $\cJ$ must be a closed six-form in superspace.
Because of the twisted biholomorphic nature of $\mathscr{L}^{(2,2)}$, one can show that
the closure of $\cJ_P$, interpreted as a five-form in projective superspace, implies the
closure of $\cJ$. One observes for the leading term of $\cJ_P$ (and similarly for the subleading
terms)
\begin{align}
0 = \rd \cJ_P 
	&= \rd \hat e^{++} \,(\nabla^-)^4 \mathscr{L}^{++}
	+ \hat e^{++} \wedge (E^A \nabla_A + E^{\trv \ul a} \nabla_{\trv \ul a} + \cdots) \,(\nabla^-)^4 \mathscr{L}^{++}
	+ \cdots~.
\end{align}
The second term in parentheses is the expansion of the exterior derivative in
projective superspace; we have exhibited the vielbeins but suppressed the other
connections. When we formally replace $\mathscr{L}^{++}$ with
$\mathscr{L}^{(2,2)}$, the second expression turns out to be missing
the $\SU{2}_\trw$ vielbeins. Adding and subtracting these gives
\begin{align}
0 &= \rd \hat e^{++} \,(\nabla^-)^4 \mathscr{L}^{(2,2)}
	+ \hat e^{++} \wedge (E^A \nabla_A + E^{\trv \ul a} \nabla_{\trv\ul a}
	+ E^{\trw \bar a} \nabla_{\trw\bar a} + \cdots)\,(\nabla^-)^4 \mathscr{L}^{(2,2)}
	\eol & \quad
	- 2 \,\hat e^{++} \wedge E^{\trw 0} (\nabla^-)^4 \mathscr{L}^{(2,2)}
	- \hat e^{++} \wedge E^{\trw --} \nabla_{\trw}^{++} (\nabla^-)^4 \mathscr{L}^{(2,2)}
	+ \cdots~.
\end{align}
The first line is the exterior derivative of the leading term of
$\cJ_P[\mathscr{L}^{(2,2)}]$. Taking the wedge product with $E^{\trw --}$
gives for this leading term
\begin{align}
0 &= \rd \cJ_P[\mathscr{L}^{(2,2)}] \wedge E^{\trw --}
	- 2 \cJ_P[\mathscr{L}^{(2,2)}] \wedge E^{\trw 0} \wedge E^{\trw --}
	= - \rd \cJ
\end{align}
and so $\cJ$ is closed. The subleading terms more or less go the same way.
A similar line of argument shows that if $\cJ_P$
is gauge-invariant in projective superspace up to an exact form, then so is
$\cJ$ in harmonic superspace.

It should go without saying that the central basis is to be preferred for
component actions. In that gauge, one finds
\begin{align}\label{eq:CompAction}
S &= \int \rd^4x\, e\, \cL~, \qquad
\cL = \frac{i}{2\pi} \int_{\cS} \Big(
		\cV^{++} \wedge \cW^{--} \cL^{(-2,2)}
		- \cV^{--} \wedge \cW^{--} \cL^{(2,2)}
	\Big)
\end{align}
where $\cL^{(-2,2)}$ and $\cL^{(2,2)}$ coincide with the component Lagrangians given
in \cite{Butter:CSG4d.Proj} with the replacement $\mathscr{L}^{++} \rightarrow \mathscr{L}^{(2,2)}$.
They are (with projection to $\q=0$ understood)
\begin{align}\label{eq:HCLag1}
\cL^{(-2,2)} &= \frac{1}{16} (\nabla^-)^2 (\bar\nabla^-)^2 \mathscr{L}^{(2,2)}
	- \frac{i}{8} (\bar\psi_m^- \bsigma^m)^\alpha \nabla_\alpha^- (\bar\nabla^-)^2 \mathscr{L}^{(2,2)}
	- \frac{i}{8} (\psi_m^- \sigma^m)_\dalpha \bar\nabla^{\dalpha -} (\nabla^-)^2 \mathscr{L}^{(2,2)}
	\eol & \quad
	+ \frac{1}{4} \Big(
	(\psi_n^- \sigma^{nm})^\alpha \bar\psi_m{}^\dalpha{}^-
	+ \psi_n{}^\alpha{}^-  (\bsigma^{nm}\bar\psi_m^-)^\dalpha
	- i \cV_m^{--} \sigma^m_{\alpha \dalpha} \Big) [\nabla_\alpha^{-}, \bar\nabla_\dalpha^{-}]  \mathscr{L}^{(2,2)}
	\eol & \quad
	+ \frac{1}{4} (\psi_m^- \sigma^{mn} \psi_n^-) (\nabla^-)^2 \mathscr{L}^{(2,2)}
	+ \frac{1}{4} (\bar\psi_m^- \bsigma^{mn} \bar\psi_n^-) (\bar \nabla^-)^2 \mathscr{L}^{(2,2)}
	\eol & \quad
	- \Big(
	\frac{1}{2} \eps^{mnpq} (\psi_m^- \sigma_n \bar\psi_p^-) \psi_q^{\alpha -}
	- 2 \,(\psi_m^- \sigma^{mn})^\alpha \cV_n^{--} \Big) \nabla_\alpha^- \mathscr{L}^{(2,2)}
	\eol & \quad
	+ \Big(
	\frac{1}{2} \eps^{mnpq} (\bar\psi_m^- \bsigma_n \psi_p^-) \bar\psi_{q\dalpha}^{-}
	- 2 \,(\bar\psi_m^- \bsigma^{mn})_\dalpha \cV_n^{--} \Big) \bar\nabla^{\dalpha -} \mathscr{L}^{(2,2)}
	\eol & \quad
	+ 3 \,\eps^{mnpq} (\psi_m^- \sigma_n \bar\psi_p^-) \cV_q^{--} \mathscr{L}^{(2,2)}
\end{align}
and
\begin{align}\label{eq:HCLag2}
\cL^{(2,2)}
	&= -\Big[3 D 
	+ \frac{3i}{2} (\bar \psi_m^- \bsigma^m \chi^+) 
	- \frac{3i}{2} (\psi_m^{-} \sigma^m \bar\chi^+) 
	+ 4 f_a{}^a 
	\eol & \qquad \qquad
	- 4 (\bar \psi_m^- \bsigma^{mn} \bar\phi_n^+) 
	+ 4 (\psi_m^{-} \sigma^{mn}\phi_n^{+}) 
	- 3\, \eps^{mnpq}
		 (\psi_m^{-} \sigma_n \bar\psi_p^{-}) \cV_q^{++}
	\Big] \mathscr{L}^{(2,2)}
	\eol & \quad
	+ \Big[\frac{3}{2} \chi^{\alpha +}
	- i (\bar \phi_m^{+} \bsigma^m)^\alpha 
	+ 2 (\psi_m^- \sigma^{mn})^\alpha \cV_n^{++} \Big] \nabla_\alpha^- \mathscr{L}^{(2,2)}
	\eol & \quad
	- \Big[\frac{3}{2} \chi_\dalpha^{+}
	- i (\phi_m^{+} \sigma^m)_\dalpha 
	+ 2 (\bar\psi_m^- \bsigma^{mn})_\dalpha \cV_n^{++}\Big] \bar\nabla^{\dalpha -} \mathscr{L}^{(2,2)}
	\eol & \quad
	- \frac{i}{4} \cV_m^{++} (\bsigma^m)^{\dalpha \alpha} [\nabla_\alpha^-, \bar\nabla_\dalpha^-] \mathscr{L}^{(2,2)}~.
\end{align}
Above we have the component fields as defined in \cite{CSG4d_2}
corresponding to the content of $\cN=2$ conformal supergravity.
These consist of (i) the five fundamental connections --
the vierbein $e_m{}^a$, the gravitini $\psi_m{}^{\alpha}{}_i$,
the $\rm SU(2)_R$ and $\rm U(1)_R$ connections $\cV_m{}^i{}_j$
and $A_m$, and the dilatation connection $b_m$;
(ii) covariant auxiliary fields $W_{ab} = \frac{1}{4} T_{ab}^-$, $\chi_{\alpha i}$, and $D$; and
(iii) composite connections $\omega_m{}^{ab}$, $\phi_m{}^{\alpha i}$ and $f_m{}^a$,
given in terms of the other fields, which are
associated with Lorentz, $S$-supersymmetry and special conformal gauge symmetries.
In the expressions \eqref{eq:HCLag1} and \eqref{eq:HCLag2}, these fields are contracted
with $\SU{2}_\trv$ harmonics, e.g.
$\psi_m^\pm = \psi_m^i v_i^\pm$, $\chi^\pm = \chi^i v_i^\pm$,
$\cV_m{}^{\pm\pm} = \cV_m{}^{ij} v_i^\pm v_j^\pm$, and so forth.

Now observe that interchanging the order of integration gives
\begin{align}\label{eq:AnalyticCompActionSwap}
S &= \frac{i}{2\pi} \int_\cS \cV^{++} \wedge \cW^{--} \omega^{(-2,2)}
	- \frac{i}{2\pi} \int_\cS \cV^{--} \wedge \cW^{--} \omega^{(2,2)}~, \eol
\omega^{(-2,2)} &= \int \rd^4x\, e\, \cL^{(-2,2)}~, \qquad
\omega^{(2,2)} = \int \rd^4x\, e\, \cL^{(2,2)}~.
\end{align}
The two-form
$\omega = \cV^{++} \wedge \cW^{--} \omega^{(-2,2)} - \cV^{--} \wedge \cW^{--} \omega^{(2,2)}$
is of the type discussed in section \ref{sec:CompS2}: it
is closed on $\cS$ and ensures that the action is insensitive
to small deformations of $\cS$. We may then restrict to $\cS = \rS$,
where the second integral in \eqref{eq:AnalyticCompActionSwap}
drops out, leaving the more conventional expression
\begin{align}
S = \int \rd^4x\, e\, \int_{\rS} \rd u\, \cL^{0}~, \qquad \cL^0 = \cL^{(-2,2)}\vert_\rS~.
\end{align}
This yields the most compact form for the component Lagrangian of a general curved
harmonic superspace action and constitutes another of our major results.

\subsection{Examples of complex harmonic superspace actions}
Below we will briefly summarize how the most common harmonic superspace actions
can be written in the covariant formulation here. Each of these cases is
a straightforward extension of a well-known action in conventional
harmonic superspace.

\subsubsection*{The $Q^+$ hypermultiplet and general self-couplings}
Introduce a family of hypermultiplets
$Q^{\ra +}$ with $\ra = 1, \cdots, 2n$, and impose a pseudoreality condition
$\widetilde{Q^{\ra +}} = - Q_\ra^+ \equiv - Q^{\rb +} \Omega_{\rb\ra}$
using the canonical symplectic form $\Omega_{\rb\ra}$ of $\Sp(n)$.
Interactions may be introduced in the form of a
twisted biholomorphic function $H^{(2,2)}$, so that the Lagrangian is
\begin{align}\label{eq:HLagSigma}
\mathscr{L}^{(2,2)} = \frac{1}{2} Q_{\ra}^+ \nabla_\trw^{++} Q^{\ra +} + H^{(2,2)}~.
\end{align}
We have denoted the potential term by $H^{(2,2)}$ as in \cite{Butter:HarmProj}
to emphasize its interpretation as a Hamiltonian \cite{GO:SSMHM}.
In flat space, it is natural to require $H^{(2,2)} = H^{(2,2)} (Q^+, v^+, w^-)$
to be analytic and twisted biholomorphic; however, when coupled to conformal
supergravity, it must have Weyl weight 2 and so
\begin{align}\label{eq:HamConds}
Q^{\ra +} \pa_{\ra+} H^{(2,2)} = 2 H^{(2,2)} \quad \implies \quad H^{(2,2)}= H^{(2,2)}(Q^+, w^-)~.
\end{align}
In other words, the Hamiltonian must not explicitly depend
on $v^{i+}$ \cite{Ketov:SC_Hypers, GIOS}.  The component sigma model corresponding
to this action describes a hyperk\"ahler cone.
The condition that $H^{(2,2)}$ cannot depend on $v^{i+}$ can also be interpreted
as requiring the Lagrangian to be a scalar function under $\gSU{2}_\trv$ diffeomorphisms.
A similar condition is required in projective superspace. In contrast, it is permissible
for the $w_i^-$ to appear because one may avoid ever using non-trivial $\gSU{2}_\trw$
diffeomorphisms.

\subsubsection*{The $\omega$ hypermultiplet action}
The free $\omega$ hypermultiplet action can be constructed from the
free $Q^{\ra+}$ hypermultiplet action with $\ra=1,2$
by making the change of variables 
$Q^{\ra+} = w^{\ra+} \omega^{(1,-1)} - w^{\ra-} f^{(1,1)}$.
One must employ the $w_i^\pm$ harmonics so that $\omega^{(1,-1)}$
and $f^{(1,1)}$ remain covariant under $\gSU{2}_\trv$ diffeomorphisms.
It is evident from this equation that $f^{(1,1)}$ is not
a twisted biholomorphic superfield but rather obeys the constraint
$\nabla_{\trw}^{--} f^{(1,1)} = \omega^{(1,-1)}$.
This can be remedied by replacing
$f^{(1,1)} \rightarrow f^{(1,1)} + \nabla_\trw^{++} \omega^{(1,-1)}$, so that
the new fundamental superfields are each twisted biholomorphic.
The Lagrangian remains twisted biholomorphic,
\begin{align}
\frac{1}{2} Q_\ra^+ \nabla_\trw^{++} Q^{\ra+} =
	\frac{1}{2} \omega^{(1,-1)} (\nabla_\trw^{++})^2 \omega^{(1,-1)}
	+ \frac{1}{2} (f^{(1,1)})^2
	+ \frac{1}{2} \nabla_\trw^{++} (\omega^{(1,-1)} f^{(1,1)})~.
\end{align}
The last term is a total derivative, and the second
term can be integrated out, leaving the free
$\omega$ hypermultiplet Lagrangian.

\subsubsection*{The improved tensor multiplet action}
As our last example, let us generalize the construction of the
improved tensor multiplet \cite{GIO:Duality} to complex harmonic superspace.
Starting with a free hypermultiplet action with a complex $Q^+$, one
makes the complicated change of variables
\begin{align}\label{eq:ImpTensorVars}
Q^{+} &= e^{i \omega} \Big(
	Q_0^+ - i \frac{w_\2^-}{\Omega_0} g^{++}
	\Big)~, \qquad
\widetilde Q^{+} = e^{-i \omega} \Big(
	\widetilde Q_0^+ - i \frac{w_\1^-}{\Omega_0} g^{++}
	\Big)~, \eol
\Omega_0 &\equiv \Omega_0^{(1,-1)} := Q_0^+ w_\1^- + \widetilde Q_0^+ w_\2^-~,
\end{align}
where $\omega$ and $g^{++}$ are two real dynamical variables and $Q_0^+$
is a new complex hypermultiplet. None of the fields $\omega$, $g^{++}$, or $Q_0^+$
carries $\gU{1}_\trw$ charge, and each is analytic, so $g^{++}$ must be Weyl weight two and
$Q^+$ Weyl weight one. However, because we have traded two real degrees of freedom for four,
this must be a redundant description, independent of some combination of the
new fields, which will become apparent in due course.
It is convenient to group $Q_0^+$ and $\widetilde Q_0^+$
into the pseudoreal doublet $Q_0^{i+} = (Q_0^+, \widetilde Q_0^+)$,
so that $\Omega_0$ can be written simply as $\Omega_0 = Q_0^{i+} w_i^-$.
Note that $\Omega_0$ is actually a weight $(1,-1)$ superfield,
but we have suppressed the charges for notational simplicity.

The above construction differs in two ways from the rigid version given in
\cite{GIO:Duality}. First, that version would correspond to choosing
a fixed $Q_0^{i+} = v^{i+}$. However, this choice is not possible
in any gauge other than the analytic one,
as it is generally inconsistent with the analytic condition
$\nabla_{\ul\alpha}^+ Q_0^{i+} = 0$ because of the presence of the non-vanishing
$\gSU{2}_\trv$ connection.\footnote{Actually, the precise gauge
choice made in conventional harmonic superspace is slightly different,
because $Q_0^{i+}$ (like all analytic superfields) is chosen to be a scalar
density.} The second difference is that
the change of variables in \cite{GIO:Duality} was more general,
involving an isotriplet $c_{ij}$. The simplifying choice we have made corresponds to
taking $c_{\1\2} = i/2$ and $c_{\1\1} = c_{\2\2} = 0$, with the non-canonical
normalization $c^2 := c^{ij} c_{ij} / 2 = 1/4$. Below we 
restore a more general $c_{ij}$ (but keeping this normalization).

After making the redefinition \eqref{eq:ImpTensorVars}, the free hypermultiplet
Lagrangian becomes
\begin{align}
\mathscr{L}^{(2,2)}
	&= \frac{1}{2} \frac{(g^{++})^2}{\Omega_0^2}
	- L^{++} \nabla_\trw^{++} \omega
	+ \frac{1}{2} Q_{0i}^+ \nabla_\trw^{++} Q_0^{i+}
	\eol & \quad
	- \frac{2g^{++} c_{ij} w^{i-} }{\Omega_0} \nabla_\trw^{++} Q_0^{j+}
	+ \nabla_\trw^{++} (g^{++} C^{+-})
\end{align}
where we have defined
\begin{align}
C^{++} &:= c_{ij} \,Q_0^{i+} Q_0^{j+}~, \qquad
C^{+-} := c_{ij} \,Q_0^{i+} w^{j-} / \Omega_0~, \qquad
C^{--} := c_{ij} \, w^{i-} w^{j-} / \Omega_0^2~, \eol
L^{++} &:= C^{++} + g^{++} + C^{--} g^{++}~.
\end{align}
The last term in the Lagrangian is a total derivative and can be
discarded. The second term, which involves $\omega$ as a Lagrange
multiplier, sets $L^{++}$ to be an $\cO(2)$ multiplet. This
determines $g^{++}$ in the usual form
\begin{align}
g^{++} = \frac{2 (L^{++} - C^{++})}{1+\sqrt{1+4 \,C^{--} (L^{++} - C^{++})}}~,
\end{align}
in terms of which the Lagrangian can be written
\begin{align}\label{eq:ImpTensorHarm}
\mathscr{L}^{(2,2)}
	= \frac{1}{2} \frac{(g^{++})^2}{\Omega_0^2}
	+ \frac{1}{2} Q_{0i}^+ \nabla_\trw^{++} Q_0^{i+}
	- 2 \frac{g^{++} c_{ij} w^{i-} }{\Omega_0} \nabla_\trw^{++} Q_0^{j+}~.
\end{align}

As already mentioned, the change of variables we have made is
equivalent to the standard choice in harmonic superspace,
except for the appearance of the new hypermultiplet $Q_0^{i+}$
rather than its frozen value.
It is possible to show that the action is actually independent
of this hypermultiplet. The proof is equivalent
to showing the conformal invariance of the conventional harmonic
action. (The conformal invariance of \eqref{eq:ImpTensorHarm} is manifest.)
Under an arbitrary variation of $Q_0^{i+}$, one simply shows that
$\mathscr{L}^{(2,2)}$ transforms as a total derivative,
\begin{align}
\delta \mathscr{L}^{(2,2)} &=
	\nabla_\trw^{++} \Big(
	\frac{1}{2} Q_{0i}^+ \delta Q_0^{i+}
	- \frac{2}{\Omega_0} g^{++} (\delta Q_0^{i+} w_i^- C^{+-} - \delta Q_0^{i+} Q_{0i}^+ C^{--})
	\Big)~.
\end{align}
When constructing the component action, the multiplet $Q_0^{i+}$ must drop out.

The necessity of this additional hypermultiplet can also be understood
by comparing the above construction to its curved projective superspace analogue
where similar features occur \cite{Kuzenko:ProjDualForms}.
Beginning with the free hypermultiplet Lagrangian
$\mathscr{L}^{++} = i \, \U^+ \breve \U^+$, with $\U^+$ an arctic multiplet
and $\breve \U^+$ its antarctic conjugate,
one introduces a redundant parametrization $\U^+ = \U_0^+ e^\L$
analogous to \eqref{eq:ImpTensorVars},
with $\U_0^+$ an arbitrary weight-one arctic superfield and $\L$ a weight-zero
arctic superfield. The free hypermultiplet Lagrangian can be rewritten as
\begin{align}
\mathscr{L}^{++} = i \U_0^+ \breve \U_0^+ e^{\L + \breve \L} - L^{++} (\L + \breve \L)~,
\end{align}
after relaxing the requirement that $\L + \breve \L$ is the sum of an arctic
and an antarctic superfield, enforcing it instead via the Lagrange multiplier
$L^{++}$. Integrating out the unconstrained real
analytic superfield $\L + \breve \L$,
one finds the improved tensor Lagrangian
\begin{align}
\mathscr{L}^{++} = L^{++} - L^{++} \log (L^{++} / i \U_0^+ \breve \U_0^+)~.
\end{align}
It is easy to show that this is actually independent of the choice of $\U_0^+$
in precise analogy to the spurious dependence of \eqref{eq:ImpTensorHarm} on
$Q_0^{i+}$.\footnote{One method is to use the argument of
\cite{KT-M:DiffReps} for the gauge invariance of the vector-tensor coupling.}
Its presence is necessary to ensure covariance of the Lagrangian, and in the appropriate
analytic basis it can be set to a constant.

\section{The analytic basis and conventional harmonic superspace} \label{sec:HSS}
In the previous sections, we have constructed a covariant formulation of
complex harmonic superspace on the supermanifold
$\cM^{4|8} \times \gSU{2}_\trv \times \gSU{2}_\trw$, with explicit
gauging of the superconformal group.
There is one major task which remains: we must explain how this
formulation is related to the prepotential approach of
\cite{GIOS:Sugra} and the covariant formulation discussed in
\cite{DelamotteKaplan:HHS, GKS:Sugra}.
The connection with \cite{GIOS:Sugra} is the easiest to elucidate
as it arises naturally upon going to the analytic basis (or analytic
gauge). After reviewing the analytic basis in the rigid limit, we
will construct it for the general curved supermanifold.
Afterwards, we will describe how to recover \cite{GIOS:Sugra}:
the key step will be to trade analytic scalar fields (which we have
used up until now) for analytic scalar densities.
In the final part of this section, we will sketch the relationship with
\cite{DelamotteKaplan:HHS, GKS:Sugra}.

\subsection{Rigid harmonic superspace and its analytic basis}
Understanding the structure of the analytic basis requires a brief
discussion of superconformal isometries. This material
largely follows \cite{GIOS:Conformal, GIOS}
and is inspired by the related construction in projective superspace \cite{Kuzenko:SPH}.
A corresponding discussion for the general class of $(\cN, p, q)$ superspaces was
given in \cite{HartwellHowe:Npq, HoweHartwell:Survey}.

In flat $\cN=2$ superspace, the superconformal transformation of
any covariant superfield $\Psi$ is
\begin{align}\label{eq:FlatdeltaPsi}
\delta \Psi &= \xi^A D_A \Psi
	+ \frac{1}{2} \l^{ab} M_{ba} \Psi
	+ \L_\trD \bbD \Psi + \L_\trA \bbA \Psi
	+ \l^i{}_j I^j{}_i \Psi
	+ \eta^{\ul\alpha i} S_{\ul\alpha i} \Psi
	+ \eps^a K_a \Psi~.
\end{align}
We are interested only in transformations that preserve a rigid background, meaning
that the fixed vielbeins and (vanishing) connections of flat $\cN=2$ superspace
must be preserved, $[\delta, D_A] = 0$,
from which a number of properties follow.
The parameters $\xi^A$ describe superconformal Killing vectors, with 
$\xi^a$ obeying the so-called \emph{master equation}
\begin{align}
D_{(\beta}^i \xi_{\alpha) \dalpha} = 0~, \qquad \bar D^{(\dbeta}_i \xi^{\dalpha) \alpha}=0~.
\end{align}
The other quantities in \eqref{eq:FlatdeltaPsi} turn out to be derived from $\xi^a$ \cite{Kuzenko:SPH}.

For a superfield $\Psi$ that depends on the harmonics $v_i^{\pm}$ and $w_i^\pm$,
the action of the $\SU{2}_R$ generators in the central basis is given by
\begin{align}
\l^i{}_j I^j{}_i \Psi
	&= -\l_\trv^{++} D_\trv^{--} \Psi
	+ \l_\trv^{0} D_\trv^0 \Psi
	+ \l_\trv^{--} D_\trv^{++} \Psi~,
\end{align}
with $\l_\trv^{\pm\pm} := \l^{ij} v_i^\pm v_j^\pm$ and $\l_\trv^0 := \l^{ij} v_i^+ v_j^-$.
The most important example is an analytic twisted biholomorphic
conformal primary $\cF^{(n,m)}$, whose full transformation law is
\begin{align}
\delta \cF^{(n,m)} = \xi^a D_a \cF^{(n,m)}
	- \xi^{\ul\alpha +} D_{\ul\alpha}^- \cF^{(n,m)}
	- \l_\trv^{++} D_\trv^{--} \cF^{(n,m)}
	+ n (\L_\trD + \l_\trv^0) \cF^{(n,m)}~,
\end{align}
where $\xi^{\ul \alpha \pm} = v_i^\pm \xi^{\ul\alpha i}$ and
$D_{\ul\alpha}^\pm := v_i^\pm D_{\ul\alpha}{}^i$.
In order for $\delta \cF^{(n,m)}$ to be analytic, one must have
\begin{align}\label{eq:KVRels}
\bar D^{\dalpha +} \xi^b = 2i (\bsigma^b)^{\dalpha \alpha} \xi_\alpha^+~, \quad
D_\alpha^+ \xi^{\beta +} = \delta_\alpha{}^\beta \l_\trv^{++}~, \quad
\bar D_\dalpha^+ \xi^{\beta +} = 0~, \quad
D_\alpha^+ (\L_\trD + \l_\trv^0) = 0~,
\end{align}
which are consequences of the master equation. The analytic quantity
$\Lambda := \L_\trD + \l_\trv^0$ may also be written
\begin{align}\label{eq:LambdaDef1}
\Lambda = \frac{1}{2} \Big(D_a \xi^a + D_{\ul\alpha}^- \xi^{\ul\alpha+}
	- D_\trv^{--} \l_\trv^{++} \Big)~.
\end{align}

Let us now introduce the analytic basis of rigid complex harmonic superspace.
We choose complex harmonic coordinates $u^{i\pm}$, $z^{\pm\pm}$ and $z^0$ defined as in
section \ref{sec:CompS2} and take
\begin{align}\label{eq:RigidABCoords}
\hat\q_{\ul\alpha}^+ &= \frac{\q_{\ul\alpha}^i v_i^+}{(v^+, w^-)} = \q_{\ul\alpha}^i u_i^+~, \qquad
\hat\q_{\ul\alpha}^- = \q_{\ul\alpha}^i w_i^- = \q_{\ul\alpha}^i u_i^-~, \eol
\hat x^m &= x^m - 2i \q^{(i} \sigma^m \bar\q^{j)} \frac{v_i \bar w_j}{(v,\bar w)}
	= x^m - 2i \q^{(i} \sigma^m \bar\q^{j)} u_i^+ u_j^-~.
\end{align}
The coordinates $\hat x^m$, $\hat \q_{\ul\alpha}^+$, and $u^\pm$ parametrize
the analytic superspace in the analytic basis.
Note that the Grassmann coordinates carry $\gU{1}_\trw$ charge, whereas the spinor derivatives
carry $\gU{1}_\trv$ charge. In this coordinate system, the superspace derivatives become
\begin{align}
D_a &= \hat \pa_m~, \qquad
D_{\ul\alpha}^+ = z_0 \,\frac{\pa}{\pa \hat \q^{\ul\alpha -}}~, \eol
D_{\alpha}^- &= - \frac{1}{z^0} \Big(\frac{\pa}{\pa \hat \q^{\alpha+}}
	- 2i (\sigma^m \hat{\bar \q}^-)_{\alpha} \,\hat\pa_m 
	- z^{--} \frac{\pa}{\pa\hat \q^{\alpha -}} \Big)~, \eol
\bar D_{\dalpha}^- &= - \frac{1}{z^0} \Big(\frac{\pa}{\pa \hat {\bar\q}^{\dalpha+}}
	+ 2i (\hat \q^{-}\sigma^m)_{\dalpha} \,\hat\pa_m 
	- z^{--} \frac{\pa}{\pa\hat \q^{\dalpha -}} \Big)~.
\end{align}
Three of the harmonic derivatives remain quite simple,
\begin{align}
D_\trv^{++} = (z^0)^2 \frac{\pa}{\pa z^{--}}~, \qquad
D_\trv^0 = z^0 \frac{\pa}{\pa z^0}~,\qquad
D_\trw^{--} = \frac{\pa}{\pa z^{++}}~,
\end{align}
with the others are moderately more complicated,
\begin{align}
D_\trw^{++} &= \pa_\tru^{++}
	- 2i (\hat \q^+ \sigma^m \hat {\bar\q}^+) \hat \hat\pa_m
	+ \hat \q^{\ul\alpha +} \pa_{\ul\alpha-}
	- z^{++} \pa_\tru^0
	\eol & \quad
	- (z^{++})^2 \frac{\pa}{\pa z^{++}}
	+ z^{++} z^0 \frac{\pa}{\pa z^0}
	+ (2 z^{++} z^{--} - 1) \frac{\pa}{\pa z^{--}}~, \eol
D_\trv^{--} &= \frac{1}{(z^0)^2} \Big(\pa_\tru^{--}
	- 2i (\hat \q^- \sigma^m \hat {\bar\q}^-) \hat \pa_m
	+ \hat \q^{\ul\alpha -} \pa_{\ul\alpha+}
	- \frac{\pa}{\pa z^{++}}
	+ z^{--} z^0 \frac{\pa}{\pa z^0}
	+ (z^{--})^2 \frac{\pa}{\pa z^{--}}\Big)~, \eol
D_\trw^0 &= \pa_\tru^0
	+ \hat \q^{\ul \alpha+} \hat \pa_{\ul \alpha +}
	- \hat \q^{\ul \alpha-} \hat \pa_{\ul \alpha -}
	+ 2 z^{++} \pa_{z^{++}} - 2 z^{--} \pa_{z^{--}}
	- z^0 \frac{\pa}{\pa z^0}~.
\end{align}

Any twisted biholomorphic superfield $\cF^{(n,m)}$ corresponds to a complex
harmonic superfield $\cF^{(n+m)}$ via
$\cF^{(n,m)} = (z^0)^n\, \cF^{(n+m)}(\hat x, \hat \q^+\!, u^\pm)$,
which transforms as
\begin{align}\label{eq:ABdeltaF}
\delta \cF^{(n+m)} &= \lambda^m \pa_m \cF^{(n+m)}
	+ \lambda^{\ul\alpha +} \pa_{\ul\alpha +} \cF^{(n+m)}
	- \l_\tru^{++} \pa_\tru^{--} \cF^{(n+m)}
	+ n \L \cF^{(n+m)}~, \eol
\lambda^m &= \xi^m -  2i (\xi^i \sigma^m \hat{\bar \q}^-) u_i^+
	- 2i (\hat \q^- \sigma^m \bar \xi^i) u_i^+
	+ 2i \lambda_\tru^{++} (\hat\q^- \sigma^m \hat{\bar \q}^-)~, \eol
\lambda^{\ul\alpha +} &= \xi^{\ul\alpha i} u_i^+ - \l_\tru^{++} \hat \q^{\ul\alpha -}~.
\end{align}
The parameters $\l^m$, $\l^{\ul\alpha+}$, $\l_\tru^{++}$, and $\L$ are
independent of $\q^{\ul\alpha-}$, $z^{++}$, $z^{--}$ and $z^0$.
They are interpreted as arising from the analytic general coordinate
transformation
\begin{align}\label{eq:RigidACT}
\delta^* u^{i+} = \l_\tru^{++}\, u^{i-}~, \qquad
\delta^* u_i^- =0~, \qquad
\delta^* \hat x^m = -\lambda^m~, \qquad
\delta^* \hat\q^{\ul\alpha+} = -\lambda^{\ul\alpha +}~,
\end{align}
on the analytic space $(\hat x, \hat \q^+, u^\pm)$,
with $\L$ given by half of the infinitesimal Berezinian,
$\L = \frac{1}{2}\Big(\pa_m \l^m - \pa_{\ul\alpha +} \l^{\ul \alpha +}
	- \pa_{\tru}^{--} \l_\tru^{++} \Big)$, equivalent to \eqref{eq:LambdaDef1}.
Any analytic superfield transforming as \eqref{eq:ABdeltaF} is called a
primary analytic scalar of weight $n$.
From the above results, one may show that the action
\begin{align}\label{eq:RigidABAction}
S = \frac{i}{2\pi} \int_\cS  \cU^{++} \wedge \cU^{--} \int \rd^4 \hat x\, \rd^4\hat\q^+\, \mathscr{L}^{+4}
\end{align}
is a superconformal invariant provided $\mathscr{L}^{+4}$ is an analytic scalar
of weight two. This is just the rigid formulation of conventional harmonic superspace
on the complexified $S^2$, and reproduces the rigid limit of the analytic
superspace action principle \eqref{eq:AnalyticSuperAction} after taking
\begin{align}
\mathscr{L}^{(2,2)} = (z^0)^2 \mathscr{L}^{+4}~, \qquad
\cE^{(-2,-2)} = (z^0)^{-2} \Big(\cU^{++}_\z \cU^{--}_{\tilde \z}
	- \cU^{++}_{\tilde \z} \cU^{--}_{\tilde \z}\Big)~.
\end{align}

\subsection{The analytic basis in complex harmonic superspace}\label{sec:CHS.AB}
In order to construct the curved analogue of \eqref{eq:RigidABAction}, we must
introduce the analytic basis in a curved geometry.
We assume that this analytic basis (or analytic gauge) is accomplished using a twisted biholomorphic
gauge transformation with gauge parameters that are homogeneous of degree zero in
$v^i$ and $\bar w_i$ separately. Schematically, the analytic gauge arises as
$\hat \nabla_\cA = e^{\cB} \nabla_\cA e^{-\cB}$
where $\nabla_\cA$ is the covariant derivative in central gauge
and $e^\cB$ is a bridge operator.
For the coordinates themselves, the bridge can be represented as a twisted
biholomorphic coordinate transformation,
taking the central basis $z^\cM$ to analytic basis $\hat z^\cM$
given by the generalization of \eqref{eq:RigidABCoords},
\begin{alignat}{3}\label{eq:ACoords}
\hat x^m &= x^m + b^m~, &\qquad
\hat \q^{\ul \mu +} &= \frac{\q^{\ul \mu i} v_i^+}{(v^+, w^-)} + b^{\ul \mu +}~, &\qquad
\hat \q^{\ul \mu -} &= \q^{\ul \mu i} w_i^- + b^{\ul \mu -}~, \eol
\hat v^{i\pm} &= v^{j\pm} b_j{}^i~, &\qquad
\hat w^{i \pm} &= w^{i \pm}~,
\end{alignat}
in terms of twisted biholomorphic bridges $b$ that depend only on the
harmonics $v^{i+}$ and $w_i^-$.  These bridges carry vanishing $\SU{2}_\trv$ charge
and their $\SU{2}_\trw$ charge is indicated; in addition, $b_j{}^i$ must have unit
determinant.
Because $\nabla_{\ul \alpha}^+$ annihilates $w^{i\pm}$ in the central
gauge (i.e. there is no $\SU{2}_\trw$ connection), it is evident that no bridge needs
to be introduced for those harmonics.
However, the presence of the $\SU{2}_\trv$ connection in the central basis requires
the bridges $b_j{}^i$ to be nonzero, because in the central basis
$\nabla_{A} v^{i+} = \cV_A{}^i{}_j \,v^{j+} \neq 0$.

As a consequence of the twisted biholomorphy of the bridges, one can see that
a number of harmonic derivatives take simple forms in the analytic basis:
\begin{align}
\hat \nabla_\trw^0 = \hat \pa_\trw^0
	 + \hat\q^{\ul \mu +} \hat\pa_{\ul\mu+}
	- \hat\q^{\ul \mu -} \hat\pa_{\ul \mu -}~, \qquad
\hat \nabla_\trw^{--} = \hat \pa_\trw^{--},  \qquad
\hat\nabla_\trv^0 = \hat\pa_\trv^0~, \qquad
\hat \nabla_\trv^{++} = \hat \pa_{\trv}^{++}~.
\end{align}
In addition, there is no obstruction to choosing $\hat\nabla_{\ul\alpha}^+$ to
simply be given by
\begin{align}\label{eq:StrongABCond}
\hat \nabla_{\ul\alpha}^+ = (\hat v^+, w^-)\, \delta_{\ul\alpha}{}^{\ul\mu} \,
	\frac{\pa}{\pa \hat \q^{\ul \mu-}}~.
\end{align}
This involves not just a choice of analytic coordinates but also a choice of
all the other gauges as well, both to trivialize the vielbein terms and to
eliminate the connections; this is possible as a consequence of
\eqref{eq:AnalyticCommutes}.\footnote{We took in \cite{Butter:CSG4d.Proj}
the analytic gauge $\nabla_{\ul\alpha}^+ = \pa_{\ul\alpha}^+$, choosing the $\q^+$ coordinates to
carry $\gU{1}_\trv$ charge. This choice is also possible
here at the cost of breaking with standard harmonic superspace conventions.}
The expression for $\hat \nabla_\trw^{++}$ is more elaborate:
\begin{align}\label{eq:ABDw1}
\hat \nabla_\trw^{++} &\!= \hat \pa_\trw^{++}
	\!- H^{(2,2)} \hat \pa_{\trv}^{--}
	\!+ H^{(0,2)} \hat \pa_\trv^0
	\!+ H^{(-2,2)} \hat \pa_{\trv}^{++}
	\!+ H^{(0,2)m } \hat \pa_m
	\!+ H^{(0, 3) \ul \mu} \hat \pa_{\ul \mu +}
	\!+ H^{(0, 1) \ul \mu} \hat \pa_{\ul \mu -}
	\eol & \quad
	- \frac{1}{2} \Omega_\trw^{++ ab} M_{ba}
	- A_\trw^{++} \mathbb A
	- B_\trw^{++} \mathbb D
	- F^{(0,2) a} K_a
	- F^{(1,2) \ul\alpha} S_{\ul\alpha}^-
	+ F^{(-1,2) \ul\alpha} S_{\ul\alpha}^+~.
\end{align}
The contributions to the vielbein, denoted by $H$,
coincide with the similarly-named objects
in the conventional harmonic superspace approach \cite{GIOS:Sugra}
(and which were introduced earlier in \cite{GIOS:Conformal})
up to redefinitions to be discussed in a moment.

A key observation one should make about \eqref{eq:ABDw1} is the absence of
additional contributions involving $\hat \pa_\trw^0$ and $\hat\pa_\trw^{--}$.
This is a consequence of the simple form for the analytic basis
for the $\gSU{2}_\trw$ harmonics.
However, there is a complication hidden within \eqref{eq:ABDw1}:
not all of the connections are twisted biholomorphic.
In particular, 
\begin{gather}
\hat \pa_\trv^{++} F^{(-1, 2) \ul\alpha} = F^{(1, 2) \ul\alpha}~, \qquad
\hat \pa_\trv^{++} F^{(1, 2) \ul\alpha} = 0~, \eol
\hat \pa_\trv^{++} H^{(-2,2)} = 2 H^{(0,2)}~, \qquad
\hat \pa_\trv^{++} H^{(0,2)} = H^{(2,2)}~, \qquad
\hat \pa_\trv^{++} H^{(2,2)} = 0 ~,
\label{eq:ABConnsDpp}
\end{gather}
while the vielbeins $H^{(0,3)\ul\mu}$ and $H^{(0,1)\ul\mu}$ obey
\begin{align}
\hat \pa_\trw^{--} H^{(0,3)\ul\mu} = -\hat \q^{\ul\mu +}~, \qquad
\hat \pa_\trw^{--} H^{(0,1)\ul\mu} = -\hat \q^{\ul\mu -}~.
\end{align}
To make these features explicit, it will be useful to adopt a change of coordinates
to the complex variables $u^{i\pm}$, $z^{\pm\pm}$, and $z^0$ defined in
section \ref{sec:HA_emergence}. Analogous definitions in the analytic basis
of $\hat u^{i\pm}$, $\hat z^{\pm\pm}$, and $\hat z^0$ in terms of
$\hat v^{i\pm}$ and $\hat w^{i\pm}$ leads to
\begin{alignat}{3}\label{eq:ACoords2}
\hat z^{++} &= z^{++} + b^{++}~, &\qquad
\hat z^0 &= z^0\, b^0~, &\qquad
\hat z^{--} &= (b^0)^2 (z^{--} + b^{--})~, \eol
\hat u^{i+} &= u^{i+} - b^{++} u^{i-}~, &\qquad
\hat u_i^- &= u_i^-~.
\end{alignat}
The three bridges $b^{\pm\pm}$ and $b^0$ are nonlinearly related to the
$\gSU{2}_\trv$ bridge $b_i{}^j$. The bridges $b^{++}$, $b^m$, $b^{\ul\mu+}$
and $b^{\ul\mu -}$ appeared in the harmonic superspace context in \cite{GIOS:Sugra};
the bridges $b^0$ and $b^{--}$ did not appear there because the additional complex
coordinates $\hat z^0$ and $\hat z^{--}$ are not needed to describe analytic
superfields.\footnote{These bridges match those given
in \cite{GIO:QK} for the complex harmonic description of
quaternionic sigma models. This is a natural consequence of
the biholomorphic analyticity assumed there, which we follow.}
In terms of these coordinates, we now have
\begin{align}
\hat \nabla_\trw^0 &= \hat \pa_\tru^0
	+ 2 \hat z^{++} \pa_{\hat z^{++}} - 2 \hat z^{--} \hat \pa_{z^{--}}
	- \hat z^0 \hat \pa_{z^0}
	+ \hat \q^{\ul \mu +} \hat \pa_{\ul\mu+}
	- \hat \q^{\ul \mu -} \hat \pa_{\ul \mu -}~, \eol
\hat \nabla_\trw^{--} &= \frac{\pa}{\pa \hat z^{++}}~, \qquad
\hat \nabla_\trv^{++} = (\hat z^0)^2 \frac{\pa}{\pa \hat z^{--}}~, \qquad
\hat\nabla_\trv^0 = \hat z^0 \frac{\pa}{\pa \hat z^0}~, \eol
\hat \nabla_{\ul\alpha}^+ &= \hat z^0 \delta_{\ul\alpha}{}^{\ul\mu}\,
	\frac{\pa}{\pa \hat \q^{\ul\mu -}}~.
\end{align}
To simplify the conditions \eqref{eq:ABConnsDpp} on the connections for $\nabla_\trw^{++}$,
it will be useful to introduce new vielbeins $\cH$ that are independent of $\hat z^{\pm\pm}$
and $\hat z^0$:
\begin{align}
H^{m(0,2)} &\equiv \cH^{++ m}~, \quad
H^{(0,3) \ul\mu} \equiv \cH^{+++ \ul\mu} - \hat z^{++} \,\hat \q^{\ul\mu +}~, \quad
H^{(0,1) \ul\mu} \equiv \cH^{+ \ul\mu} - \hat z^{++} \,\hat \q^{\ul\mu -}~, \eol
\frac{H^{(2,2)}}{(\hat z^0)^2} &\equiv \cH^{+4}~, \qquad
H^{(0,2)} \equiv \cH^{++} + \hat z^{--} \cH^{+4}~, \eol
(\hat z^0)^2 H^{(-2,2)} &\equiv \cH^{0} + 2 \hat z^{--} \cH^{++} + (\hat z^{--})^2 \cH^{+4}~.
\end{align}
It is convenient to similarly modify the $S$-supersymmetry connections as well:
\begin{align}
F^{(1,2) \ul\alpha} \equiv \hat z^0\, \cF^{+++ \ul \alpha}~, \qquad
F^{(-1,2) \ul\alpha} \equiv \frac{1}{\hat z^0}\, \Big(\cF^{+ \ul\alpha} + \hat z^{--} \cF^{+++\ul\alpha}\Big)~.
\end{align}
Then \eqref{eq:ABDw1} may be rewritten
\begin{align}\label{eq:ABDw2}
\hat \nabla_\trw^{++} &=
	\hat \pa_\tru^{++}
	- \hat z^{++} \hat\nabla_\trw^0
	+ (\hat z^{++})^2 \hat\nabla_\trw^{--}
	+ \cH^{+4} \Big(\hat \pa_{z^{++}} -\hat \pa_\tru^{--} \Big)
	+ \cH^{++} \Big( \hat\nabla_\trv^0 + 2 \hat z^{--} \hat \pa_{z^{--}} \Big)
	\eol & \quad
	+ (\cH^0 - 1) \hat \pa_{z^{--}}
	+ \cH^{++ m} \hat \pa_m
	+ \cH^{+++ \ul\mu} \hat \pa_{\ul \mu +}
	+ \cH^{+ \ul\mu} \hat \pa_{\ul \mu -}
	- \frac{1}{2} \Omega_\trw^{++ ab} M_{ba}
	\eol & \quad
	- A_\trw^{++} \mathbb A
	- B_\trw^{++} \mathbb D
	- F_\trw^{++ a} K_a
	+ \cF^{+\ul\alpha} S_{\ul\alpha}^+
	- \cF^{+++ \ul\alpha}
		\Big(\hat z^0 S_{\ul\alpha}^- - \frac{\hat z^{--}}{\hat z^0} S_{\ul\alpha}^+\Big)~.
\end{align}
Each of the connection superfields above is twisted biholomorphic
with vanishing $\gU{1}_\trv$ charge, depending only on the complex harmonics.
Up to sign conventions, the superfields $\cH^{+4}$, $\cH^{+++ \ul\mu}$, $\cH^{++ m}$, and
$\cH^{+\ul\mu}$
coincide with similarly named objects in the conventional harmonic superspace description
of conformal supergravity \cite{GIOS:Conformal, GIOS:Sugra}. The importance of the connections
$B_\trw^{++}$ and $\cH^{++}$ will be addressed in the next subsection.
The absence of the remaining connections in \cite{GIOS:Sugra}
is apparent when one recalls that $\hat\nabla_\trw^{++}$
always acts on analytic twisted biholomorphic primary scalars, and so these
connections tend to drop out.

At this stage, we could proceed further and analyze the full structure of the
analytic basis, including the connections within $\hat\nabla_\trv^{--}$. However,
this will not be directly necessary: our main motivation is to provide a covariant
description, and a full construction of the analytic basis connections is not
necessary for that purpose. However, it may be useful to note the consequences
of $[\hat \nabla_\trw^{++}, \hat \nabla_{\ul\alpha}^+] = 0$.
One finds (using $\pa_{\ul\alpha}^+ \equiv \pa_{\ul\alpha -}$) that
several of the vielbeins and the combination $\cH^{++} - B_\trw^{++}$
are analytic,
\begin{align}
\pa_{\ul\alpha}^+ \cH^{+4} = \pa_{\ul\alpha}^+ \cH^{++m} = \pa_{\ul\alpha}^+ \cH^{+++\ul\mu} = 
\pa_{\ul\alpha}^+ (\cH^{++} - B_\trw^{++}) = 0~,
\end{align}
while the other vielbeins are less strongly constrained,
\begin{align}\label{eq:ABOtherH}
\pa_\alpha^+ \cH^{+ \dot\mu} = 0~, \qquad
(\pa^+)^2 \cH^0 = (\bar \pa^+)^2 \cH^0 = 0~.
\end{align}
The other connections are entirely determined in terms of the vielbeins as
\begin{alignat}{2}\label{eq:ABOtherC}
B_\trw^{++}  &= 2 \,\cH^{++} - \frac{1}{2} (\pa_{\alpha}^+ \cH^{\alpha +} + \bar\pa_\dalpha^+ \cH^{\dalpha +})~, &\qquad
\Omega_\trw^{++}{}_{\alpha\beta} &= -\pa_{(\alpha}^+ \cH_{\beta)}^+~, \eol
A_\trw^{++} &= -\frac{i}{4} (\pa_{\alpha}^+ \cH^{\alpha +} - \bar\pa_\dalpha^+ \cH^{\dalpha +})~, &\qquad
F_\trw^{++}{}_{\alpha\dbeta} &= -\frac{i}{4} \pa_{\alpha}^+ \bar\pa_\dbeta^+ \cH^0~, \eol
\cF^{+++ \alpha} &= -\frac{1}{8} (\pa^+)^2 \cH^{\alpha +}~, &\qquad
\cF^{+\alpha} &= -\frac{1}{4} \pa^{\alpha +} \cH^0~.
\end{alignat}
The conditions \eqref{eq:ABOtherH} and \eqref{eq:ABOtherC}
are a consequence of the strong gauge choice \eqref{eq:StrongABCond} made for
$\nabla_{\ul\alpha}^+$. These can be simplified still further by imposing the additional
gauges $\cH^{++} = \cH^0 = 0$ and $\cH^{\ul\mu +} = \hat \q^{\ul \mu+}$,
which imply that $\hat z^{0}$, $\hat z^{--}$, and $\hat \q^{\ul \mu-}$ are respectively
chosen to solve $\nabla_\trw^{++} \log \hat z^0 = \hat z^{++}$,
$\nabla_\trw^{++} \Big(\hat z^{--} / (\hat z^0)^2 \Big) = -1 / (\hat z^0)^2$, and
$\nabla_\trw^{++} (\hat \q^{\ul \mu-} / \hat z^0) = \hat \q^{\ul \mu+} / \hat z^0$.

\subsection{Curved harmonic superspace and analytic densities}
In the conventional formulation of curved harmonic superspace \cite{GIOS:Sugra}, one remains
in the analytic gauge while generalizing the transformations \eqref{eq:RigidACT} and the
action \eqref{eq:RigidABAction}. That is, on the analytic superspace, the coordinates
$\hat x$, $\hat \q^+$, and $\hat u^{\pm}$ transform as
\begin{align}\label{eq:CurvedACT}
\delta^* \hat u^{i+} = \hat \l^{++} \hat u^{i-}~, \qquad \delta^* \hat u^{i-} = 0~, \qquad
\delta^* \hat x^m = -\hat \l^m~, \qquad \delta^* \hat \q^{\ul \alpha +} = -\hat \l^{\ul\alpha +}~,
\end{align}
involving unconstrained analytic parameters $\hat\l$. The action
\begin{align}\label{eq:ActionAB}
S = \frac{i}{2\pi} \int_\cS  \hat \cU^{++} \wedge \hat \cU^{--}
	\int \rd^4 \hat x\, \rd^4\hat \q^+\, \hat{\mathscr{L}}^{+4}
\end{align}
is invariant provided $\hat{\mathscr{L}}^{+4}$ transforms as a primary analytic scalar of weight two,
\begin{align}\label{eq:deltaLagAB}
\delta \hat {\mathscr{L}}^{+4} = \Big(\hat \l^m \hat \pa_m
	+ \hat \l^{\ul \alpha+} \hat \pa_{\ul\alpha+}
	- \hat \l^{++} \hat \pa_{\tru}^{--} \Big) \hat{\mathscr{L}}^{+4}
	+ 2 \hat \L \hat{\mathscr{L}}^{+4}~,
\end{align}
where
$2\hat \Lambda := \hat \pa_m \hat \l^m - \hat \pa_{\ul\alpha +} \hat \l^{\ul \alpha +} - \hat \pa_{\tru}^{--} \hat \l^{++}$.

To establish the action above, it helps to slightly modify our formulation
\eqref{eq:AnalyticSuperAction} for the action in a general gauge. Instead of the measure
factor $\rd^2\z$ built out of the coordinates $\z$ and $\tilde\z$
parametrizing $\cS$ directly, we wish to use the complex harmonic measure
$\cU^{++} \wedge \cU^{--}$, with the complex harmonics $u^\pm$
implicitly depending on $\z$ and $\tilde\z$. So we first reexpand the vielbein
one-forms using the complex harmonic coordinates:
\begin{align}
E^{\cA} &= \rd z^M E_M{}^{\cA}
		+ \cU^{++} E_{\tru ++}{}^{\cA}
		+ \cU^{0} E_{\tru 0}{}^{\cA}
		+ \cU^{--} E_{\tru --}{}^{\cA}
		\eol & \quad
		+ \rd z^{++} E_{z^{++}}{}^{\cA}
		+ \rd z^{--} E_{z^{--}}{}^{\cA}
		+ \rd z^{0} E_{z^{0}}{}^{\cA}~.
\end{align}
The one-forms $\cU^{++}$, $\cU^{--}$ and $\cU^0$ constitute a fixed anholonomic frame.
Then the analytic superspace action \eqref{eq:AnalyticSuperAction} becomes
\begin{align}
&\frac{i}{2\pi} \int_\cS \cU^{++} \wedge \cU^{--}
	\int \rd^4x\, \rd^4\q^+\, \cE^{(-2, -2)} \mathscr{L}^{(2,2)}~, \eol
&\cE^{(-2,-2)} = 
\sdet \begin{pmatrix}
E_m{}^a & E_m{}^{\ul \alpha +} & E_m{}^{\trv ++} & E_m{}^{\trw --}\\[0.3em]
E_{\ul\mu +}{}^a & E_{\ul\mu +}{}^{\ul \alpha +} & E_{\ul\mu +}{}^{\trv ++} & E_{\ul\mu +}{}^{\trw --} \\[0.3em]
E_{\tru ++}{}^a & E_{\tru ++}{}^{\ul \alpha +} & E_{\tru ++}{}^{\trv ++} & E_{\tru ++}{}^{\trw --} \\[0.3em]
E_{\tru --}{}^a & E_{\tru --}{}^{\ul \alpha +} & E_{\tru --}{}^{\trv ++} & E_{\tru --}{}^{\trw --}
\end{pmatrix}~.
\end{align}
Under a diffeomorphism, one still finds
$\delta E_{\cM}{}^{\cA} = \xi^{\cN} \pa_{\cN} E_{\cM}{}^{\cA}
	+ \pa_{\cM} \xi^{\cN} E_{\cN}{}^{\cA}$
but must interpret $\pa_{\cM}$ as including the non-commuting derivatives
$\pa_{\tru}^{\pm\pm}$ and $\pa_\tru^0$. In particular, the measure $\cE^{(-2,-2)}$
transforms under diffeomorphisms of the coordinates $x$, $\q^+$, $u_i^\pm$ as
\begin{align}
\delta \cE^{(-2,-2)} = (-)^{\bf m} \pa_{\bf m} \Big(\xi^{\bf m} \cE^{(-2,-2)}\Big)~, \qquad
{\bf m} = (m, \ul\mu +, {\rm u} +\!\!+, {\rm u} -\!- )~
\end{align}

Now let us go to the analytic basis. Using the results in Appendix \ref{App:AI&D}, one can
show that within the analytic gauge, $\hat \cE^{(-2,-2)}$ is independent of
$\hat\q^{\ul\mu-}$ and $\hat z^{\pm\pm}$,
\begin{align}
{\hat\nabla}_{\ul\alpha}^+ \hat\cE^{(-2,-2)} =
	{\hat\nabla}_\trv^{++} \hat\cE^{(-2,-2)} =
	{\hat\nabla}_\trw^{--} \hat\cE^{(-2,-2)} = 0
\end{align}
and possesses charge $(-2, +2)$ under the action of the $\gU{1}_\trv \times \gU{1}_\trw$ derivatives:
\begin{align}
\hat\nabla_\trv^0 \hat\cE^{(-2,-2)} = -2 \,\hat\cE^{(-2,-2)}~, \qquad
\hat\nabla_\trw^0 \hat\cE^{(-2,-2)} = +2 \,\hat\cE^{(-2,-2)}~.
\end{align}
Note that these differential conditions are distinct from the transformation
properties of the measure under covariant diffeomorphisms. The above
conditions hold only in the analytic gauge and arise as a consequence of the
explicit way we have chosen the Grassmann coordinates.
Now combine the scalar Lagrangian with the measure to give the Lagrangian density
with (analytic) charge $(0, +4)$,
\begin{align}
\hat {\mathscr{L}}^{+4} = \hat \cE^{(-2,-2)} \, \mathscr{L}^{(2,2)}~. \qquad
\hat\nabla_\trv^0 \hat\cL^{+4} = 0~, \qquad
\hat\nabla_\trw^0 \hat\cL^{+4} = +4 \,\hat\cL^{+4}~.
\end{align}
It transforms under analytic coordinate transformations precisely as in \eqref{eq:deltaLagAB}.
For the invariant action, we recover \eqref{eq:ActionAB}.

\subsection{Examples in the analytic basis}
Let us verify agreement with \cite{GIOS:Sugra} by constructing two simple actions
in the analytic basis.

\subsubsection*{The $Q^+$ hypermultiplet action}
We start with the general form of the $Q^{\ra+}$ hypermultiplet action
\begin{align}
\mathscr{L}^{(2,2)} = \frac{1}{2} L_\ra^+ \nabla_\trw^{++} Q^{\ra +} + H^{(2,2)}~.
\end{align}
$Q^{\ra+}$ is a scalar multiplet of weight 1, while $L_\ra^+$ and $H^{(2,2)}$
are functions of the appropriate homogeneities for their weights.
Let us denote $\hat \cE \equiv \hat \cE^{(-2,-2)}$ in this section to keep
notation relatively simple. Then we make the change of variables
to the scalar density $\widehat Q^{\ra+} = (\hat \cE)^{1/2}\, Q^{\ra+}$.
Using the conformal properties of the fields, we find
\begin{align}
L_\ra^+ (Q, w^-) = (\hat \cE)^{-1/2} L_\ra^{+} (\widehat Q^+, w^-)~, \qquad
H^{(2,2)} = (\hat \cE)^{-1} H^{+4} (\widehat Q^+, w^-)~.
\end{align}
Then the analytic Lagrangian $\hat {\mathscr{L}}^{+4} \equiv \hat\cE \mathscr{L}^{(2,2)}$ can be written
\begin{align}
\hat {\mathscr{L}}^{+4}
	&= \frac{1}{2} L_\ra^+\, \hat \cE^{1/2}
	\Big(\hat\cD^{++}
	+ \,(\cH^{++} - B_\trw^{++})\Big) \frac{\widehat Q^{\ra+}}{\hat \cE^{1/2}}
	+ H^{+4}~, \eol
\hat \cD^{++} &\equiv 
	\hat \pa_\tru^{++} 
	- \cH^{+4} \hat \pa_\tru^{--}
	+ \cH^{++ m} \hat \pa_m
	+ \cH^{+++ \mu} \hat \pa_{\mu +}~.
\end{align}
To proceed further, we require the following identity in the analytic basis:
\begin{align}
\hat\cD^{++} \hat \cE
	- 2 (\cH^{++} - B_\trw^{++}) \hat \cE
	= \Big(\hat \pa_\tru^{--} \cH^{+4} 
	- \hat \pa_m \cH^{++ m} 
	+ \hat \pa_{\mu +} \cH^{+++ \mu}  \Big)\hat \cE
	\equiv -\Gamma^{++} \hat \cE~.
\end{align}
Its proof follows by noting that the harmonic superspace integral of
\begin{align}
\hat \cE \hat \nabla_\trw^{++} \cF^{(2,0)}
	&= \hat \cE \Big(
	\hat \cD^{++} \cF^{(2,0)}
	+ 2 \,(\cH^{++} - B_\trw^{++}) \,\cF^{(2,0)} \Big)
\end{align}
is a total derivative for any $\cF^{(2,0)}$. Then we recover
\begin{align}
\hat {\mathscr{L}}^{+4}
	= \frac{1}{2} L_\ra^+ (\hat\cD^{++} + \frac{1}{2} \G^{++}) \widehat Q^{\ra+} + H^{+4}~.
\end{align}
This is the correct analytic basis action, expressed in terms of scalar
densities $\widehat Q^{\ra+}$. As anticipated, the connection
$(\cH^{++} - B_\trw^{++})$ has been replaced
by the analytic quantity $\frac{1}{2} \Gamma^{++}$.

\subsubsection*{The improved tensor multiplet action}
As a similar exercise, we now show how to recover the improved tensor multiplet action
in the analytic basis. Beginning with \eqref{eq:ImpTensorHarm},
we must make the change of variables
$\widehat Q_0^{i+} = (\hat \cE)^{1/2} \, Q_0^{i+}$ and
$\widehat L^{++} = \hat \cE\, L^{++}$. Then the analytic Lagrangian density becomes
\begin{align}
\hat {\mathscr{L}}^{+4}
	= \frac{1}{2} \frac{(\widehat g^{++})^2}{(\widehat  \Omega_0)^2}
	+ \frac{1}{2} \widehat Q_{0i}^+ \,\hat\cD^{++} \widehat Q_0^{i+}
	- 2 \frac{\widehat g^{++} c_{ij} w^{i-} }{\widehat \Omega_0} (\hat \cD^{++} + \frac{1}{2} \G^{++}) \widehat Q_0^{j+}~.
\end{align}
We know that $\widehat Q_0^{i+}$ may be any analytic function: in the analytic
gauge, we can simply choose it to be $\hat u^{i+}$. Doing so, it is not hard to
show
\begin{align}
\hat {\mathscr{L}}^{+4}
	= \frac{1}{2} (\widehat g^{++})^2
	- \G^{++} C^{-+} \widehat g^{++}
	+ \frac{1}{2} \cH^{+4} (1 + 4 \,\widehat g^{++} C^{--})~.
\end{align}
This matches the action of \cite{GIO:Duality} up to the
redefinitions $C^{++} \rightarrow \frac{1}{2} C^{++}$
and $L^{++} \rightarrow \frac{1}{2} L^{++}$, the difference in sign
$\cH^{+4} \rightarrow -\cH^{+4}$, and an overall normalization.

\subsection{Harmonic superspace on $\cM^{4|8} \times S^2$}

Finally, let us address the relationship between the covariant formulation
presented here and that of Galperin, Ky, and Sokatchev \cite{GKS:Sugra}
(see also \cite{DelamotteKaplan:HHS}). These papers established that within
a superspace gauging only the Lorentz group and
superdiffeomorphisms, the integrability conditions for analytic superfields
coincided with the torsion constraints of the simplest version of $\cN=2$
Einstein supergravity \cite{FV:N2Sugra, dWvH:N2Sugra}.
Because of absence of $\SU{2}_R$ connections,
one can introduce harmonics as external coordinates annihilated by the
covariant spinor derivatives. The harmonics are identified simply with
the external automorphism group $\SU{2}_A$, and so a single set of real
harmonics and a real auxiliary $S^2$ is sufficient.
Within this framework, the analytic prepotentials of Einstein supergravity
were uncovered and explicit formulae for connections and vielbeins
were worked out. However, certain features were not explored.
A general component action analogous to \eqref{eq:CompAction}
was not given, nor were the superspace actions cast in a manifestly covariant form.
Although it is clear that such results could be constructed directly
as we have done using conformal superspace,
it is instructive to consider instead how to recover this
Poincar\'e harmonic superspace framework directly from the conformal one.

Recall that this $\cN=2$ Einstein supergravity corresponds to conformal
supergravity coupled to two compensators: a vector multiplet $W$ and a non-linear multiplet
$L_\ra{}^i$.  Their lowest components fix the dilatation
and the $\SU{2} \times \gU{1}$ $R$-symmetries and lead to an off-shell
supergravity involving only gauged Lorentz transformations, diffeomorphisms,
and supersymmetry. One of the lowest fermions fixes the $S$-supersymmetry
gauge and the other becomes a dimension-1/2 matter field.

To describe the same supergravity at the superfield level, one can introduce
superfields $W$ and $L_\ra{}^i$ in conformal superspace and adopt the
dilatation-$\gU{1}_R$ gauge $W=1$ and the $\gSU{2}_R$ gauge $L_\ra{}^i = \delta_\ra{}^i$.
Actually, it is instructive to construct the new superspace not by gauge-fixing
explicitly but by using the superfields to redefine the vielbein and spin connection
to compensate the symmetries that should be fixed.
We may do this in two steps. First, using the superfield $W$, convert conformal
superspace derivatives $\nabla_A$ to $\SU{2}$ superspace derivatives $\cD_A$
(see \cite{BN:CR} for the $\cN=2$ case or \cite{BK:Dual} for a pedagogical
discussion in $\cN=1$). The $\cD_A$ carry no dilatation weight or $\gU{1}_R$ charge
and are conformally inert. Provided one acts only on
primary superfields whose Weyl and $\gU{1}_R$ weights have been removed with
the compensator, the dilatation, $\gU{1}_R$ and special conformal connections
become inert.

Next, using $\SU{2}$ superspace, we can repeat the construction of
covariant harmonic superspace. The main details do not change; in particular,
because the supergeometry includes the $\SU{2}$ $R$-symmetry group,
two sets of real harmonics are still necessary. The algebra of spinor covariant derivatives
$\cD_{\ul\alpha}^+$ is \cite{KLRT-M1}
\begin{align}
\{\cD_\alpha^+, \cD_\beta^+\} = 2 S^{++} M_{\alpha\beta} + 4 Y_{\alpha\beta} \cD_\trv^{++}~,\qquad
\{\cD_\alpha^+, \bar\cD_\dbeta^+ \} = G_{\alpha \dbeta} \cD_\trv^{++}~,
\end{align}
consistent with the existence of twisted biholomorphic analytic scalars.
Now we may introduce the second compensator and build $\gSU{2}_R$-inert derivatives
in the central basis:
\begin{align}\label{eq:WTcD}
\widetilde \cD_{\ul \alpha \ra} = L_\ra{}^{i} \Big(\cD_{\ul\alpha i}
	+ (\cD_{\ul\alpha i} L_\rb{}^j) L^\rb{}_k I^k{}_j \Big)~.
\end{align}
The second term modifies the $\gSU{2}_R$ connection so that $L_\ra{}^i$
is covariantly constant. In the gauge where $L_\ra{}^i = \delta_\ra{}^i$,
its effect is to eliminate the $\gSU{2}_R$ connection entirely.
An important check is to verify that the algebra of covariant derivatives still
permits analytic multiplets. One finds that\footnote{If $L_\ra{}^i$ were a
general superfield with $\det L_\ra{}^i = 1$
but not obeying the constraint of a non-linear superfield,
more complicated torsion terms would forbid analytic multiplets. This is one
way of understanding why the $\gSU{2}$ compensator must be a non-linear
multiplet.}
\begin{align}
\{\widetilde \cD_\alpha^{(\ra}, \widetilde \cD_\beta^{\rb)}\}
	= \Psi_{(\alpha}^{(\ra} \widetilde \cD_{\beta)}^{\rb)}
	+ \text{curvatures}~,
\end{align}
where the fermion $\Psi_\alpha^\ra$ is the dimension-1/2 fermion present in the
non-linear compensator, now reinterpreted as a torsion superfield.
We refer to the superspace associated with $\widetilde \cD_A$ as
``Einstein superspace''.\footnote{It is also possible to further modify the definition \eqref{eq:WTcD}
by an additional spin connection piece to eliminate the dimension-1/2 torsion.
Doing so recovers precisely the supergeometry employed in \cite{DelamotteKaplan:HHS, GKS:Sugra}
(except for the central charge, which is easy to introduce).
This supergeometry was introduced in \cite{BreitenlohnerSohnius, Gates:SdSWN2Sugra}.}

This shows how to arrive at the right supergeometry from the central
basis, but what about from a more general gauge?
Let the general gauge covariant derivatives in complex harmonic $\gSU{2}$ superspace be denoted
$\cD_\cA = (\cD_A, \cD_{\trv \ul a}, \cD_{\trw \bar a})$.
The non-linear multiplet superfields $L^{\pm \pm}$ and $L^{\mp \pm}$
carry both $\gU{1}_\trw$ and $\gU{1}_\trv$ charge
(in the central basis these are $L^{\ra i} w_\ra^\pm v_i^\pm$ and $L^{\ra i} w_\ra^\mp v_i^\pm$).
We introduce the compensated harmonic derivatives
\begin{align}
\widetilde \cD^{++} &:= \cD_\trw^{++} + \frac{1}{(L^{-+})^2} \cD_\trv^{++}
	+ N^{++} \cD_\trw^0
	- (N^{++})^2 \cD_\trw^{--}~, \eol
\widetilde \cD^0 &:= \cD_\trv^0 + \cD_\trw^0 + 2 \frac{L^{--}}{L^{-+}} \cD_\trv^{++}
	- 2 N^{++} \cD_\trw^{--}~, \eol
\widetilde \cD^{--} &:= (L^{-+})^2 \cD_\trv^{--} + \cD_\trw^{--}
	- (L^{--})^2 \cD_\trv^{++}
	- L^{--} L^{-+} \cD_\trv^0~.
\end{align}
These act on twisted biholomorphic superfields, preserving their twisted biholomorphy.
They also annihilate $L^{\pm\pm}$ and $L^{\mp\pm}$, and on the field dependent combinations
$U^{\ra+} := \frac{L^{\ra +}}{L^{-+}}$ and 
$U^{\ra -} := w^{\ra -}$
the compensated harmonic derivatives formally act as
if they were simple harmonic derivatives, i.e.
$\widetilde\cD^{++} U^{\ra-} = U^{\ra+}$, etc.
Furthermore, because $L^{-+}$ is covariantly constant and twisted biholomorphic,
one may use it to trade $\gU{1}_\trv$ for $\gU{1}_\trw$ charge of any superfield.
These observations are clear in the central gauge where $L_\ra{}^i = \delta_\ra{}^i$.
There $L^{-+}$ is just $z^0$, $U^{\ra\pm} = u^{\ra\pm}$,
and $\widetilde \cD^{\pm\pm}$ and $\widetilde \cD^0$ reduce to
$\pa_\tru^{\pm\pm}$ and $\pa_\tru^0$.

The corresponding spinor derivatives in a general gauge are built from the $\gSU{2}$
superspace derivatives as
\begin{align}
\widetilde \cD_\alpha^+ &= 
	\frac{1}{L^{-+}}\, \Big(
	\cD_\alpha^+
	+ (\cD_\alpha^+ L_\rb^+) L^{\rb -} D_\trv^0
	+ (\cD_\alpha^+ L_\rb^-) L^{\rb -} D_\trv^{++}	
	\Big)~, \eol
\widetilde \cD_\alpha^- &= 
	L^{-+} \Big(
	\cD_\alpha^-
	- (\cD_\alpha^- L_\rb^+) L^{\rb +} D_\trv^{--}
	+ (\cD_\alpha^- L_\rb^+) L^{\rb -} D_\trv^0
	\Big)
	- L^{--} L^{-+} \widetilde \cD_\alpha^+~.
\end{align}
The new derivatives $\widetilde \cD_\alpha^\pm$ carry only $\gU{1}_\trw$ charge
and are themselves twisted biholomorphic:
\begin{align}
[\cD_\trv^{++}, \widetilde \cD_\alpha^\pm] = 
[\cD_\trw^{--}, \widetilde \cD_\alpha^\pm] =
[\cD_\trv^0, \widetilde \cD_\alpha^\pm] = 0~.
\end{align}
They also obey the relations
$[\widetilde \cD^{++}, \widetilde \cD_\alpha^+] = 0$ and
$[\widetilde \cD^{--}, \widetilde \cD_\alpha^+] = \widetilde\cD_\alpha^-$
and annihilate the harmonics and $L^{-+}$.
It is not hard to see that in the central basis
$\widetilde \cD_\alpha^\pm = U_\ra^\pm \widetilde \cD_\alpha^\ra$.
These reduce in the gauge $L_\ra{}^i = \delta_\ra{}^i$ to
the harmonic derivatives constructed directly in \cite{DelamotteKaplan:HHS, GKS:Sugra}.

To summarize: starting with the conformal superspace description, one can first translate
all formula to $\gSU{2}$ superspace, and then translate again to the Einstein superspace.
Using the non-linear compensator, one can rewrite all formulae, \emph{including the component
reduction formula}, to be manifestly twisted biholomorphic. This completely eliminates
any dependence on $z^{\pm\pm}$ and $z^0$, so that all fields and operators depend purely on the
complex harmonics $u^{i\pm}$. Going to the gauge where $L_\ra{}^i = \delta_\ra{}^i$,
the field dependent compensator $U^{\ra+}$ can then be identified with the complex harmonics
themselves and one recovers the covariant formulation of \cite{DelamotteKaplan:HHS, GKS:Sugra}.
However, as has been found in a number of recent publications (see e.g.
\cite{CSG3d2, CSG5d} and references therein), the conformal superspace
approach is often a more efficient scheme for analyzing superspace actions;
therefore, we will refrain from giving the translations explicitly and remain
with the manifestly superconformal framework.

There is one last issue which we would like to address within this section.
Even within the superconformal framework, it is possible to interpret a non-linear
multiplet as underlying the complex harmonic description \cite{GIOS:Sugra}.
Begin by choosing a gauge where $\cH^{+4}$ vanishes.
Because the independent auxiliary field $D$ of conformal supergravity is naturally found as the
highest component of $\cH^{+4}$, this gauge must amount to one where $D$ is composite --
exactly the indicator that one is employing a non-linear compensator.
So where is the compensator if we haven't introduced it explicitly?
As explained in \cite{GIOS:Sugra}, the vanishing of $\cH^{+4}$
implies that $\nabla_\trw^{++} \hat z^{++} = - (\hat z^{++})^2$.
Provided one has chosen $\nabla_\alpha^+ \hat z^{++} = 0$ \emph{without} fixing any
gauges but the coordinate choice,
$\hat z^{++}$ can be interpreted as a covariant non-linear primary superfield.
What we wish to add to this old observation is that upon rewriting
$\hat z^{++}$ as
\begin{align}
\hat z^{++} = \frac{\hat v^{i +} w_i^+}{\hat v^{i+} w_i^-} =
	\frac{v^{j+} b_j{}^i w_i^+}{v^{j+} b_j{}^i w_i^-}
\end{align}
in terms of the bridge superfield $b_j{}^i$, one can show that the
bridge $b_j{}^i$ is actually harmonic-independent --
it is precisely the non-linear
multiplet compensator $L^{\ra i} = -\delta^\ra_k \,\eps^{ij} b_j{}^k$.

\section{Superconformal sigma models from curved harmonic superspace}\label{sec:SigmaModels}
As an application of the covariant harmonic superspace methods presented in earlier
sections, we will derive the general component action for a hyperk\"ahler sigma model
coupled to conformal supergravity \cite{dWKV} using harmonic superspace methods. Because of the
covariance of the approach, we can largely follow the same scheme used in
rigid harmonic superspace.\footnote{See the pedagogical discussion in chapter 11 of the
monograph \cite{GIOS}.}
A similar calculation was performed by Ivanov and Valent using prepotential methods
in \cite{IV:QuatMetrics}.\footnote{Ref. \cite{IV:QuatMetrics} addressed
quaternionic sigma models coupled to supergravity with a hypermultiplet
compensator. This is equivalent to hyperk\"ahler sigma models coupled to
conformal supergravity after gauge-fixing.}
The projective superspace version of this calculation appeared
in \cite{Butter:HKP}.

Our starting point is the superconformal harmonic superspace Lagrangian \eqref{eq:HLagSigma}.
The conditions \eqref{eq:HamConds} are exactly those associated with a superconformal sigma
model in harmonic superspace and will lead to a hyperk\"ahler cone in the
target space upon elimination of the auxiliary fields.
Indices are raised and lowered with an $\Sp(n)$ matrix $\Omega_{\ra\rb}$,
\begin{align}
Q^{\ra+} = \Omega^{\ra\rb} Q_\rb^+~, \qquad Q_{\ra}^+ = Q^{\rb +} \Omega_{\rb\ra}~, \qquad
\Omega^{\ra \rb} \Omega_{\rb\rc} = -\delta^\ra_\rc~.
\end{align}
In order to eliminate the auxiliary fields, we must solve their equations of motion.
In the rigid approach, these were analyzed at the component level, but we will
analyze them directly at the superfield level.
Varying the Lagrangian \eqref{eq:HLagSigma} and discarding a total derivative, one finds
\begin{align}\label{eq:HSigmaEOM}
\delta \mathscr{L}^{(2,2)} = -\delta Q^{\ra+} \Big(
\nabla_\trw^{++} Q_\ra^+ - \pa_{\ra+} H^{(2,2)}
\Big) = 0~.
\end{align}
The term in parentheses must vanish, leading to the superfield equation of motion.
It vanishes component-by-component in its $\q^+$ expansion as
we place each component of $Q^{\ra+}$ on-shell.
Because we only wish to set the auxiliary fields on-shell, we should impose
$\nabla_\trw^{++} Q_\ra^+ = \pa_{\ra+} H^{(2,2)}$ component-by-component
\emph{except} for the highest two components, which contain the equations of
motion for the physical bosons and fermions.

Before proceeding with calculation, we first review how the hyperk\"ahler
geometry emerges, including the structure of gauged isometries, superconformal
isometries, and supersymmetry transformations of the physical component fields.
These properties are insensitive to whether the physical equations of motion
are imposed or not, so we will be able to impose the full superfield equations of
motion to aid our discussion.
In addition, all of the formulae from here on will be taken in the central basis
to simplify matters.

\subsection{Hyperk\"ahler geometry and on-shell $\cN=2$ superfields}

\subsubsection*{Harmonic superspace and hyperk\"ahler geometry}
The harmonic superspace approach to hyperk\"ahler geometry was introduced in
\cite{GIOS:HK, GIOS:Geo2_HK}. Here we provide only a concise summary of the results
of these papers necessary for the evaluation of the component action of \eqref{eq:HLagSigma}.
The main difference with \cite{GIOS:HK, GIOS:Geo2_HK} is that we will employ a
complex harmonic formulation: that is, we complexify the real harmonics to
$u_i^+ = v_i^+ / z^0$ and $u_i^- = w_i^-$, with various factors of $z^0 = (v^+, w^-)$
appearing as needed.

The lowest component of the superfield equation of motion \eqref{eq:HSigmaEOM} takes the form
\begin{align}\label{eq:HarmSigmaEOM}
D_\trw^{++} q_\ra^+(\phi,v^+, w^-) = \pa_{\ra+} H^{(2,2)}(q^+, v^+, w^-)
\end{align}
and possesses a solution with $4n$ real fields $\phi^\mu$ parametrizing a hyperk\"ahler
manifold $\cM$. For now we will allow $H^{(2,2)}$ to depend also on $v^{i+}$; later on,
we will discuss the specific features associated with cones.
A special choice of coordinates is given by taking $\phi^\mu$ as
the leading terms $f^{\ra i}$ in the harmonic expansion
$q^{\ra+} = z_0 (f^{\ra i} u_i^+ + \cdots)$; however, the specific choice of
$\phi^\mu$ will not be relevant here.
Once some choice is made, the geometry of the target space $\cM$ can be summarized as follows.
The solutions $q^{\ra+}(\phi, v^+, w^-)$ to \eqref{eq:HarmSigmaEOM} can be used
to construct a closed two-form
\begin{align}
\Omega^{++} = \frac{1}{2} \rd q^{\ra+} \wedge \rd q^{\rb +} \Omega_{\ra\rb}
	= \frac{1}{2} \rd q_\ra^{+} \wedge \rd q^{\ra+}~,
\end{align}
which is annihilated by $D_\trw^{++}$.
This means it possesses a terminating harmonic expansion
$\Omega^{++} = (z_0)^2 \Omega^{ij} u_i^+ u_j^+ = \Omega^{ij} v_i^+ v_j^+$. The two-forms $\Omega_{\mu\nu}{}^{ij}$ are the three hyperk\"ahler two-forms. Introducing $E_\mu{}^{\ra +}$ via $\rd q^{\ra +} = \rd \phi^\mu\, E_\mu{}^{\ra +}$, we find
\begin{align}\label{eq:HarmOmega}
\Omega^{++} = \frac{1}{2} \rd \phi^\mu \wedge \rd \phi^\nu\, E_\mu{}^{\ra +} E_\nu{}^{\rb +} \Omega_{\ra \rb} \quad \implies \quad
\Omega_{\mu\nu}{}^{ij} u_i^+ u_j^+ = E_\mu{}^{\ra +} E_\nu{}^{\rb +} \Omega_{\ra \rb} ~.
\end{align}
The harmonic-dependent function $E_\mu{}^{\ra +}$ can be interpreted as (half of) a local vielbein on $\cM \times T \mathbb CP^1$. Its local analytic $\Sp(n)$ structure group leaves the two-form $\Omega^{++}$ invariant.

The form of \eqref{eq:HarmOmega} suggests the existence of a
local \Sp(n) transformation $L_\rb{}^\ra(\phi, u^\pm)$ whereby
\begin{align}
E_\mu{}^{\rb +} L_\rb{}^\ra = e_\mu{}^{\ra i} v_i^+ ~, \qquad e_\mu{}^{\ra i} = e_\mu{}^{\ra i}(\phi)~, \qquad
\Omega_{\mu\nu}{}^{ij} = e_\mu{}^{\ra (i} e_\nu{}^{\rb j)} \,\Omega_{\ra \rb}~.
\end{align}
Noting that $D_\trw^{++} e_\mu{}^{\ra +} = 0$, one can show that
\begin{align}
\cD_\trw^{++} E_\mu{}^{\ra +} :=
	D_\trw^{++} E_\mu{}^{\ra +} - E_\mu{}^{\rb+} \omega^{(0,2)}{}_\rb{}^\ra = 0~, \qquad
\omega^{(0,2)}{}_\rb{}^\ra = L_\rb{}^\rc D_\trw^{++} (L^{-1})_\rc{}^\ra~,
\end{align}
where $\cD_\trw^{++}$ is interpreted as an $\Sp(n)$-covariant derivative in the analytic $\Sp(n)$ gauge with connection $\omega^{(0,2)}{}_\rb{}^\ra$. This connection can alternatively
be specified in terms of $H^{(2,2)}$,
\begin{align}
\omega^{(0,2)}{}_{\ra \rb} = \pa_{\ra +} \pa_{\rb +} H^{(2,2)}~.
\end{align}
From the two different expressions for $\omega^{(0,2)}{}_{\ra\rb}$, the \Sp(n)
transformation $L_\rb{}^\ra$ may be determined up to a harmonic-independent piece.
This permits one to find the vielbein $e_\mu{}^{\ra i}$, from which the hyperk\"ahler
metric $g_{\mu \nu}$ is constructed as
\begin{align}
g_{\mu\nu} = - e_\mu{}^{\ra i} e_{\nu}{}^{\rb j} \eps_{ij} \Omega_{\ra\rb}
	= e_\mu{}^{\ra i} e_\nu{}_{\ra i}~.
\end{align}
The usual vielbein postulate
$\nabla_\mu e_\nu{}^{\ra i} = \pa_\mu e_\nu{}^{\ra i} - \Gamma_{\mu \nu}{}^\rho e_\rho{}^{\ra i}
	- e_\nu{}^{\rb i} \omega_\mu{}_\rb{}^\ra = 0$
(equivalently using the vanishing of the torsion tensor)
allows one to determine the target space \Sp(n) connection $\omega_\mu{}_\rb{}^\ra$ in the
central \Sp(n) gauge.

There is one additional piece of necessary information. The hyperk\"ahler Riemann tensor
$R_{\mu\nu\rho\sigma}$ possesses the tangent space decomposition
\begin{align}
R_{\ra i\, \rb j\, \rc k\, \rd l} :=
	e_{\ra i}{}^\mu\, e_{\rb j}{}^\nu\, e_{\rc k}{}^\rho\, e_{\rd l}{}^\sigma\,
	R_{\mu \nu \rho \sigma}
	= \eps_{ij} \eps_{kl}\, R_{\ra\rb\rc\rd}
\end{align}
where $R_{\ra\rb\rc\rd}$ is a totally symmetric \Sp(n) tensor. Applying the local \Sp(n)
transformation $L_\ra{}^\rb$, we find a new tensor
\begin{align}
\cR_{\ra\rb\rc\rd} =
	L_\ra{}^{\ra'} \,L_\rb{}^{\rb'} \,L_\rc{}^{\rc'} L_\rd{}^{\rd'}
	R_{\ra'\rb'\rc'\rd'}~, \qquad \cD_\trw^{++} \cR_{\ra\rb\rc\rd} = 0~.
\end{align}
One can show that $\cR_{\ra\rb\rc\rd}$ obeys
\begin{align}\label{eq:cR}
(z_0)^{-2} \cR_{\ra \rb \rc \rd}
	&= H^{(-2,2)}_{\ra\rb\rc\rd}
	+ 3 \,\omega^-_{(\ra \rb}{}^{\re} H^{(-1,2)}_{\rc \rd) \re}
	- \cD_\trw^{++} \cB_{\ra\rb\rc\rd}^{--}~, \eol
H^{(-1,2)}_{\ra\rb\rc} &:= \pa_{\ra+} \pa_{\rb+} \pa_{\rc+} H^{(2,2)}~, \qquad
H^{(-2,2)}_{\ra\rb\rc\rd} := \pa_{\ra+} \pa_{\rb+} \pa_{\rc+} \pa_{\rd+} H^{(2,2)}~,
\end{align}
where $\omega^-_{\ra\rb\rc}$ is the solution to the equation
\begin{align}\label{eq:omega-ab}
\cD_\trw^{++} \omega^-_{\ra\rb\rc} = H^{(-1,2)}_{\ra\rb\rc}~,
\end{align}
and $\cB_{\ra\rb\rc\rd}^{--}$ is a twisted biholomorphic quantity.\footnote{In the framework
of \cite{GIOS:HK, GIOS:Geo2_HK}, $\cB_{\ra\rb\rc\rd}^{--} = \cD_\ra^- \omega^-_{\rb\rc\rd}$
and $\omega^-_{\ra\rb\rc}$ is the $\Sp(n)$ connection for the
covariant target space derivative $\cD_\ra^-$ in the analytic gauge.}

Because these results follow from the equation of motion \eqref{eq:HarmSigmaEOM}, they
must hold as superfield equations provided \eqref{eq:HSigmaEOM} holds.
Then we should find the superfields $Q^{\ra+}$ are completely
determined in terms of $4n$ real superfields $\Phi^\mu$, so that any variation
becomes
\begin{align}\label{eq:VaryQtoPhi}
\delta Q^{\ra+} = \delta \Phi^\mu \, E_\mu{}^{\ra+}~.
\end{align}
In particular, this means that the two-form $\Omega^{++}$ defined on the target space,
\begin{align}
\Omega^{++} = \frac{1}{2} \rd Q^{\ra +} \wedge \rd Q^{\rb+}\, \Omega_{\ra\rb}
	&= \frac{1}{2} \rd \Phi^\mu \wedge \rd \Phi^\nu\, e_\mu{}^{\ra +} e_\nu{}^{\rb +} \Omega_{\ra \rb}~,
\end{align}
may be generalized to any antisymmetrized variation
\begin{align}\label{eq:OmegaVar}
\frac{1}{2} \delta Q^{\ra +} \wedge \delta Q^{\rb+}\, \Omega_{\ra\rb}
	= \frac{1}{2} \delta \Phi^\mu \wedge \delta \Phi^\nu\, e_\mu{}^{\ra +} e_\nu{}^{\rb +} \Omega_{\ra \rb}~.
\end{align}
Replacing $\delta$ with various local symmetry operations leads to useful
results, some of which we will come across in the next few subsections.

\subsubsection*{Gauged isometries from harmonic superspace}
We will be including the possibility of gauged isometries in the action. These were
originally described in harmonic superspace in \cite{BGIO}, whose results we summarize
here. Suppose that the harmonic superspace Lagrangian possesses some isometries under which
\begin{align}
\delta Q^{\ra+} = \l^{r} \cJ_{r}^{\ra+}(Q^+, v^+, w^-)~,
\end{align}
for constant parameters $\l^r$. In the superconformal case, $\cJ_r^{\ra+}$ must actually be
independent of $v^{i+}$, but we will remain with this more general case for the moment.
In order for this to be an invariance of the action, it must obey
\begin{align}\label{eq:J_Conds}
\cJ_{r}^{\ra +} = \Omega^{\ra \rb} \pa_{\rb +} D_r^{++}~, \qquad
\cJ_r^{\ra+} \pa_{\ra +} H^{(2,2)} = \pa_\trw^{++} D_r^{++}~.
\end{align}
for some biholomorphic function $D_r^{++}$ of charge $(2,0)$, defined up to a shift by
constant $c_r^{++} = c_r^{ij} v_i^+ v_j^+$. The second equation in \eqref{eq:J_Conds}
determines the \emph{explicit} dependence of $D_r^{++}$ on $w_i^-$.
When the equations of motion are imposed, so that $Q^{\ra+}$ is determined
as a function of the harmonics,
the first condition of \eqref{eq:J_Conds} is interpreted as the requirement
of invariance of the hyperk\"ahler two-form $\Omega^{++}$,
while the second relation of \eqref{eq:J_Conds} implies that $D_r^{++}$ is $w_i^-$-independent,
\begin{align}
D_\trw^{++} D_r^{++} = 0 \quad \implies \quad D_r^{++} = D_r^{ij} v_i^+ v_j^+~.
\end{align}
The function $D_r^{++}$ is the Killing potential (or moment map) of the hyperk\"ahler manifold.
On-shell, the gauge transformations of $Q^{\ra+}$ must manifest on the fields
$\phi^\mu$ of the target space as $\delta \phi^\mu = \l^r J_r{}^\mu(\phi)$.
As a consequence of \eqref{eq:VaryQtoPhi}, one finds
$\cJ_r^{\ra+} = J_r{}^\mu E_\mu{}^{\ra+}$.
From \eqref{eq:J_Conds}, one finds
$\pa_\mu D_r{}^{ij} = -(\Omega^{ij})_{\mu\nu} J_r{}^\nu$.

We are interested in situations where the Lagrangian (and not just the action) is gauge
invariant and where it can be gauged in a manifestly covariant way -- that is, by
simply adding the appropriate connection to $D_\trw^{++}$. In these cases, the
Killing potential may be chosen in special form
$D_r^{++} = -\frac{1}{2} Q_\ra^+ \cJ_r^{\ra+}$.\footnote{This choice for the Killing potential
is actually always possible provided we do not adopt the special gauge $L_\ra^+ = Q_\ra^+$.
As is familiar from K\"ahler target spaces in $\cN=1$ theories \cite{HuKLR},
it is possible to introduce non-dynamical multiplets with vanishing kinetic terms
whose sole purpose is to render the Lagrangian completely gauge
invariant, so that minimal substitution may proceed.
The cost of this approach in harmonic superspace would be
zero eigenvalues of the harmonic ``kinetic matrix''
$\pa_{[\ra +} L_{\rb]}^+$.}
When the Lagrangian is superconformal, this is always possible.
By taking \eqref{eq:OmegaVar} and replacing one $\delta$ with
a gauge transformation and the other with a dilatation, one may find the special
form of the moment map on a hyperk\"ahler cone,
$D_{r}{}^{ij} = -\frac{1}{2} \chi^\mu (\Omega^{ij})_{\mu\nu} J_r^\nu$.

\subsubsection*{Superconformal isometries and the hyperk\"ahler potential}
Now let us analyze the superconformal properties of the target space.
The hypermultiplet $Q^{\ra+}$ transforms locally under dilatations
and $\SU{2}_R$ as
\begin{align}
\delta Q^{\ra+} &= \L_\trD \bbD Q^{\ra+} + \l^i{}_j I^j{}_i Q^{\ra+} =
(\L_\trD + \l_\trv^{-+}) Q^{\ra+} - \l_\trv^{++} \nabla_{\trv}^{--} Q^{\ra+}~, \eol
\l_\trv^{++} &= \l^{ij} v_i^+ v_j^+~, \quad \l_\trv^{+-} = \l^{ij} v_i^+ v_j^-~,
\end{align}
and is inert under $\gU{1}_R$. On the target space,
$\delta \Phi^\mu = \L_{\trD} k_{\trD}^\mu + \l^i{}_j k^j{}_i{}^\mu$
for some choice of vectors $k_\trD^\mu := \bbD \Phi^\mu$ and
$k_{ij}{}^\mu := I_{ij} \Phi^\mu$. Using \eqref{eq:VaryQtoPhi}, one may show that
\begin{alignat}{2}
k_\trD^\mu E_\mu{}^{\ra+} &= Q^{\ra+}~, &\quad
(k_\trv^{+-})^\mu E_\mu{}^{\ra+} &= \frac{1}{2} Q^{\ra+}~, \eol
(k_\trv^{--})^\mu E_\mu{}^{\ra+} &= D_\trv^{--} Q^{\ra+}~, &\quad
(k_\trv^{++})^\mu E_\mu{}^{\ra+} &= 0~.
\end{alignat}
These conditions imply that 
\begin{align}\label{eq:defKij}
I^i{}_j \Phi^\mu \equiv (k^i{}_j)^\mu = (\Omega^i{}_j)^\mu{}_\nu k_\trD^\nu~.
\end{align}
Now using the transformation properties of the hyperk\"ahler two-form,
one can prove that $k_\trD^\mu$ is a homothetic conformal Killing vector on the target space,
$\nabla_\mu k_\trD^\nu = \delta_\mu{}^\nu$.
In particular, there exists a globally-defined hyperk\"ahler potential $K$
for which
\begin{align}\label{eq:HKP1}
K = \frac{1}{2} k_{\trD}^\mu k_{\trD\mu}~, \qquad \pa_\mu K = k_{\trD \mu}~.
\end{align}
There is a very useful alternative form for the hyperk\"ahler potential:
\begin{align}\label{eq:HKP2}
K = -Q_\ra^+ D_\trv^{--} Q^{\ra+}~.
\end{align}
To prove this agrees with \eqref{eq:HKP1}, observe that when
$Q^{\ra+}$ obeys its equation of motion, $K$ obeys $D_\trw^{++} K = 0$
and so is harmonic independent. Now we use
\begin{align}
-Q_\ra^+ D_\trv^{--} Q^{\ra+}
	= - 2\, (k_\trv^{+-})^\mu (k_\trv^{--})^\nu E_{\mu \ra}^+ E_\nu{}^{\ra+}
	= - 2\, (k_\trv^{+-})^\mu (k_\trv^{--})^\nu \Omega_{\mu\nu}{}^{++}
	= \frac{1}{2} k_\trD^\mu k_{\trD \mu}
\end{align}
after applying \eqref{eq:defKij}, and so the equality between
\eqref{eq:HKP1} and \eqref{eq:HKP2} follows.
This expression also follows by replacing the variations in
\eqref{eq:OmegaVar} with $D_\trv^0$ and $D_\trv^{--}$: in other words,
the function $K$ is a component of the pullback of $\Omega^{++}$ to the complex
harmonic manifold.

\subsubsection*{Supersymmetry and fermion transformations}
Up until now, we have discussed only the physical bosonic field $\phi^\mu$ and
its superfield lift $\Phi^\mu$. The sigma model also involves an $\Sp(n)$ fermion
$\z_{\ul\alpha}{}^{\rb}$ related to $\phi$ by supersymmetry. Here we will review
how it emerges from the harmonic structure. Following \cite{GIOS}, we define the fermion
$\Psi_{\ul\alpha}{}^\ra := \nabla_{\ul\alpha}^- Q^{\ra+} \loco$, which is
twisted biholomorphic and obeys
\begin{align}
\cD_\trw^{++} \Psi_{\ul\alpha}{}^\ra := D_\trw^{++} \Psi_{\ul\alpha}{}^\ra
	- \Psi_{\ul\alpha}{}^\rb \,\omega^{(0,2)}{}_\rb{}^{\ra} = 0
\end{align}
when $Q^{\ra+}$ is on-shell. This implies that we can introduce the harmonic-independent
physical fermions $\z_{\ul\alpha}{}^\ra$ via
$\Psi_{\ul\alpha}{}^\ra \equiv \zeta_{\ul\alpha}{}^\rb (L^{-1})_\rb{}^\ra$
In accordance with \eqref{eq:VaryQtoPhi}, one finds that
\begin{align}
\nabla_{\alpha}^- \Phi^\mu \loco \,E_\mu{}^{\ra+} = \zeta_{\ul\alpha}{}^\rb (L^{-1})_\rb{}^\ra~ \quad\implies\quad
\nabla_{\ul\alpha i} \Phi^\mu \loco = \z_{\ul\alpha}{}^\ra e_{\ra i}{}^\mu~,
\end{align}
which implies the supersymmetry transformations
\begin{align}
\delta_{\trQ} \phi^\mu = \xi_i \z^\rb \, e_\rb{}^i{}^\mu
	+ \bar \xi^i \bar\z^\rb \, e_{\rb i}{}^\mu
\end{align}
where $\xi_i^\alpha$ and $\bar\xi^i_\dalpha$ are the SUSY parameters.
The target space vielbeine $e_\mu{}^{\ra i}$ and their inverses $e_{\ra i}{}^\mu$ coincide
with corresponding quantities introduced in \cite{SierraTownsend1, SierraTownsend2}
(see also \cite{BaWi:QK} and \cite{dWKV}). The fermions obey the condition
$(\z_{\alpha}{}^\rb)^* = \bar \z_{\dalpha \rb}$.

In a similar way, we can calculate the supersymmetry and $S$-supersymmetry
transformations of the fermions,
\begin{align}
\delta \z_\alpha^\ra &=
	- 2 i \,\nabla_{\alpha \dbeta} \phi^\mu\, e_{\mu i}{}^\ra \,\bar\xi^{\dbeta i} 
	- 2 \bar W^r J_r{}^{i\ra} \xi_{\alpha i}
	+ 4 \eta_\alpha^i A_i{}^\ra
	+ \delta \phi^\mu \, \z_\alpha^\rc\, \omega_{\mu \rc}{}^\ra ~, \eol
\delta \bar \z^\dalpha_\ra &=
	+ 2 i \,\nabla^{\dalpha \beta} \phi^\mu\, e_{\mu}{}^{i}{}_\ra \,\xi_{\beta i} 
	- 2 W^r J_r{}_{i\ra} \bar \xi^{\dalpha i}
	- 4 \bar\eta^\dalpha_i A^i{}_\ra
	- \delta \phi^\mu \, \bar\z^\dalpha_\rc\, \omega_{\mu \ra}{}^\rc ~,
\end{align}
where $\eta^{i\alpha}$ and $\bar \eta_{i \dalpha}$ are the $S$-supersymmetry parameters
and $W^r$ is the complex scalar of the vector multiplet. We employ the same conventions
for the vector multiplet as \cite{Butter:CSG4d.Proj}.
We have introduced the pseudoreal $\rm Sp(n) \times Sp(1)$ sections
$A^{i \ra}$ associated with the conformal Killing vectors \cite{dWKV}
\begin{align}
A^{i \ra} := k_\trD^\mu \,e_\mu{}^{i \ra}~, \qquad
(A^{i \ra})^* = A_{i \ra}~, \qquad
A^{\ra +} = Q^{\rb +} L_\rb{}^\ra~.
\end{align}
It is helpful to note that
$\nabla_a \phi^\mu e_\mu{}^{\ra i} = \hat \nabla_a A^{\ra i}$
where $\hat\nabla_a$ also carries the \Sp(n) connection.
For reference, we also give the transformations of the fermions under gauged isometries,
\begin{align}
\delta_g \z_\alpha^\ra = \frac{1}{2} \l^r \z_\alpha^\rb e_{\rb j}{}^\mu (\nabla_\mu J_r^\nu)
	e_\nu{}^{j \ra}
	+ \delta_g \phi^\mu \, \z_\alpha^\rc\, \omega_{\mu \rc}{}^\ra ~, \eol
\delta_g \bar\z^\dalpha_\ra = \frac{1}{2} \l^r \bar\z^\dalpha_\rb e^{\rb j}{}^\mu (\nabla_\mu J_r^\nu)
	e_\nu{}_{j \ra}
	- \delta_g \phi^\mu \, \bar\z^\dalpha_\rc\, \omega_{\mu \ra}{}^\rc ~.
\end{align}

\subsubsection*{Some useful identities}
Finally, we will need some useful identities arising from \eqref{eq:OmegaVar}.
For example, taking the spinor part of the pullback of $\Omega^{++}$ to $\cM^{4|8}$ leads to
\begin{align}
\Omega_{\ul\alpha \ul\beta} &= \nabla_{\ul\alpha}^- Q_{\ra}^+ \loco \,\nabla_{\ul\beta}^- Q^{\ra +} \loco
	= \z_{\ul\alpha \ra} \,\z_{\ul \beta}{}^{\ra}~.
\end{align}
Each of the quantities $\Omega_{\alpha \beta}$, $\Omega_{\alpha \dbeta}$
and $\Omega_{\dalpha \dbeta}$ are harmonic-independent at lowest order in their
$\q$ expansion. Similar expressions arise using the vector derivatives
or the gauge generator,
\begin{align}
\Omega_{\ul\alpha\, b}^+ := \nabla_{\ul\alpha}^- Q_\ra^+ \loco\, \nabla_{b} Q^{\ra +} \loco
	= \z_{\ul\alpha \ra} \widehat \nabla_b A^{\ra +}~, \qquad
\Omega_{\ul\alpha\, r}^+ := \nabla_{\ul\alpha}^- Q_\ra^+ \loco\, X_r Q^{\ra +}\loco
	= \z_{\ul\alpha \ra} J_r{}^{\ra +}~.
\end{align}
Other expressions such as $\Omega_{r b}^{++}$ or $\Omega_{ab}^{++}$ could be introduced,
but we won't need them.
We will however need spinor derivatives of the hyperk\"ahler potential and moment map:
\begin{align}
\nabla_{\ul\alpha}^+ K \loco &= Q_\ra^+ \nabla_{\ul\alpha}^- Q^{\ra +} \loco
	= A_\ra^+ \z_{\ul\alpha}{}^{\ra}~, \qquad
\nabla_{\ul\alpha}^- K \loco = A_\ra^- \z_{\ul\alpha}{}^{\ra}
	= D_\trv^{--} Q_\ra^+ \nabla_{\ul\alpha}^- Q^{\ra +} \loco~, \eol
\nabla_{\ul\alpha}^- D_r^{++} \loco &= -\Omega_{\ul \alpha r}^+~.	
\end{align}

\subsection{The component fields of $Q^{\ra+}$ and auxiliary equations of motion}
Now we are prepared to set up the component calculation.
Begin by defining the components of $Q^{\ra +}$, largely following \cite{GIOS}:
\begin{subequations}
\begin{alignat}{2}
\Psi_{\alpha}^\ra &:= \nabla_{\alpha}^- Q^{\ra+} \loco~, &\qquad
\Psi_\dalpha^\ra &:= \bar\nabla_\dalpha^- Q^{\ra+} \loco~, \\
M^{\ra-} &:= -\frac{1}{4} (\nabla^{-})^2 Q^{\ra+} \loco~, &\qquad
N^{\ra-} &:= -\frac{1}{4} (\bar\nabla^{-})^2 Q^{\ra+}\loco~, \\
A_{\alpha\dbeta}^{\ra-} &:= -i \nabla_\alpha^- \nabla_\dbeta^- Q^{\ra+}\loco~, \\
\Xi_\alpha^{\ra--} &:= \frac{1}{8} \nabla_\alpha^- (\bar\nabla^{-})^2 Q^{\ra+}\loco~, &\qquad
\Xi_\dalpha^{\ra--} &:= \frac{1}{8} \bar\nabla_\dalpha^- (\nabla^{-})^2 Q^{\ra+}\loco~, \\
P^{\ra (-3)} &:= \frac{1}{16} (\nabla^{-})^2 (\bar\nabla^{-})^2 Q^{\ra+}\loco~.
\end{alignat}
\end{subequations}
These will be the most convenient definitions for the component calculation
we will soon undertake, but they possess one important disadvantage that
must be kept in mind. Each of the terms aside from $\Psi_{\ul\alpha}^\ra$ is
not twisted biholomorphic. In particular, one can show that
\begin{align}\label{eq:DvppHC1}
D_\trv^{++} A_b{}^{\ra -} = -2 \nabla_b q^{\ra+}~, \qquad
D_\trv^{++} M^{\ra -} = -\bar W^r \cJ_r{}^{\ra +}~, \qquad
D_\trv^{++} N^{\ra -} = -W^r \cJ_r{}^{\ra +}~.
\end{align}
The expressions for $D_\trv^{++} \Xi_{\ul\alpha}^{\ra --}$ and
$D_\trv^{++} P^{\ra (-3)}$ are more complicated and can be derived
from similar formulae given in \cite{Butter:CSG4d.Proj}
\begin{subequations}\label{eq:DvppHC2}
\begin{align}
D_\trv^{++} \Xi_{\alpha}^{\ra --} &=
		- \frac{i}{2} \nabla_{\alpha \dbeta} \bar\nabla^{\dbeta -} Q^{\ra+}
		+ 2 \lambda_\alpha^{r-} \cJ_r^{\ra+}
		+ \frac{1}{2} W^{r} \nabla_\alpha^- \cJ_r^{\ra+}
		\eol & \quad
		- \frac{1}{2} W_\alpha{}^\beta \nabla_\beta^- Q^{\ra+}
		- \frac{3}{2} \chi_\alpha^- Q^{\ra+}
		+ \frac{3}{2} \chi_\alpha^+ D^{--} Q^{\ra+}~, \\
D_\trv^{++} P^{\ra (-3)} &=
	- \frac{i}{2} \nabla^{\dalpha \alpha} \nabla_{\alpha}^- \nabla_\dalpha^- Q^{\ra+}
	- 3 D \,D^{--} Q^{\ra+}
	\eol & \quad
	+ \frac{3}{2} \chi^{\alpha +} D^{--} \nabla_\alpha^- Q^{\ra+}
	- \frac{3}{2} \bar\chi_\dalpha^+ D^{--} \bar\nabla^{\dalpha -} Q^{\ra+}
	\eol & \quad
	+ 2 \lambda^{\alpha r -}\nabla_\alpha^- \cJ_r^{\ra+}
	- 2 {\bar \lambda}_\dalpha^{r-} \bar\nabla^\dalpha{}^- \cJ_r^{\ra+}
	\eol & \quad
	+ \frac{1}{4} \bar W^{r} (\bar\nabla^-)^2 \cJ_r^{\ra+}
	+ \frac{1}{4} W^{r} (\nabla^-)^2 \cJ_r^{\ra+}
	+ 3 Y^{r--} \cJ_r^{\ra+}~.
\end{align}
\end{subequations}
The vector multiplet gaugino is denoted $\l_{\alpha i}{}^r$ and the pseudo-real
auxiliary is $Y_{ij}{}^r$, again following \cite{Butter:CSG4d.Proj}.
To keep notation simple, we have omitted the explicit component projection.

\begin{table*}[t]
\centering
\renewcommand{\arraystretch}{1.4}
\begin{tabular}{@{}ccc@{}} \toprule
component field & $\implies$ & equation of motion\\ \midrule
$P^{\ra(-3)}$ & $\implies$& $D_\trw^{++} q_\ra^+ = \pa_{\ra+} H^{(2,2)}$ \\
$\Xi_{\ul\alpha}^{\ra--}$ & $\implies$&
	$D_\trw^{++} \Psi_{\ul\alpha \ra} = \nabla_{\ul\alpha}^- (\pa_{\ra+} H^{(2,2)})\loco$ \\
$N^{\ra-}$ & $\implies$&
	$D_\trw^{++} M_\ra^- = -\frac{1}{4} (\nabla^-)^2 (\pa_{\ra+} H^{(2,2)}) \loco$  \\
$A_{\alpha\dalpha}^{\ra-}$ & $\implies$& 
	$D_\trw^{++} A_{\alpha \dalpha \rb}^- =
		-i \nabla_\alpha^- \bar\nabla_\dalpha^- (\pa_{\rb+} H^{(2,2)}) \loco$  \\
$M^{\ra-}$ & $\implies$&
	$D_\trw^{++} N_\ra^- = -\frac{1}{4} (\bar \nabla^-)^2 (\pa_{\ra+} H^{(2,2)}) \loco$  \\
$\Psi_{\alpha}^\ra$ & $\implies$& 
	$D_\trw^{++} \Xi_{\alpha \rb}^{--} = \frac{1}{8} \nabla_{\alpha}^{-} (\bar \nabla^-)^2 (\pa_{\ra+} H^{(2,2)}) \loco$  \\
$q^{\ra+}$ & $\implies$& 
	$D_\trw^{++} P_{\ra}^{(-3)} =  (\nabla^-)^4 (\pa_{\ra+} H^{(2,2)}) \loco$  \\
\bottomrule
\end{tabular}
\caption{Component field equations of motion}\label{tab:N2Comp}
\end{table*}

The equations of motion of the component fields
correspond to the action of $D_\trw^{++}$ and can be derived by
successively taking spinor derivatives of the superfield equation of motion \eqref{eq:HSigmaEOM}.
They are summarized in Table \ref{tab:N2Comp}.
We impose these only through the $\q^2$ level:
\begin{subequations}
\begin{align}
D_\trw^{++} q_{\ra}^+ &= \pa_{\ra+} H^{(2,2)}~, \\
\cD_\trw^{++} \Psi_{\ul \alpha \ra} &\equiv D_\trw^{++} \Psi_{\ul \alpha \ra}
	- \Psi_{\ul\alpha}^{\rb} H_{\rb\ra}^{(0,2)} = 0~, \\
\cD_\trw^{++} M_\ra^- &\equiv D_\trw^{++} M_\ra^- - M^{\rb -} H_{\rb\ra}^{(0,2)}
	= -\frac{1}{4} \Psi^\rb \Psi^\rc H_{\ra\rb\rc}^{(-1,2)}~, \\
\cD_\trw^{++} N_\ra^- &\equiv D_\trw^{++} N_\ra^- - N^{\rb -} H_{\rb\ra}^{(0,2)}
	= -\frac{1}{4} \bar \Psi^\rb \bar\Psi^\rc H_{\ra\rb\rc}^{(-1,2)}~, \\
\cD_\trw^{++} A_{\alpha \dalpha\, \ra}^- &\equiv
	D_\trw^{++} A_{\alpha \dalpha\, \ra}^- - A_{\alpha\dalpha}^{\rb -} H_{\rb \ra}^{(0,2)}
	= - i\, \Psi_\alpha^\rb \bar\Psi_\dalpha^\rc H_{\ra\rb\rc}^{(-1,2)}~.
\end{align}
\end{subequations}
The solutions completely determine the harmonic expansions of $q^{\ra+}$,
$\Psi^\ra_{\ul\alpha}$, $M^{\ra-}$, $N^{\ra-}$, and $A_b{}^{\ra-}$
in terms of the physical fields $\phi^\mu$ and $\z_{\ul\alpha}^{\rb}$.
We have already explained how these arise in $q^{\ra+}$ and $\Psi^\ra_{\ul\alpha}$.
For the others, we find
\begin{subequations}
\begin{align}
M_\ra^- &= - \bar W^r \cJ_{r \ra}^- - \frac{z^{--}}{(z^0)^2} \bar W^r \cJ_{r \ra}^+
	- \frac{1}{4} \Psi^\rb \Psi^\rc \,\omega^-_{\ra\rb\rc}~, \\
N_\ra^- &= - W^r \cJ_{r \ra}^- - \frac{z^{--}}{(z^0)^2} W^r \cJ_{r \ra}^+
	- \frac{1}{4} \bar\Psi^\rb \bar\Psi^\rc \,\omega^-_{\ra\rb\rc}~, \\
A_{\alpha \dalpha}^{\ra -} &= -2 \,\nabla_{\alpha \dalpha} \phi^\mu\, E_{\mu}{}^{\ra -}
	- 2 \,\frac{z^{--}}{(z^0)^2} \nabla_{\alpha \dalpha} q^{\ra+}
	- i \Psi_{\alpha}^\rb \bar\Psi_\dalpha^{\rc} \,\omega^-_{\rb \rc}{}^\ra~,
\end{align}
\end{subequations}
where $\cJ_r^{\ra -}$ and $E_\mu{}^{\ra-}$ are twisted biholomorphic solutions
to the equations
\begin{align}
(z^0)^2 \cD_\trw^{++} \cJ_r^{\ra -} = \cJ_r^{\ra+}~, \qquad
(z^0)^2 \cD_\trw^{++} E_\mu{}^{\ra -} = E_\mu{}^{\ra+}~.
\end{align}
Note that they are chosen to carry $\gU{1}_\trv$ charge.
This means that
\begin{align}
E_\mu{}^{\rb +} L_\rb{}^\ra = e_\mu{}^{\ra i} v_i^+~, \qquad
E_\mu{}^{\rb -} L_\rb{}^\ra = e_\mu{}^{\ra i} \,\frac{w_i^-}{z^0} ~, \qquad
\cJ_r{}^{\ra \pm} = J_r{}^\mu E_\mu{}^{\ra \pm}~.
\end{align}

\subsection{Summary of the component calculation}
All the pieces are now in place to work out the component reduction.
Here one important observation will drastically simplify the analysis:
when $Q^{\ra+}$ is placed fully on-shell, the superconformal Lagrangian $\mathscr{L}^{(2,2)}$
\emph{completely vanishes} as a consequence of the homogeneity of $H^{(2,2)}$.
Now we are only imposing the equations of motion to the $\q^2$ level,
so this means that in evaluating the component Lagrangian we only need to keep
the terms at higher order than $\q^2$. (This same observation simplified
the projective superspace calculation of \cite{Butter:HKP}.)
Thus the component reduction formula 
\eqref{eq:CompAction} simplifies to evaluating two terms:
\begin{align}\label{eq:T0T1Act}
S &= \frac{i}{2\pi} \int \rd^4x\, e\, \int_{\cS}
		\cV^{++} \wedge \cW^{--} \Big(T_0^{(-2,2)} + T_1^{(-2,2)}\Big)~, \eol
T_0^{(-2,2)} &= \frac{1}{16} (\nabla^-)^2 (\bar\nabla^-)^2 \mathscr{L}^{(2,2)} \loco ~, \quad
T_1^{(-2,2)} = - \frac{i}{8} (\bar\psi_m^- \bsigma^m)^\alpha \nabla_\alpha^- (\bar\nabla^-)^2 \mathscr{L}^{(2,2)} \loco + \HC
\end{align}

Let us begin with the leading order term. Imposing the superfield equation of
motion through the $\q^2$ level, one can show
(recalling that $D_\trw^{++}$ commutes with $\nabla_{\ul\alpha}^-$)
\begin{align}
T_0^{(-2,2)}
	&=
	\frac{1}{2} D_\trw^{++} \big(q_\ra^+ P^{\ra(-3)}\big)
	+ \frac{1}{2} D_\trw^{++} \big(\Psi^{\alpha}_\ra \Xi_\alpha^{\ra --}
	+ \HC \big)
	\eol & \quad
	- \frac{1}{8} H^{(-1,2)}_{\ra\rb\rc} \,\Big(\Psi^{\ra} \Psi^{\rb} N^{\rc -}
		+ \bar\Psi^{\ra} \bar\Psi^{\rb} M^{\rc -}\Big)
	+ \frac{i}{8} H^{(-1,2)}_{\ra\rb\rc} \, \Psi_\alpha^{\ra} \bar\Psi_\dalpha^{\rb} A^{\dalpha \alpha\, \rc -}
	\eol & \quad
	+ \frac{1}{16} H^{(-2,2)}_{\ra\rb\rc\rd} \Psi^{\ra} \Psi^\rb \bar\Psi^\rc \bar\Psi^\rd~.
\end{align}
Using \eqref{eq:cR} and \eqref{eq:omega-ab}, this can be rewritten
\begin{align}
T_0^{(-2,2)}
	&=
	\frac{1}{2} D_\trw^{++} \big(q_\ra^+ P^{\ra(-3)}\big)
	+ \frac{1}{2} D_\trw^{++} \big(\Psi^{\alpha}_\ra \Xi_\alpha^{\ra --}
	+ \HC \big)
	\eol & \quad
	+ \frac{1}{16} (z^0)^{-2} \cR_{\ra\rb\rc\rd} \,\Psi^{\ra} \Psi^\rb \bar\Psi^\rc \bar\Psi^\rd
	+ \frac{1}{16} D_\trw^{++} \Big(
		\cB_{\ra\rb\rc\rd} \, \Psi^{\ra} \Psi^\rb \bar\Psi^\rc \bar\Psi^\rd
		\Big)
	\eol & \quad
	+ \frac{1}{8} D_\trw^{++} \Big[\omega^-_{\ra\rb\rc} \Big(
		i \Psi_\alpha^{\ra} \bar\Psi_\dalpha^{\rb} A^{\dalpha \alpha\, \rc -}
		- \Psi^{\ra} \Psi^{\rb} N^{\rc -}
		- \bar\Psi^{\ra} \bar\Psi^{\rb} M^{\rc -}\Big)\Big]
	\eol & \quad
	- \frac{1}{8} \omega^-_{\ra\rb\rc} \Big(
		i \Psi_\alpha^{\ra} \bar\Psi_\dalpha^{\rb} \cD_\trw^{++} A^{\dalpha \alpha\, \rc -}
		- \Psi^{\ra} \Psi^{\rb} \cD_\trw^{++} N^{\rc -}
		- \bar\Psi^{\ra} \bar\Psi^{\rb} \cD_\trw^{++} M^{\rc -}\Big)
	\eol & \quad
	-\frac{1}{32} \,\omega^-_{\ra \rb}{}^{\re} H^{(-1,2)}_{\rc \rd \re} \Psi^\ra \Psi^\rb \bar\Psi^\rc \bar\Psi^\rd
	-\frac{1}{32} \,\omega^-_{\ra \rb}{}^{\re} H^{(-1,2)}_{\rc \rd \re} \bar \Psi^\ra \bar \Psi^\rb \Psi^\rc \Psi^\rd	
	\eol & \quad
	-\frac{1}{8} \,\omega^-_{\ra \rc}{}^{\re} H^{(-1,2)}_{\rb \rd \re} \Psi^\ra \Psi^\rb \bar\Psi^\rc \bar\Psi^\rd~.
\end{align}
Imposing the equations of motion for the auxiliaries, we find that the last three lines cancel.
Furthermore, $\cR_{\ra\rb\rc\rd} \,\Psi^{\ra} \Psi^\rb \bar\Psi^\rc \bar\Psi^\rd$
simplifies to the harmonic-independent
$R_{\ra\rb\rc\rd} \,\z^{\ra} \z^\rb \bar\z^\rc \bar\z^\rd$.
All of these manipulations correspond so far (as they must) to the rigid harmonic superspace
calculation \cite{GIOS}.
Now we need to integrate by parts. Using \eqref{eq:HarmTD1b}, we can write
\begin{align}\label{eq:T0toT0'}
\frac{i}{2\pi} \int \cV^{++} \wedge \cW^{--}\, T_0^{(-2,2)}
	= \frac{i}{2\pi} \int \cV^{++} \wedge \cV^{--}\, T_0'
\end{align}
for some new quantity $T_0'$ given by
\begin{align}
T_0' &=
	- \frac{1}{4} q_\ra^+ \nabla^{\dalpha \alpha} A_{\alpha \dalpha}^{\ra -}
	+ \Big(\frac{i}{4} \Psi^\alpha_\ra \widehat \nabla_{\alpha \dalpha} \bar\Psi^{\dalpha \ra} + \HC \Big)
	+ \frac{1}{16} R_{\ra\rb\rc\rd} \,\z^{\ra} \z^\rb \bar\z^\rc \bar\z^\rd
	\eol & \quad
	- \frac{i}{4} \omega^-_{\ra\rb\rc}
		\Psi_\alpha^{\ra} \bar\Psi_\dalpha^{\rb} \nabla^{\dalpha \alpha} q^{\rc+}
	- \frac{1}{8} \omega^-_{\ra\rb\rc} \Big(\Psi^\ra \Psi^\rb W^r J_r^{\rc +} + \HC \Big)
	\eol & \quad
	+ \frac{3}{2} D \,q_\ra^+ D_\trv^{--} q^{\ra +}
	- \frac{1}{4} W^{\alpha \beta} \Psi_{\ra \alpha} \Psi^\ra_\beta
	- \frac{1}{4} \bar W_{\dalpha \dbeta} \bar\Psi_{\ra}^\dalpha \bar\Psi^{\dbeta \ra}
	\eol & \quad
	+ 3 Y^{-- r} D_{r}^{++}
	+ \Big(2 \lambda^{r \alpha -} \nabla_\alpha^- D_{r}^{++} + \HC \Big)
	\eol & \quad
	+ \Big(\frac{1}{4} W^r (\nabla^-)^2 D_r^{++}
	- \frac{1}{2} W^r M_\ra^- J_r^{\ra +} + \HC\Big)
	\eol & \quad
	+ \Big(\frac{3}{2} \chi^{\alpha +} D^{--} q_\rb^+ \Psi_\alpha^{\rb}
	- \frac{3}{2} D_\trv^{--} \big(
		\chi^{\alpha +} q_\rb^+ \Psi_\alpha^\rb
	\big)
	+ \HC \Big)~.
\end{align}
In arriving at this result, we have extensively used \eqref{eq:DvppHC1} and \eqref{eq:DvppHC2}
and have written the fermion kinetic term with the analytic basis $\Sp(n)$ connection
\begin{align}
\widehat \nabla_{\alpha \dalpha} \bar \Psi^{\dalpha \ra}
	= \nabla_{\alpha \dalpha} \bar\Psi^{\dalpha \ra} + \nabla_{\alpha\dalpha} q^{\rc +} \omega^-_{\rc}{}^{\ra \rb} \bar\Psi^{\dalpha}_\rb~.
\end{align}
Now we can reconstruct the kinetic terms. The one for the physical fermions
emerges after rewriting
$\Psi^\alpha_\ra \widehat \nabla_{\alpha \dalpha} \bar\Psi^{\dalpha \ra} =
	\z^\alpha_\ra \widehat \nabla_{\alpha \dalpha} \bar\z^{\dalpha \ra}$
where now $\widehat \nabla_a$ is in the central basis.
To find the correct bosonic kinetic term, we use
$\nabla_{\alpha \dalpha} q^{\ra +} = \nabla_{\alpha \dalpha} \phi^\mu\, E_\mu{}^{\ra +}$
and $g_{\mu\nu} = 2 E_{(\mu}{}^{\ra +} E_{\nu)}{}_{\ra}^-$ to give
\begin{align}
- \frac{1}{4} q_\ra^+ \nabla^{\dalpha \alpha} A_{\alpha \dalpha}^{\ra -}
	&= -\frac{1}{4} \nabla^{\dalpha \alpha} (q_\ra^+ A_{\alpha \dalpha}^{\ra -})
	- \frac{1}{4} \nabla^a \phi^\mu\, \nabla_a \phi^\nu \, g_{\mu\nu}
	+ \frac{i}{4} \nabla^{\dalpha \alpha} q^{\ra+} \Psi_\alpha^\rb \Psi_\dalpha^\rc
		\,\omega^-_{\ra\rb\rc} \eol
	&= \frac{1}{2} K_\mu \widehat \nabla^a \nabla_a \phi^\mu
	+ \nabla^{\dalpha \alpha} \Big(
	\frac{1}{4} K_\mu \nabla_{\alpha \dalpha} \phi^\mu
	- \frac{1}{4} q_\ra^+ A_{\alpha \dalpha}^{\ra-}
	\Big)
	+ \frac{i}{4} \nabla^{\dalpha \alpha} q^{\ra+} \Psi_\alpha^\rb \Psi_\dalpha^\rc\,
		\omega^-_{\ra\rb\rc}~,
\end{align}
where $K_\mu = k_{\trD\mu} = \pa_\mu K$ is the derivative of the hyperk\"ahler potential.
The covariant derivative $\widehat \nabla_a$ should be understood to also carry the
target space affine connection.
To simplify the remaining terms, we need the solution for $M_\ra^-$ and the expression
for $(\nabla^-)^2 D_r^{++}\loco$:
\begin{align}
(\nabla^-)^2 D_r^{++} \loco
	&=
	- 4 \bar W^s f_{sr}{}^t D_t^{+-}
	- 2 \bar W^s J_s^\mu J_{r\mu}
	+ \frac{1}{2} \z^\ra \z^\rb \,\cD_{\ra j} J_{r \rb}{}^j
\end{align}
where $D_r^{+-} \equiv D_r{}^{ij} v_i^+ v_j^- = \frac{1}{2} D_\trv^{--} D_r^{++}$.
This leads to
\begin{align}
T_0' &=
	\frac{1}{2} K_\mu \widehat \nabla^a \nabla_a \phi^\mu
	+ \Big(\frac{i}{4} \z^\alpha_\ra \widehat \nabla_{\alpha \dalpha} \bar\z^{\dalpha \ra} + \HC \Big)
	+ \frac{1}{16} R_{\ra\rb\rc\rd} \,\z^{\ra} \z^\rb \bar\z^\rc \bar\z^\rd
	\eol & \quad
	+ \Big(\frac{1}{4} W^r \z^\ra \z^\rb \cD_{\ra j} J_{r \rb}{}^j
	+ 2 \lambda^{r \alpha -} \Omega_{r \alpha}^+
	- \frac{1}{4} W^{\alpha \beta} \z_{\ra \alpha} \z^\ra_\beta 
	+ \HC\Big)
	\eol & \quad
	- \frac{1}{2} W^r \bar W^s J_s^\mu J_{r\mu}
	+ 3 Y^{-- r} D_{r}^{++}
	- \frac{3}{2} D \,K
	\eol & \quad
	+ \Big(\frac{3}{2} \chi^{\alpha +} D_\trv^{--} q_\rb^+ \Psi_\alpha^{\rb}
	- \frac{3}{2} D_\trv^{--} \big(
		\chi^{\alpha +} q_\rb^+ \Psi_\alpha^\rb
	\big)
	+ \HC\Big)
	\eol & \quad
	+ \nabla^{\dalpha \alpha} \Big(
	\frac{1}{4} K_\mu \nabla_{\alpha \dalpha} \phi^\mu
	- \frac{1}{4} q_\ra^+ A_{\alpha \dalpha}^{\ra-}
	\Big)~.
\end{align}
Each of the expressions in the first three lines depends on the $v^{i\pm}$ harmonics and not on
$w_i^\pm$. Because we are integrating against the
$\gSU{2}_\trv / \gU{1}_\trv$ measure $\cV^{++} \wedge \cV^{--}$, the naive rules of
harmonic integration apply now for the $v_i^\pm$ harmonics, leading to
\begin{align}
T_0' &=
	\Big[\frac{1}{2} K_\mu \widehat \nabla^a \nabla_a \phi^\mu
	+ \Big(\frac{i}{4} \z^\alpha_\ra \widehat \nabla_{\alpha \dalpha} \bar\z^{\dalpha \ra} + \HC \Big)
	+ \frac{1}{16} R_{\ra\rb\rc\rd} \,\z^{\ra} \z^\rb \bar\z^\rc \bar\z^\rd
	\eol & \quad
	+ \Big(\frac{1}{4} W^r \z^\ra \z^\rb \cD_{\ra j} J_{r \rb}{}^j
	- \lambda^{r \alpha}{}_i \,\z_{\alpha \ra} J_r{}^{\ra i}
	- \frac{1}{4} W^{\alpha \beta} \z_{\ra \alpha} \z^\ra_\beta
	- \frac{3}{4} \chi^{\alpha}_i A_\rb{}^i \z_\alpha{}^\rb
	+ \HC\Big)
	\eol & \quad
	- \frac{1}{2} W^r \bar W^s J_s^\mu J_{r\mu}
	+ Y^{ij r} D_{r\, ij}
	- \frac{3}{2} D \,K
	\Big]
	+ \nabla^{\dalpha \alpha} \Big(
	\frac{1}{4} K_\mu \nabla_{\alpha \dalpha} \phi^\mu
	- \frac{1}{4} q_\ra^+ A_{\alpha \dalpha}^{\ra-}
	\Big)~.
\end{align}
The bracketed terms will end up in the final Lagrangian.
The next step is to integrate by parts the final term.
We can swap the order of integration to yield
\begin{align}
\frac{i}{2\pi} \int \rd^4x\, e\, 
	\nabla^{\dalpha \alpha} \int_\cS \cV^{++} \wedge \cV^{--} \Big(
	\frac{1}{4} \nabla_{\alpha \dalpha} K
	- \frac{1}{4} q_\ra^+ A_{\alpha \dalpha}^{\ra-}
	\Big)~.
\end{align}
The term in parentheses actually vanishes at this order in its $\theta$-expansion
because it can be rewritten as
$\dfrac{i}{8} D_\trv^{--} \nabla_\alpha^+ \bar\nabla_\dalpha^+ K$
where to this order $K$ is harmonic independent. This expression now vanishes
using usual rules of harmonic integration.
The lacunae in this argument is that the gravitino contributions
lead to an $\cO(\q^3)$ term, which must be separately addressed.
Denote these by $T_{0|Q}'$:
\begin{align}
T_{0|Q}'
	&= 
	\frac{1}{4} \psi^{\dalpha \alpha \beta -} q_\rb^+ \nabla_{\beta \dalpha} \Psi_\alpha^{\rb}
	+ i \psi^{\dalpha \alpha}{}_\alpha^- \bar {\lambda}_\dalpha^{r -} D_r^{++}
	+ \frac{i}{4} (\psi_m^- \sigma^m \bar\Psi^{\rc}) \bar W^r q_\rb^+
		\pa_{\rc +} \cJ_r{}^{\rb+}
	\eol & \quad
	- \frac{i}{2} \psi^{\dalpha \alpha}{}_\alpha^+ q_\rb^+ \Xi_\dalpha^{\rb --}
	- \frac{1}{8} \psi^{\dalpha \alpha \beta +} \Psi_{\beta \rb} A_{\alpha \dalpha}^{\rb-}
	- \frac{1}{8} \psi^{\dalpha \alpha \beta}{}_j \nabla_{\alpha \dalpha} \nabla_\beta{}^j K\loco
	\eol & \quad
	- \frac{i}{8} \psi_{\alpha \dalpha}{}^\alpha_j \, \bar W^\dalpha{}_\dbeta\,
		\bar \z^{\dbeta \ra} A_\ra{}^j
	+ \frac{3i}{8} (\psi_{m j} \sigma^m \bar\chi^j)\, K
	+ \HC
\end{align}
Now we return to the explicit gravitino terms in \eqref{eq:T0T1Act}. They can be written
\begin{align}
T_1^{(-2,2)}
	&= -\frac{i}{2} D_\trw^{++} \Big(
		(\psi_m^- \sigma^m)_\dalpha \, q_\rb^+ \,\Xi^{\dalpha \rb --}
		- \frac{1}{8} (\psi_m^- \sigma^m \bar \Psi^\ra) \Psi^\rb \Psi^\rc \omega^{-}_{\ra\rb\rc} \Big)
		+ \HC
\end{align}
where we imposed the equations of motion through $\cO(\q^2)$.
Proceeding as in \eqref{eq:T0toT0'} gives
\begin{align}
T_1'
	&= \frac{i}{2} (\psi_m^+ \sigma^m)_\dalpha \, q_\rb^+ \,\Xi^{\dalpha \rb--}
	- \frac{i}{16} (\psi_m^+ \sigma^m \bar \Psi^\ra) \Psi^\rb \Psi^\rc \omega^{-}_{\ra\rb\rc}
	\eol & \quad
	- \frac{1}{4} (\psi_a^- \sigma^a \bsigma^b)^\beta q_\rb^+ \nabla_b \Psi_\beta^{\rb +}
	- \frac{i}{4} (\psi_m^- \sigma^m)_\dalpha \bar W^\dalpha{}_\dbeta \,q_\rb^+ \bar\Psi^{\dbeta \rb +}
	+ \frac{3i}{4} (\psi^-_m \sigma^m \bar\chi^+)\, K
	\eol & \quad
	+ 2 i (\psi_m^- \sigma^m \bar {\l}^{r -}) \, D_r^{++}
	+ \frac{i}{4} (\psi_m^- \sigma^m \bar\Psi^{\rc}) q_\rb^+ \bar W^r
		 \pa_{\rc +} \cJ_r^{\rb +}
	+ \HC
\end{align}
Combining terms gives
\begin{align}
T_{0|Q}' + T_1' &=
	- (\psi_a^- \sigma^{ab} )^\beta q_\rb^+ \nabla_b \Psi_\beta^{\rb +}
	+ 3 i (\psi_m^- \sigma^m \bar {\l}^{r -}) \, D_r^{++}
	- \frac{i}{2} (\psi_m^- \sigma^m \bar\Psi^{\rb}) \bar W^r \cJ_{r\rb}{}^{+}
	\eol & \quad
	- \frac{1}{4} \psi^{\dalpha \alpha \beta -} \Psi_{\beta \rb} \nabla_{\alpha \dalpha} q^{\rb+}
	- \frac{1}{8} \psi^{\dalpha \alpha \beta}{}_j \nabla_{\alpha \dalpha} \nabla_\beta{}^j K
	- \frac{i}{8} \psi_{\alpha \dalpha}{}^\alpha_j \, \bar W^\dalpha{}_\dbeta\,
		\bar \z^{\dbeta \ra} A_\ra{}^j
	\eol & \quad
	- \frac{i}{4} (\psi_m^- \sigma^m)_\dalpha \bar W^\dalpha{}_\dbeta \,q_\rb^+ \bar\Psi^{\dbeta \rb}
	+ \frac{3i}{8} (\psi_{m j} \sigma^m \bar\chi^j)\, K
	+ \frac{3i}{4} (\psi^-_m \sigma^m \bar\chi^+)\, K
	\eol & \quad
	+ D_\trv^{++} \Big[- \frac{1}{8} \psi^{\dalpha \alpha \beta -} \Psi_{\beta \rb} A_{\alpha \dalpha}^{\rb-}
		- \frac{i}{16} (\psi_m^- \sigma^m \bar \Psi^\ra) \Psi^\rb \Psi^\rc \omega^{-}_{\ra\rb\rc}
		\Big] + \HC
\end{align}
The argument of $D_\trv^{++}$ in the last line is actually independent of the $w_i^\pm$
harmonics, so it vanishes under the harmonic integral. Other terms can be simplified
in analogous ways.

Putting everything together, the full component Lagrangian 
for the sigma model of a hyperk\"ahler cone coupled to conformal supergravity is
\begin{align}
\cL &=
	\frac{1}{2} K_\mu \widehat \Box \phi^\mu
	+ \Big(\frac{i}{4} \z^\alpha_\ra \widehat \nabla_{\alpha \dalpha} \bar\z^{\dalpha \ra} + \HC \Big)
	+ \frac{1}{16} R_{\ra\rb\rc\rd} \,\z^{\ra} \z^\rb \bar\z^\rc \bar\z^\rd
	\eol & \quad
	+ Y^{ij r} D_{r\, ij}
	+ \lambda^{r \alpha}_i \,\z_{\alpha}^\ra J_r{}_\ra{}^{i}
	- \lambda^r_\dalpha{}^i \,\bar\z^{\dalpha}_\ra J_r{}^\ra{}_{i}
	- \frac{1}{2} W^r \bar W^s J_s^\mu J_{r\mu}
	\eol & \quad
	+ \frac{1}{4} W^r \z^\ra \z^\rb \cD_{\ra j} J_{r \rb}{}^j
	+ \frac{1}{4} \bar W^r \bar \z^\ra \bar \z^\rb \cD_{\ra j} J_{r \rb}{}^j
	\eol & \quad
	- \frac{3}{2} D \,K
	- \frac{1}{4} (W^{\alpha \beta} \z_{\ra \alpha} \z^\ra_\beta
		+ \bar W_{\dalpha \dbeta} \bar \z^{\dalpha}_\ra \bar \z^{\dbeta \ra})
	- \frac{3}{4} (\chi^{\alpha}{}_i  \z_\alpha^\rb \,A_\rb{}^i
		- \bar \chi_\dalpha{}^i \bar \z^\dalpha_\rb \,A^\rb{}_i)
	\eol & \quad
	+ \!(\psi_{m j} \sigma^m)_\dalpha \!\Big(
	\frac{1}{4} \widehat \nabla^{\dalpha\alpha} \z_{\alpha \rb} \, A^{\rb j}
	- \frac{i}{4} \bar W^\dalpha{}_\dbeta \bar \z^{\dbeta \rb} A_\rb{}^j
	+ \frac{3i}{4} \bar \chi^{\dalpha j} K
	- \frac{i}{4} \bar \z^{\dalpha \rb} \bar W^r J_r{}_\rb{}^j
	+ i \bar {\l}^{\dalpha r}_k D_r{}^{jk}
	\Big)
	\eol & \quad
	+ \!(\bar\psi_{m}^j \bsigma^m)^\alpha \!\Big(
	\frac{1}{4}  \widehat \nabla_{\alpha\dalpha} \bar\z^\dalpha_\rb \, A^\rb{}_j
	\!- \frac{i}{4} W_\alpha{}^\beta \z_{\beta \rb} \!A^\rb{}_j
	\!+ \!\frac{3i}{4} \chi_{\alpha j} K
	\!- \frac{i}{4} \z_{\alpha \rb} W^r J_r^\rb{}_j
	\!- i {\l}_{\alpha r}^k D_r{}_{jk}\!
	\Big).
\end{align}
This agrees with the same result derived in projective superspace \cite{Butter:CSG4d.Proj}
and can be compared with the original component reference \cite{dWKV}.

\section{Further applications and outlook}

The main goal of this paper has been to construct a covariant formulation of harmonic
superspace based on the geometry of conformal superspace. This included
(i) the construction of invariant actions in a general gauge; (ii) the specification
of how to pass covariantly between full harmonic and analytic superspaces; and (iii) the explicit
component reduction formula in the central basis. One important task which we have not
pursued is the complete specification of the vielbeins and connections in the analytic
basis in terms of the prepotentials. This was the main result of
\cite{DelamotteKaplan:HHS, GKS:Sugra}. Although that approach used a different
superspace (equivalent to conformal supergravity coupled to vector and non-linear
multiplet compensators), one could follow essentially identical steps within conformal
superspace to arrive at comparable results. The reverse is also true: within the
framework of \cite{DelamotteKaplan:HHS, GKS:Sugra}, one could derive e.g. the
explicit component reduction rule in the central basis. We have sketched how this could
be accomplished by introducing compensated derivatives. The explicit expressions are
expected to be substantially more complicated due to the presence of
additional torsion tensors, including a dimension-1/2 spinor superfield.
Of course, when the action in question is superconformal, all dependence on the compensators
must drop out of the final component action; the advantage of conformal
superspace is that such extraneous objects are avoided from the beginning,
leading to vastly simpler computations.

Although we have chosen to limit the scope of this investigation to the above goal
and the sample calculation in section \ref{sec:SigmaModels},
a number of applications become immediately apparent.
In the remainder of this concluding section, we will briefly sketch several proposals.

\subsection*{Projective superspace and prepotentials}
One particular aspect we have not explored is the full connection between
the curved harmonic and projective superspaces, which was in part a driving motivation
of this research. Certainly in the central basis
the mapping sketched in the introduction between complex harmonic
and projective superfields must hold.
A more interesting question is how to use this mapping to convert harmonic supergravity prepotentials into
projective ones. For gauge prepotentials, this was sketched by Jain and Siegel \cite{JainSiegel:HarmProj}.
The analytic gauge prepotential $V$ in projective superspace is given by \cite{LR90} 
\begin{align}\label{eq:ProjV}
e^{V} = e^{-V_S} e^{V_N} 
\end{align}
in terms of two bridge superfields.\footnote{See also
\cite{KT-M:N4_SYM_AdS3} and \cite{CSG5d} for recent discussions in curved 3D and 5D projective
superspace.} $V_N$ is the arctic multiplet bridge, converting
analytic arctic multiplets into covariant ones, while $V_S$ is the antarctic bridge.
$V_N$ is well-defined near the north pole and $V_S$ near the south pole, but neither is
itself analytic. Using the mapping between harmonic and projective multiplets, 
$V_N$ and $V_S$ can be written in terms of the complex harmonic bridge $\cB$ as
\begin{align}
V_N(v) = i \cB(v,w)\vert_{w_i^- = (1,0)}~, \qquad
V_S(v) = i \cB(v,w)\vert_{w_i^- = (0,1)}~.
\end{align}
It is easy to confirm that while $V_N$ and $V_S$ are not themselves analytic, the
combination \eqref{eq:ProjV} is and transforms appropriately under the arctic and
antarctic $\lambda$-groups.
Extending this idea to the supergravity prepotentials would seem to be reasonably
straightforward. For example, the harmonic bridges for the coordinates should similarly
decompose into separate arctic and antarctic bridges. However, a more explicit
investigation is warranted.

\subsection*{Quaternion-K\"ahler superspace}
One of the many successes of harmonic superspace is its very elegant
description of general hypermultiplet systems, including supergravity-matter couplings.
Rigid harmonic superspace admits general off-shell hypermultiplet actions,
with the harmonic potential encoding the sigma model geometry.
In a similar way, curved harmonic superspace admits off-shell hypermultiplet
actions, with the potential function(s) encoding the sigma
model \cite{GIOS:Sugra, BGIO, GIO:QK}.

There are two existing formulations.
The first, following the same conventions as in section \ref{sec:SigmaModels}, involves the Lagrangian
\begin{align}\label{eq:HarmHKC}
\mathscr{L}^{(2,2)} = \frac{1}{2} Q_\ra^+ \nabla_\trw^{++} Q^{\ra+} + H^{(2,2)}(Q^+, w^-)
\end{align}
with superfields $Q^{\ra+}$ with $\ra = 1, \cdots, 2(n+1)$
describing the sigma model of a $4(n+1)$-dimensional hyperk\"ahler cone
coupled to conformal supergravity. Each of the superfields has Weyl weight one and
charge $(1,0)$ under $\gU{1}_\trv \times \gU{1}_\trw$. As is well-known, a $4(n+1)$-dimensional
hyperk\"ahler cone is in one-to-one correspondence with a $4n$-dimensional quaternion-K\"ahler
manifold \cite{Swann} (see also \cite{dWRV}).
This structure can be made more apparent in superspace, which leads to the second formulation
\cite{BGIO}.
One makes the field redefinition $Q^{\ra+} \rightarrow (\omega \cQ^{\ra+}, q^{i+})$,
where the superfields $q^{i+}$ have Weyl weight one and charge $(1,0)$,
$\cQ^{\ra+}$, $\ra=1, \cdots, 2n$, have vanishing Weyl
weight and charge $(0,1)$, and $\omega := q^{i+} w_i^-$.
The Lagrangian becomes
\begin{align}\label{eq:HarmQK}
\mathscr{L}^{(2,2)} = \frac{1}{2} q_i^+ \nabla_\trw^{++} q^{i +}
	+ \omega^2 \Big[
	\frac{1}{2} \cQ_\ra^+ \nabla_\trw^{++} \cQ^{\ra+}
	+ H^{+4}(\cQ^+, q^{i+} / \omega, w_i^-)
	\Big]~.
\end{align}
The term within square braces has charge $(0,4)$ and vanishing Weyl weight.
The hypermultiplet $q^{i+}$ is interpreted as a compensating multiplet while $\cQ^{\ra+}$ are the matter
multiplets. This is equivalent (up to a gauge choice and a sign change in the compensator
kinetic term) to the original form given in \cite{GIOS:Sugra}. It was shown
in \cite{GIO:QK} that the general quaternion-K\"ahler sigma model
is described by two prepotentials ${\cL_\ra}^+$ and ${\cL}^{+4}$, and
the action above is just the special gauge $\cL_\ra^+ = \cQ_\ra^+$
(with $\cL^{+4} \rightarrow H^{+4}$) of this general form.

The two expressions \eqref{eq:HarmHKC} and \eqref{eq:HarmQK}
bear a striking similarity to analogous formulae involving chiral superfields in $\cN=1$ superspace.
There the action for a $2 (n+1)$-dimensional K\"ahler cone sigma model
coupled to conformal supergravity involves the superspace Lagrangian
\begin{align}
\mathscr{L} = K(\Phi^\ra, \bar \Phi^{\bar \ra})
\end{align}
with chiral superfields $\Phi^\ra$ for $\ra = 1, \cdots, n+1$. The function $K$ is the
K\"ahler potential for a K\"ahler cone. Reorganizing the chiral superfields as
$\Phi^\ra \rightarrow (\phi \,\varphi^\ra, \phi)$ where $\phi$ has Weyl weight one and
$\varphi^\ra$, $\ra=1,\cdots, n$ are weight zero, the Lagrangian can be rewritten
\begin{align}\label{eq:HK1}
\mathscr{L} = -3 \,\phi \bar\phi \,e^{-\cK/3}
\end{align}
where $\cK$ is the potential for a $2n$-dimensional Hodge-K\"ahler manifold.
The fields $\varphi^\ra$ are matter fields and $\phi$ is a chiral compensator.
The usual factor of $3$ is chosen for convenience and the negative sign is consistent
with the role of $\phi$ as a compensator.

The $\cN=1$ action actually has a third form, which is in many applications more useful.
Recall that the component reduction of \eqref{eq:HK1} still requires an inconvenient set of super-Weyl
gauge choices to be imposed to canonically normalize the graviton and gravitino
actions (see e.g. the discussion in \cite{WessBagger}). This can be directly addressed
at the superfield level by absorbing the Hodge-K\"ahler potential into the
superspace vielbein \cite{BGGM, BGG}. Equivalently, one can interpret $\cK$ as if it were the
prepotential for some additional $\gU{1}_\cK$ symmetry under which the compensator $\phi$ is charged.
Moving to covariantly chiral superfields $\phi$, the Lagrangian simplifies to
\begin{align}\label{eq:HK2}
\cL = -3 \,\phi \bar \phi~.
\end{align}
Now the matter fields are encoded within the composite $\gU{1}_\cK$ connection,
and the matter action appears within the composite auxiliary field $D$ associated
with $\gU{1}_\cK$. Adopting the gauge $\phi = 1$
fixes the Weyl gauge and a linear combination of $\gU{1}_R$ and $\gU{1}_\cK$, leaving
another linear combination as the composite K\"ahler $\gU{1}$ symmetry of the physical
action. This geometrizes the K\"ahler potential, ensuring that the $\gU{1}_R$ gauge field of
supergravity is identified with the composite $\gU{1}$ potential of the K\"ahler line
bundle while also canonically normalizing all terms in the component action.
The resulting superspace is known as \emph{K\"ahler superspace} \cite{BGGM, BGG}
and provides a general framework for handling $\cN=1$ supergravity-matter
systems, even including higher-derivative couplings in a K\"ahler covariant form.
For example, a simple class of such higher-derivative terms is given by a
full superspace integral of
\begin{align}\label{eq:HDKahler}
\mathscr L = c_1 R \bar R + c_2 G^a G_a + c_3 \cK^{\alpha \dalpha} \cK_{\alpha \dalpha}
\end{align}
for real constants $c_i$, where $R$ and $G_a$ are the torsion superfields of K\"ahler superspace and
$\cK_{\alpha\dalpha} = \cK_{\ra \bar \rb} \cD_\alpha \varphi^\ra \bar \cD_\dalpha \bar\varphi^{\bar \rb}$.
Writing such K\"ahler-covariant terms in the original superspace frame is significantly
more complicated.

In light of these observations, it is natural to conjecture a third formulation of
$\cN=2$ superspace corresponding to a slight reformulation of \eqref{eq:HarmQK},
just as \eqref{eq:HK2} reformulates \eqref{eq:HK1}. The idea is that \eqref{eq:HarmQK} can be rewritten
\begin{align}
\mathscr{L}^{(2,2)} = \frac{1}{2} q_i^+ (\nabla_\trw^{++} q^{i +} - V_\trw{}^i{}_j q^{j+})
\end{align}
where $V_\trw{}^i{}_j$ is a composite prepotential for an additional ``matter''
$\gSU{2}$ group rotating the $q^{i+}$; this is known as the Pauli-G\"ursey
group $\gSU{2}_{\rm PG}$ \cite{GIOS}.
In the central basis for $\gSU{2}_{\rm PG}$, the equations
of motion set $q^{i+}$ to be a covariant $\cO(1)$ multiplet, $q^{i+} = f^i{}_j v^{j+}$.
The Weyl gauge along with a linear combination of $\gSU{2}_R$ and $\SU{2}_{\rm PG}$
is fixed by taking $f^i{}_j = \delta^i{}_j$, and the other linear combination of
$\gSU{2}_R$ and $\gSU{2}_{\rm PG}$ survives as the composite $\gSU{2}$ symmetry
of the quaternion-K\"ahler manifold. This strongly suggests that one can construct
a \emph{quaternion-K\"ahler superspace} as the $\cN=2$ analogue of K\"ahler superspace.
We intend to explore this subject in the near future.

\subsection*{Higher-derivative terms}
The advantage of a covariant approach is the ease of component reductions, including
all couplings to supergravity. We demonstrated this by deriving the general two-derivative
hyperk\"ahler cone action, but it would be plausible to address higher-derivative actions
as well. Large classes of these have been discussed recently in projective
superspace \cite{BK:HD} and comparable calculations could undoubtedly be pursued within
the harmonic approach (see e.g. the rigid higher-derivative terms of \cite{AABE:HDHarm}).

If indeed one can construct a quaternion-K\"ahler superspace as the $\cN=2$ analogue
of K\"ahler superspace, it would undoubtedly provide the natural framework for addressing
higher-derivative terms involving hypermultiplets. For example, in the gauge-fixed
formulation of $\cN=2$ superspace, there exist torsion superfields $S^{ij}$ and $G_a{}^{ij}$.
It is plausible that from these one could construct a higher-derivative harmonic
superspace Lagrangian analogous to \eqref{eq:HDKahler}, corresponding to a new curvature-squared invariant.

It was shown recently in \cite{BudWKuLo} that a certain Ricci-squared invariant
could be constructed, given by the chiral superspace Lagrangian,
$\mathscr{L}_c = \frac{1}{6} \bar \cD^{ij} \bar S_{ij}
	+ \bar S^{ij} \bar S_{ij} + \bar Y_{\dalpha\dbeta} \bar Y^{\dalpha \dbeta}$,
in $\gSU{2}$ superspace. This is compensator-independent and, when combined with the
known Weyl-squared invariant, gives the $\cN=2$ Gauss-Bonnet.
Compensator-dependent higher-derivative invariants have been constructed e.g. using the trick of building
composite vector multiplets out of fundamental tensor multiplets
\cite{dWPvP:ImpTensor, Siegel:OffShellTensor, dWS:Tensor, BK:HD}.
Because a tensor multiplet is dual to a general $Q^+$ hypermultiplet, it is
possible that such terms may be constructed for general quaternionic-K\"ahler manifolds;
if so, one might be able to construct the $\cN=2$ analogue of one of the $\cN=1$
invariants \eqref{eq:HDKahler}. Its form in the proposed
quaternion-K\"ahler superspace might be particularly elegant.

\subsection*{Supergravity prepotentials and higher derivative terms}

Finally, we should mention that one important application of having both
a covariant form of harmonic superspace as well as the analytic basis prepotentials
is that one could more easily find \emph{harmonic-independent} prepotentials.
It was shown in \cite{KuTh} that the fundamental scalar prepotential $\cH$ of
$\cN=2$ conformal supergravity could be uncovered in this way, and one could
analyze how it appears within the central-basis superspace vielbeins, mirroring the harmonic
construction of \cite{GKS:Sugra}.
This would be useful e.g. for understanding the supercurrents of higher-derivative
Lagrangians such as the ones discussed above.

\vspace{1cm}

\section*{Acknowledgements}
I would like to thank Sergei Kuzenko, Gabriele Tartaglino-Mazzucchelli, and
Evgeny Ivanov for valuable comments on the draft.
This work was supported in part by the ERC Advanced Grant no. 246974,
{\it ``Supersymmetry: a window to non-perturbative physics''}
and by the European Commission Marie Curie International Incoming Fellowship grant no.
PIIF-GA-2012-627976.	

\vspace{1cm}

\appendix

\section{Analytic integrals and densities}\label{App:AI&D}
This appendix is a summary and continuation of appendix B of \cite{Butter:CSG4d.Proj},
which addressed covariant integration over supermanifolds.
Let us recall the basics. A supermanifold $\cM$ (without boundary) of dimension $D$ possesses
local coordinates $z^M$, $M=1, \ldots, D$, a vielbein $E_M{}^A$, and separate
connection $H_M{}^{\ul a}$ associated with internal symmetries, which we denote $\cH$.
Under diffeomorphisms both connections transform as one-forms, while under $\cH$-gauge transformations,
\begin{align}
\delta_\cH E_M{}^A = E_M{}^B g^{\ul c} f_{\ul c B}{}^A~, \qquad
\delta_\cH H_M{}^{\ul a} = \pa_M g^{\ul a} + E_M{}^B g^{\ul c} f_{\ul c B}{}^{\ul a}
	+ H_M{}^{\ul b} g^{\ul c} f_{\ul c \ul b}{}^{\ul a}~.
\end{align}
The parameters $f$ are structure constants of a soft algebra including the covariant
curvatures associated with the vielbein and $\cH$-connection (see e.g. the discussion
in \cite{CSG4d_2}). When diffeomorphisms are covariantized with the $\cH$-connection,
the full transformation rules become
\begin{align}
\delta E_M{}^A &=
	\pa_M \xi^A + H_M{}^{\ul b} \xi^{C} f_{C \ul b}{}^A
	+ E_M{}^B g^{\ul c} f_{\ul c B}{}^A
	+ E_M{}^B \xi^C T_{CB}{}^A~, \eol
\delta H_M{}^{\ul a} &= \pa_M g^{\ul a} + E_M{}^B g^{\ul c} f_{\ul c B}{}^{\ul a}
	+ H_M{}^{\ul b} g^{\ul c} f_{\ul c \ul b}{}^{\ul a}
	+ H_M{}^{\ul b} \xi^{C} f_{C \ul b}{}^{\ul a}
	+ E_M{}^B \xi^C R_{CB}{}^{\ul a}~.
\end{align}
An action over the full supermanifold,
$\int \rd^D z\, E\, \mathscr{L}$,
is invariant provided $\mathscr{L}$ is a scalar under diffeomorphisms
and transforms under $\cH$ as
$\delta_\cH \mathscr{L} = - (-)^A g^{\ul b} f_{\ul b A}{}^A\, \mathscr{L}$.

\subsection{Analytic submanifolds}
We are interested in an analytic submanifold $\mathfrak{M}$ (without boundary)
of dimension $d$ with local
coordinates $\mathfrak{z}^m$, $m=1, \ldots, d$. We have in mind a situation where
the original coordinates can be decomposed as $z^M = (\mathfrak{z}^m, y^\mu)$ with the submanifold
$\mathfrak{M}$ corresponding to the surface parametrized by $\mathfrak{z}^m$
with (for example) $y^\mu = 0$. The coordinates $\mathfrak{z}^m$ and $y^\mu$ may be bosonic
or fermionic; we denote the grading of a coordinate $z^M$ by $(-)^M$.
The vielbein and its inverse are given by
\begin{align}
E_M{}^A =
\begin{pmatrix}
\cE_m{}^a & E_m{}^\alpha \\
E_\mu{}^a & E_\mu{}^\alpha
\end{pmatrix}~, \qquad
E_A{}^M =
\begin{pmatrix}
E_a{}^m & E_a{}^\mu \\
E_\alpha{}^m & \phi_\alpha{}^\mu
\end{pmatrix}~, \qquad
\end{align}
with the assumption that both $\cE_m{}^a$ and $\phi_\alpha{}^\mu$ are invertible,
with inverses $\cE_a{}^m$ and $\phi_\mu{}^\alpha$, respectively.
This allows one to compactly specify all the remaining components of the vielbein
and its inverse in terms of these quantities, $E_m{}^\alpha$, and $E_\alpha{}^m$:
\begin{align}
E_M{}^A &=
\left(\begin{array}{c|c}
\cE_m{}^a & E_m{}^\alpha \\ \hline
- \phi_\mu{}^\beta E_\beta{}^n \cE_n{}^a & \phi_\mu{}^\alpha - \phi_\mu{}^\beta E_\beta{}^n E_n{}^\alpha
\end{array}\right)~, \eol
E_A{}^M &=
\left(\begin{array}{c|c}
\cE_a{}^m - \cE_a{}^n E_n{}^\beta E_\beta{}^m &
	-\cE_a{}^n E_n{}^\beta \phi_\beta{}^\mu \\ \hline
E_\alpha{}^m & \phi_\alpha{}^\mu
\end{array}\right)~.
\end{align}
No assumptions have been made about $E_m{}^\alpha$ or $E_{\alpha}{}^m$.
We treat $\cE_m{}^a$ as the vielbein of the submanifold $\mathfrak M$. In
particular, the class of diffeomorphisms acting on $\mathfrak{z}^m$
induce
\begin{align}\label{eq:DiffE}
\delta \cE_m{}^a = \pa_m \xi^n \cE_n{}^a + \xi^n \pa_n \cE_m{}^a~, \qquad
\xi^M = (\xi^m, 0)
\end{align}
as required for a vielbein. This formula holds even if $\xi^m$ depends on $y^\mu$.

The submanifold is an analytic submanifold (although not yet in the analytic basis)
if the following properties are satisfied.
Under $\cH$-gauge transformations and \emph{covariant} diffeomorphisms generated by
$\xi^A = (0, \xi^\alpha)$, the analytic vielbein transforms into itself via
\begin{align}\label{eq:AVBTrans}
\delta \cE_m{}^a = \cE_m{}^b \xi^\gamma T_{\gamma b}{}^a
	+ \cE_m{}^b g^{\ul c} f_{\ul c b}{}^a~, \qquad \xi^A = (0, \xi^\alpha)~.
\end{align}
These conditions derive from the transformation rules of $E_M{}^A$ assuming
the vanishing of the torsion tensor
$T_{\gamma\beta}{}^a$ and the structure constants $f_{\ul c \beta}{}^a$, which
permit the existence of superfields annihilated by $\nabla_\alpha$.
It is convenient to decompose the full set of possible transformations
into the diffeomorphisms on $\mathfrak M$ with $\xi^M = (\xi^m, 0)$,
the covariant diffeomorphisms generated by $\xi^A = (0, \xi^\alpha)$,
and the $\cH$-gauge transformations.
This is always possible to do using the invertibility of $\phi_\alpha{}^\mu$
and $\cE_m{}^a$.

We will not actually require that all analytic superfields be annihilated by each
of the $\nabla_\alpha$. We have in mind the situation where some of the
$\nabla_\alpha$ have an interpretation as charge generators. In harmonic superspace,
these would be $\nabla_\trv^0$ and $\nabla_\trw^0$.
So we instead call an analytic superfield $\Psi$ one for which
\begin{align}\label{eq:AnalyticPsi}
\nabla_\alpha \Psi = c_\alpha^{(\Psi)} \Psi
\end{align}
where $c_\alpha^{(\Psi)}$ is a (possibly vanishing) constant number.
Only bosonic covariant derivatives may possess non-vanishing $c_\alpha^{(\Psi)}$.

Suppose now we have a scalar Lagrangian $\cL$ that obeys
\begin{align}
\nabla_\alpha \cL = -(-)^b T_{\alpha b}{}^b \, \cL \equiv c_\alpha^{(\cL)} \, \cL~, \qquad
\delta_\cH \cL = -(-)^a g^{\ul b} f_{\ul b a}{}^a\,\cL~.
\end{align}
where the expression $T_{\alpha b}{}^b (-)^b$ built from the torsion tensor
is constant (and possibly vanishing).
We may define an analytic action $\cS$ over the submanifold $\mathfrak{M}$,
\begin{align}\label{eq:submanifoldS}
\cS = \int \rd^d \mathfrak{z}\, \cE\, \cL~, \qquad \cE = \sdet \cE_m{}^a~.
\end{align}
Using \eqref{eq:DiffE} and \eqref{eq:AVBTrans}, one finds that under a
general transformation parametrized as
$\delta = \xi^m \pa_m + \xi^\alpha \nabla_\alpha + \delta_\cH$,
the integrand of \eqref{eq:submanifoldS} transforms as a total derivative,
$\delta (\cE \cL) = \pa_m (\xi^m \cE \cL)$
so the action is invariant.
In particular, the action \eqref{eq:submanifoldS} is invariant even under diffeomorphisms
in $y^\mu$. These can be interpreted as arbitrary small deformations of the embedding of
$\mathfrak{M}$ in $\cM$. In other words, the particular choice of the embedding has no
effect on the action integral.

As an example of a covariant action principle, let us take
harmonic superspace on the analytic submanifold. We group the derivatives as
\begin{align}
\nabla_a = (\nabla_a, \nabla_{\ul\alpha}^-, \nabla_\trv^{--}, \nabla_\trw^{++}) ~, \qquad
\nabla_\alpha = (\nabla_{\ul\alpha}^+, \nabla_\trv^{++}, \nabla_\trw^{--}, \nabla_\trv^0, 
	\nabla_\trw^0)~,
\end{align}
with the coordinates $\mathfrak z^m = (x^m, \q^{\ul \mu +}, \z, \tilde \z)$
parametrizing the submanifold $\mathfrak M = \cM^{4|4} \times \cS$.
From the torsion constraints, the covariant Lagrangian $\cL$ must obey
\begin{align}
\nabla_{\ul\alpha}^+ \cL = \nabla_{\trv}^{++} \cL = \nabla_\trw^{--} \cL = 0~, \qquad
\nabla_\trv^0 \cL = \nabla_\trw^0 \cL = 2~.
\end{align}
The measure $\cE$ in turn transforms under covariant $\xi^\alpha$ diffeomorphisms as
\begin{align}\label{eq:cEChargeCov}
\delta \cE = -2 (\xi_\trv^0 + \xi_\trw^0) \, \cE~.
\end{align}
For these reasons we denote the Lagrangian and measure by
$\mathscr{L}^{(2,2)}$ and $E^{(-2,-2)}$, respectively. The covariant action is
then just \eqref{eq:AnalyticSuperAction}.

Actually, even ``full'' harmonic superspace is an analytic superspace in a sense,
as we always restrict to twisted biholomorphic quantities of fixed
$\gU{1}_\trv \times \gU{1}_\trw$ charge. Now the decomposition of derivatives is
\begin{align}
\nabla_a = (\nabla_a, \nabla_{\ul\alpha}^\pm, \nabla_\trv^{--}, \nabla_\trw^{++}) ~, \qquad
\nabla_\alpha = (\nabla_\trv^{++}, \nabla_\trw^{--}, \nabla_\trv^0, 
	\nabla_\trw^0)~,
\end{align}
with the coordinates $\mathfrak z^m = (x^m, \q^{\ul \mu \pm}, \z, \tilde \z)$
on $\cM^{4|8} \times \cS$. Lagrangians obey
\begin{align}
\nabla_{\trv}^{++} \cL = \nabla_\trw^{--} \cL = 0~, \qquad \nabla_\trv^0 \cL = -2 \cL~, \qquad
\nabla_\trw^0 \cL = 2 \cL~.
\end{align}
The measure $\cE$ transforms under covariant $\xi^\alpha$ diffeomorphisms as
\begin{align}
\delta \cE = 2 (\xi_\trv^0 - \xi_\trw^0 )\, \cE~.
\end{align}
For these reasons we denote the Lagrangian and measure by
$\mathscr{L}^{(-2,2)}$ and $E^{(+2,-2)}$, respectively. The covariant action is
then just \eqref{eq:FullSuperAction}.

\subsection{Analytic gauge, densities and transformation rules}
Now let us make a special choice for the embedding.
Suppose we can adopt a basis $\hat {\mathfrak z}^m$ for the analytic coordinates where
\begin{align}
\nabla_\alpha \hat {\mathfrak z}^m = c_\alpha^{(m)} \hat {\mathfrak z}^m
\end{align}
for some (possibly vanishing) bosonic constants $c_\alpha^{(m)}$.
We will call this an analytic coordinate system, and it is equivalent to
requiring $\hat E_\alpha{}^m = c_\alpha^{(m)} \hat{\mathfrak z}^m$.
Consistency requires
\begin{align}
\nabla_\beta \hat E_\alpha{}^m = c_\beta^{(m)} c_\alpha^{(m)} \hat{\mathfrak z}^m~, \quad
\nabla_b \hat E_\alpha{}^m = c_\alpha^{(m)} \hat E_b{}^m~, \quad
\delta_\cH \hat E_\alpha{}^m = -g^{\ul b} f_{\ul b \alpha}{}^\beta \hat E_\beta{}^m = 0~.
\end{align}
We will further assume that the other coordinates $\hat y^\mu$ have been
chosen so that $\phi_\alpha{}^\mu = \phi_\alpha{}^\mu(y)$ is independent of $\mathfrak{z}^m$;
this is possible using the vanishing of the torsion tensor $T_{\alpha\beta}{}^c$
and applying Frobenius' theorem.
Now an analytic diffeomorphism is defined as a diffeomorphism preserving the
above conditions. This leads to
\begin{align}
\delta^* \hat {\mathfrak{z}}^m = - \xi^m(\hat{\mathfrak z})~, \qquad
\nabla_\alpha \hat \xi^m = c_\alpha^{(n)} \hat{\mathfrak{z}}^n \pa_n \hat \xi^m =
	c_\alpha^{(m)} \hat \xi^{m}~.
\end{align}
This condition ensures that the $\nabla_\alpha$ charge of the analytic parameter $\hat \xi^m$
matches that of the coordinate $\hat{\mathfrak{z}}^m$.

Using these assumptions, one may show that the analytic measure $\cE$ is
analytic in the sense of \eqref{eq:AnalyticPsi},
\begin{align}
\nabla_\alpha \cE = \Big(T_{\alpha b}{}^b (-)^b - \sum_m c_\alpha^{(m)} (-)^m \Big) \cE~.
\end{align}
Now it is easy to show that the \emph{Lagrangian density},
$\hat \cL = \cE \cL$ is analytic, obeying
\begin{align}
\nabla_\alpha \hat \cL = -\sum_m c_\alpha^{(m)}(-)^m\, \hat \cL~.
\end{align}

It may perhaps be useful to illustrate these points using the analytic harmonic superspace as
an example. We choose the analytic basis coordinates
$\hat{\mathfrak z}^m = (\hat x^m, \hat \q^{\ul \mu +}, \hat u_i^\pm)$.
We can choose each of these to be annihilated by $\nabla_{\ul\alpha}^+$, $\nabla_\trv^{++}$,
$\nabla_\trw^{--}$, and $\nabla_\trv^0$, while
\begin{align}
\nabla_\trw^0 \hat x^m = 0~, \qquad
\nabla_\trw^0 \hat \q^{\ul \mu +} = \hat \q^{\ul \mu+}~, \qquad
\nabla_\trw^0 \hat u_i^\pm = \pm \hat u_i^\pm~.
\end{align}
Now one can show, in contrast with \eqref{eq:cEChargeCov}, that
\begin{align}
\nabla_\trv^0 \cE = -2 \cE~, \qquad \nabla_\trw^0 \cE = +2 \cE
\end{align}
and so the Lagrangian density $\hat\cL$ now obeys
\begin{align}
\nabla_\trv^0 \hat\cL = 0~, \qquad \nabla_\trw^0 \hat\cL = 4 \hat\cL~.
\end{align}

In a similar way, we can adopt an ``analytic basis'' for full harmonic superspace:
the simplest choice is actually the central basis! Now we have
$\mathfrak z^m = (x^m, \q^{\ul \mu}{}_\imath, v^{i+}, w_i^-)$,
where each is annihilated by $\nabla_\trv^{++}$ and $\nabla_\trw^{--}$. We
easily see that $\cE$ is independent of the harmonics -- in fact, it is just
the Berezinian $\sdet E_M{}^A$ in the central basis -- and the action integral
becomes \eqref{eq:FullSuperActionCentral}. From this perspective, the
measure factor $\cV^{++} \wedge \cW^{--}$ is just the anholonomic measure associated
with the constrained coordinates $v^{i+}$ and $w_i^-$.

\subsection{Rules for total derivatives}
We will need some general rules for integrating total derivatives.
It is a simple exercise to show that for the full supermanifold $\cM$,
\begin{align}\label{eq:TDRule1}
\int \rd^D z\, E\, \nabla_A \cV^A (-)^A
	= \int \rd^D z\, \Big[
		\nabla_M \big(E \cV^A E_A{}^M\big) (-)^M
		- E \,\cV^A T_{A B}{}^B (-)^B\, 
		\Big]~.
\end{align}
The term involving $\nabla_M$ may be decomposed into $\pa_M$, which may
be discarded, and a connection piece. The connection piece may be
non-trivial as $E \cV^A E_A{}^M$ might transform under some of the $\cH$ group.
We are really interested in integrals on the analytic submanifold $\mathfrak M$, where
\begin{align}\label{eq:TDRule2}
\int \rd^d \mathfrak{z}\, \cE\, \nabla_a \cV^a (-)^a
	&= \int \rd^d\, \mathfrak{z} \Big[
	\nabla_m \big(\cE \cV^a \cE_a{}^m\big) (-)^m
	- \cE \,\cV^a T_{a b}{}^b (-)^b\, 
	\eol & \qquad
	- \cE \cE_a{}^m E_m{}^\alpha \big(
	\nabla_\alpha \cV^a
	+ T_{\alpha b}{}^b \cV^a (-)^b
	+ \cV^b T_{b \alpha}{}^a
	\big) (-)^a
	\Big]~.
\end{align}
Again $\nabla_m$ may be decomposed into $\pa_m$, which may be
discarded, and a connection piece.

The expression \eqref{eq:TDRule2} can be a bit unwieldy, so a few examples should help.
Let us take full harmonic superspace, which is the analytic superspace
of twisted biholomorphic fields with fixed charges. We choose the non-vanishing components
of $\cV^A$ to be $\cV^{\trw --}$ and $\cV^{\trv ++}$. Keeping in mind
that the Lagrangian in this case must carry charge $(-2,2)$, these components
may be labeled $\cV^{(-2,0)}$ and $\cV^{(0,2)}$ (up to a sign). 
The Lagrangian is
$\mathscr{L}^{(-2,2)} = \nabla_\trw^{++} \cV^{(-2,0)} + \nabla_\trv^{--} \cV^{(0,2)}$.
These are both valid integrands provided $\cV^{(-2,0)}$
and $\cV^{(0,2)}$ are twisted biholomorphic and invariant with respect to
the gauge symmetries. Using \eqref{eq:TDRule2}, it is easy to show that
this is a total derivative.

As another example, we take analytic
superspace and choose a single non-vanishing component $\cV^{\trw --}$.
Because the Lagrangian now must have charge $(2,2)$, this component may
be labeled $\cV^{(2,0)}$. The Lagrangian
$\mathscr{L}^{(2,2)} = \nabla_\trw^{++} \cV^{(2, 0)}$
is covariant provided $\cV^{(2,0)}$ is a twisted biholomorphic
analytic primary. It is easy to check this is a total derivative.
In contrast, the expression $\nabla_\trv^{--} \cV^{(4, 2)}$ is not
even generically a covariant Lagrangian because it is not annihilated
by $\nabla_\trv^{++}$.

\bibliography{library.bib}
\bibliographystyle{utphys_mod}

\end{document}

%% file: sugracom.tex
\newcommand {\cA}{{\cal A}}
\newcommand {\cB}{{\cal B}}
\newcommand {\cC}{{\cal C}}
\newcommand {\cD}{{\cal D}}
\newcommand {\cE}{{\cal E}}
\newcommand {\cF}{{\cal F}}
\newcommand {\cG}{{\cal G}}
\newcommand {\cH}{{\cal H}}

\newcommand {\cJ}{{\cal J}}
\newcommand {\cK}{{\cal K}}
\newcommand {\cL}{{\cal L}}
\newcommand {\cM}{{\cal M}}
\newcommand {\cN}{{\cal N}}
\newcommand {\cO}{{\cal O}}

\newcommand {\cQ}{{\cal Q}}
\newcommand {\cR}{{\cal R}}
\newcommand {\cS}{{\cal S}}

\newcommand {\cU}{{\cal U}}
\newcommand {\cV}{{\cal V}}
\newcommand {\cW}{{\cal W}}

\def\G{\Gamma}

\def\l{\lambda}

\def\q{\theta}

\def\z{\zeta}

\def\L{\Lambda}

\def\U{\Upsilon}

\newcommand{\ra}{{\rm a}}
\newcommand{\rb}{{\rm b}}
\newcommand{\rc}{{\rm c}}
\newcommand{\rd}{{\rm d}}
\newcommand{\re}{{\rm e}}

\newcommand{\dalpha}{{\dot{\alpha}}}
\newcommand{\dbeta}{{\dot{\beta}}}
\newcommand{\dgamma}{{\dot{\gamma}}}

\newcommand{\1}{{\underline{1}}}
\newcommand{\2}{{\underline{2}}}

\newcommand{\eps}{\epsilon}

\newcommand{\pa}{\partial}
\newcommand{\ul}{\underline}

\newcommand{\bsigma}{{\bar \sigma}}

\DeclareMathOperator{\sdet}{sdet}

\newcommand{\HC}{{\mathrm{h.c.}}}

\newcommand{\eol}{\notag \\}

\newcommand{\sym}[1]{\stackrel{\scriptstyle#1}{\mbox{\tiny $\smile$}}}

